\documentclass[aps, a4paper,superscriptaddress, nofootinbib, showpacs, 12pt]{revtex4}

\usepackage{amsmath}
\allowdisplaybreaks
\usepackage{amssymb}
\usepackage{latexsym}
\usepackage{amsfonts}
\usepackage{graphicx}
\usepackage{ bm, natbib}
\usepackage{dsfont}
\usepackage{graphicx}
\usepackage{slashed}
\usepackage{ulem}
\usepackage{accents}
\usepackage{bigstrut}

\usepackage{epsfig}
\usepackage{color}

\usepackage{multirow} 
\usepackage{mathrsfs}

\makeatletter
\newcommand*{\rom}[1]{\expandafter\@slowromancap\romannumeral #1@}
\makeatother

\def\I{\mathrm{i}}

\newcommand{\bea}{\begin{eqnarray}}
\newcommand{\eea}{\end{eqnarray}}

\newcommand{\be}{\begin{equation}}
\newcommand{\ee}{\end{equation}}

\begin{document}

\title{High scale mixing relations as a natural explanation for large neutrino mixing}

\author{Gauhar Abbas}
\email{Gauhar.Abbas@ific.uv.es}
\affiliation{IFIC, Universitat de Val\`encia -- CSIC, Apt. Correus 22085, 
E-46071 Val\`encia, Spain}
\author{Mehran Zahiri Abyaneh}
\email{Mehran.Za@ific.uv.es}
\affiliation{IFIC, Universitat de Val\`encia -- CSIC, Apt. Correus 22085, 
E-46071 Val\`encia, Spain}
\author{Aritra Biswas}
\email{aritrab@imsc.res.in}
\affiliation{The Institute of Mathematical Sciences, Chennai 600 113, India}
\author{Saurabh Gupta}
\email{saurabh@if.usp.br}
\affiliation{Instituto de F\'\i sica, Universidade de S\~ao Paulo,
C. Postal 66318,  05314-970 S\~ao Paulo, Brazil}
\author{Monalisa Patra}
\email{monalisa@physik.uzh.ch}
\affiliation{Physik-Institut, Universit\"{a}t Z\"{u}rich, CH-8057 Z\"{u}rich, Switzerland}
\author{G. Rajasekaran}

\affiliation{The Institute of Mathematical Sciences, Chennai 600 113, India}
\affiliation{Chennai Mathematical Institute, Siruseri 603 103, India}
\author{Rahul Srivastava}
\email{rahuls@imsc.res.in}
\affiliation{The Institute of Mathematical Sciences, Chennai 600 113, India}

%%%%%%%%%%%%%%%%%%%%%%%%%%%%%%%%%%%%%%%%%%%%%%%%%%%%%%%%%%%%%%%%%%%%%%%%%%%%%%%%%%%%%%%%%%%%%%%%%%%%%%%%%%%%%%%%%%%%%%%%%%%%%%%%%%%%%%%%%%%%%%%%%%%%%%%%%%%%%%%%%%%%%%%%%%%%%%%%%%%%%%%%%%%%%%%%%%%%%%%%%%%%%%%%%%%%%%%%%%%%%%%%%%%%%%%%%%%%%%%%%%%%%%%%%%%%%%%%%%%%%%

\begin{abstract}
The origin of small mixing among the quarks and a large mixing among the neutrinos 
has been an open question in particle physics. In order to answer this question,
we postulate general relations among the quarks and the leptonic mixing angles 
at a high scale, which could be the scale of Grand Unified Theories. The central
idea of these relations is that the quark and the leptonic mixing angles can be 
unified at some  high scale either due to some quark-lepton symmetry or some other 
underlying mechanism and as a consequence, the mixing angles of the leptonic sector 
are proportional to that of the quark sector. We investigate the phenomenology of the
possible relations where the leptonic mixing angles are proportional to the quark 
mixing angles at the unification scale by  taking into account the latest experimental
constraints from the neutrino sector.  These relations are able to explain the pattern
of leptonic mixing at the low scale and thereby hint that these relations could be 
possible signatures of a quark-lepton symmetry or some other underlying
quark-lepton mixing unification mechanism at some high scale linked to Grand 
Unified Theories.

\end{abstract}

\pacs{14.60.Pq, 11.10.Hi, 11.30.Hv, 12.15.Lk}

\maketitle

%%%%%%%%%%%%%%%%%%%%%%%%%%%%%%%%%%%%%%%%%%%%%%%%%%%%%%%%%%%%%%%%%%%%%%%%%%%%%%%%%%%%%%%%%%%%%%%%%%%%%%%%%%%%%%%%%%%%%%%%%%%%%%%%%%%%%%%%%%%%%%%%%%%%%%%%%%%%%%%%%%%%%%%%%%%%%%%%%%%%%%%%%%%%%%%%%%%%%%%%%%%%%%%%%%%%%%%%%%%%%%%%%%%%%%%%%%%%%%%%%%%%%%%%%%%%%%%%%%%%%%
\section{Introduction}
The quark mixing matrix, $V_{CKM}$, parametrizes the misalignment in the 
diagonalisation of the $up$ and $down$ type quark mass matrices.  It is well known
that $V_{CKM}$ is almost close to a unit matrix.  This implies that the quark mixing
angles are small.  On the other hand, analogous misalignment in the leptonic sector
is encoded in the neutrino mixing matrix, $U_{PMNS}$.  It turns out that $U_{PMNS}$ 
is not close to a unit matrix.  The mixing angles in the neutrino 
sector are large except $\theta_{13}$ \cite{Capozzi:2013csa,Gonzalez-Garcia:2014bfa,Forero:2014bxa}.  
The origin of small mixing among quarks and a large mixing in the neutrino sector 
poses an intriguing open question.  

Among many approaches to explain the mixing pattern of the leptons, the assumption
of family or flavor symmetries  is a popular one.  These symmetries differentiate 
among the members of different families and are usually discrete, finite and 
non-abelian, for reviews see Refs.~\cite{Ishimori:2010au,Grimus:2011fk}. 
This approach has been intensively used to study the mixing in the leptonic sector~\cite{Altarelli:2010gt,King:2013eh,Ma:2014qra,Ma:2015raa}.
In addition to the leptonic mixing, there are also considerable efforts to understand the quark 
mixing through family symmetries \cite{Holthausen:2013vba,Araki:2013rkf,Ishimori:2014jwa}.
The family symmetries can also be a built-in characteristic of the Grand Unified Theories (GUT) \cite{Lam:2014kga}.

The quark-lepton unification is one of the most attractive features of the GUT 
theories \cite{Pati:1974yy,Georgi:1974sy,Fritzsch:1974nn}. The 
GUT symmetry group contains quarks and leptons in a joint representation.  
The weak interaction properties of the quarks and the leptons therefore get correlated. Hence it is possible in these theories, to derive the origin of the small and the large
mixing in the quark and the lepton sectors respectively, along with any relation 
between them, if it exists.

There are also reasons to speculate about the quark-lepton unification even on the experimental side. 
The so-called quark-lepton complementarity (QLC) relation 
\cite{Smirnov:2004ju,Raidal:2004iw} between the leptonic mixing angle $\theta_{12}$ and the Cabibbo angle $\theta_C$ 
\begin{equation}
\label{eq1}
\theta_{12} + \theta_{C}  \approx \frac{\pi}{4},
\end{equation} 
can be a footprint of a high scale quark-lepton unification 
\cite{Smirnov:2004ju,Raidal:2004iw,Minakata:2004xt,Frampton:2004vw,Kang:2005as}. 
Another interesting observation  is  due to the recent non-zero measurement of 
leptonic mixing angle $\theta_{13}$ \cite{Abe:2011sj,Adamson:2011qu,Abe:2012tg,Ahn:2012nd,An:2012eh} which is 
\begin{equation}
\label{eq2}
\theta_{13} \approx  \frac{\theta_{C} }{\sqrt{2}}. 
\end{equation} 
This relation also hints a possible link  between the quark and leptonic mixing, and it
can be an artifact of some high scale quark-lepton symmetry in an underlying GUT theory \cite{Antusch:2012fb}.

Therefore, the present state of the measured leptonic mixing angles provide the 
theoretical motivation for a common origin of the quark and leptonic mixing at
some high scale.  In fact, the idea that the quark and lepton mixing can be 
unified at some high scale, referred to as ``high scale mixing unification" (HSMU) 
hypothesis,  was first proposed in Ref. \cite{Mohapatra:2003tw,Mohapatra:2005gs,Mohapatra:2005pw,Agarwalla:2006dj}.   
In recent studies \cite{Abbas:2014ala,Abbas:2015aaa,Srivastava:2015tza} it has been 
shown that HSMU hypothesis `naturally' leads to nonzero and a small value for
the leptonic mixing angle $\theta_{13}$ and predicts a non-maximal $\theta_{23}$ 
( cf.~\cite{Abbas:2014ala,Srivastava:2015tza} for details). This hypothesis has
been studied in the context of Dirac neutrinos as well ( cf. \cite{Abbas:2013uqh}
for details). The central idea of this hypothesis is that the quark mixing
angles become identical to that of the leptons at some high scale (referred
to as the unification scale) which is typically taken as GUT scale 
(cf.~\cite{Abbas:2013uqh,Abbas:2015aaa,Abbas:2014ala} for details).  
In other words, at the unification scale
\begin{equation}
\label{eq3}
\theta_{12}=  \theta_{12}^q ,~~\theta_{13}=  \theta_{13}^q ,~~\theta_{23} =  \theta_{23}^q ,
\end{equation}  
where $\theta_{ij}$ (with $i,j=1,2,3$) are leptonic mixing angles and $\theta_{ij}^q$ 
are the quark mixing angles. This hypothesis nicely explains the pattern of mixing in the neutrino
sector including the recent observation of nonzero and a small value of $\theta_{13}$ 
\cite{Abe:2011sj,Adamson:2011qu,Abe:2012tg,Ahn:2012nd,An:2012eh}. The large 
leptonic mixing angles at the low scale are obtained through the renormalization group (RG) 
evolution of the corresponding mixing parameters from the unification scale to the low scale.

The implementation of the HSMU hypothesis requires the minimum supersymmetric
standard model (MSSM) as an extension of the standard model (SM).  The 
working of the HSMU hypothesis is as follows. We first evolve the quark 
mixing angles from the low scale (mass of the $Z$ boson) to  the 
supersymmetry (SUSY) breaking scale using the SM RG equations.  After that, 
from the SUSY breaking scale to the unification scale, evolution of 
quark mixing angles is governed by the MSSM RG equations.   In the next step, the 
quark mixing angles at the unification scale, are put equal to that of 
the neutrinos following the HSMU hypothesis.   The leptonic mixing parameters
are then run from the unification scale to the SUSY breaking scale using the 
MSSM RG equations.  From the SUSY breaking scale to the low scale, 
mixing parameters are evolved through the SM RG equations.  

In addition to SUSY, we also need a large $\tan \beta$ to realize the 
HSMU hypothesis \cite{Abbas:2014ala, Abbas:2015aaa}. The only free parameters during the 
top-down running of the leptonic mixing parameters
are masses of the three light neutrinos.  They are chosen at the unification scale 
in such a manner
that we recover all the mixing parameters at the low scale within 
the 3$\sigma$ limit of the global fit.   It turns out that the chosen masses of 
neutrinos must be quasi-degenerate (QD) and normal hierarchical \cite{Abbas:2014ala, Abbas:2015aaa}

In this work, inspired by the HSMU hypothesis, we   
postulate the most general relations among the quark and the 
leptonic mixing angles at the 
unification scale. In a compactified form the most general relation
among the leptonic and the quark mixing angles within the same generations is as following

\begin{equation}
\theta_{12} = \alpha_1^{k_1} ~\theta_{12}^q, ~~ \theta_{13} = \alpha_2^{k_2}~ \theta_{13}^q, ~~\theta_{23} =\alpha_3^{k_3}  \theta_{23}^q. 
\end{equation}
where $k_i,$ with $i=(1,2,3)$ are real exponents. We refer to this 
relation as the ``high scale mixing relation" (HSMR). We have 
chosen $(k_1,k_2,k_3)$ to be $(1,1,1)$ for the simplicity of our 
analysis. The relations within the same generations are the simplest generalization of the HSMU hypothesis.  In principle, we can also construct the most general HSMR relations among different generations completely independent of the HSMU hypothesis.  The analysis of these relations is beyond the scope of this work and could be studied elsewhere.

There will be different possibilities depending on the 
relations among the proportionality factors. We firstly list 
below the different possible cases with the maximum and the minimum 
allowed values of the three independent proportionality factors $\alpha_i$, 
\begin{align}
{\bf Case~A} &: &\theta_{12} &= \alpha_{1}^{max} ~ \theta_{12}^q, 
                &\theta_{13} &=  \alpha_{2}^{max} ~\theta_{13}^q, 
                &\theta_{23} &=\alpha_{3}^{max}  ~ \theta_{23}^q \label{caseA}, \\
{\bf Case~B} &: &\theta_{12} &=\alpha_{1}^{max} ~  \theta_{12}^q, 
                & \theta_{13} &=\alpha_{2}^{max} ~  \theta_{13}^q, 
                &\theta_{23} &= \alpha_{3}^{min} ~\theta_{23}^q \label{caseB},  \\
{\bf Case~C} &: & \theta_{12} &= \alpha_{1}^{max} ~ \theta_{12}^q,
                & \theta_{13} &=\alpha_{2}^{min}~   \theta_{13}^q, 
                & \theta_{23} &=  \alpha_{3}^{max} ~\theta_{23}^q \label{caseC}, \\
{\bf Case~D} &:& \theta_{12} &= \alpha_{1}^{max} ~ \theta_{12}^q,
               & \theta_{13} &=\alpha_{2}^{min}~   \theta_{13}^q, 
               &\theta_{23} &=  \alpha_{3}^{min} ~\theta_{23}^q \label{caseD}, \\
{\bf Case~E} &: &\theta_{12} &= \alpha_{1}^{min} ~ \theta_{12}^q, 
                & \theta_{13} &=\alpha_{2}^{max}  ~\theta_{13}^q, 
                & \theta_{23} &= \alpha_{3}^{max} ~ \theta_{23}^q \label{caseE}, \\
{\bf Case~F} &: &\theta_{12} &=\alpha_{1}^{min}  ~ \theta_{12}^q,
                & \theta_{13} &=  \alpha_{2}^{max}  ~\theta_{13}^q, 
                &\theta_{23} &=  \alpha_{1}^{min} ~\theta_{23}^q 
\label{caseF}, \\
{\bf Case~G} &: &\theta_{12} &=\alpha_{1}^{min} ~  \theta_{12}^q,
                & \theta_{13}&=  \alpha_{2}^{min} ~ \theta_{13}^q, 
                &\theta_{23} &= \alpha_{3}^{max} ~ \theta_{23}^q \label{caseG}, \\
{\bf Case~H} &: &\theta_{12} &= \alpha_{1}^{min}~  \theta_{12}^q,
                & \theta_{13}&= \alpha_{2}^{min} ~ \theta_{13}^q, 
                &\theta_{23} &= \alpha_{3}^{min} ~ \theta_{23}^q \label{caseH}. 
\end{align}
\noindent
In this work, we have presented
our results for the maximum and minimum allowed values of 
$\alpha_i$ for all the above cases, Eqs.~(\ref{caseA}-\ref{caseH}).
We then move on to scenarios, assuming relations among the
$\alpha_i$'s.  
There can be more general HSMR where two proportionality constants can be identical and the third one is different. However
we will discuss in this work more  simplified scenerios, where the 
three proportionality constants are equal.
\begin{equation}\label{eq:hsmr}
\theta_{12} = \alpha^{k_1} ~\theta_{12}^q, ~~ \theta_{13} = \alpha^{k_2}~ \theta_{13}^q, ~~\theta_{23} =\alpha^{k_3}  \theta_{23}^q. 
\end{equation}
As explained before we have restricted to values of $k_i$ as
either $0$ or $1$.  We note that the value $(k_1,k_2,k_3)=(0,0,0)$ 
will reduce HSMR to HSMU hypothesis making Eq.~(\ref{eq3}) a 
specific form of HSMR, Eq.~(\ref{eq:hsmr}). We present below 
the seven different possible cases, where the quark mixing angles 
are assumed to be proportional to the corresponding leptonic mixing angles.

\begin{align}
{\bf Case~1} &: &\theta_{12} &= \alpha ~ \theta_{12}^q, & \theta_{13} 
&=  \theta_{13}^q, &~~\theta_{23} &=  \theta_{23}^q \label{case1}, \\
{\bf Case~2} &: &\theta_{12}&=  \theta_{12}^q, & \theta_{13}&= \alpha~ \theta_{13}^q, 
&~~\theta_{23} &=  \theta_{23}^q \label{case2},  \\
{\bf Case~3} &: &\theta_{12} &=  \theta_{12}^q, & \theta_{13} &=  \theta_{13}^q, 
&~~\theta_{23} &=  \alpha~\theta_{23}^q \label{case3}, \\
{\bf Case~4} &: &\theta_{12} &=\alpha~  \theta_{12}^q, & \theta_{13} 
&=\alpha~  \theta_{13}^q, &~~\theta_{23} &=  \theta_{23}^q \label{case4}, \\
{\bf Case~5} &: &\theta_{12} &=  \theta_{12}^q, & \theta_{13} &=  \alpha ~\theta_{13}^q, 
&~~\theta_{23} &=  \alpha~\theta_{23}^q \label{case5}, \\
{\bf Case~6} &: &\theta_{12} &= \alpha~ \theta_{12}^q, & \theta_{13} &=  \theta_{13}^q, 
&~~\theta_{23} &= \alpha~ \theta_{23}^q \label{case6}, \\
{\bf Case~7} &: &\theta_{12} &=\alpha~  \theta_{12}^q, & \theta_{13} 
&=  \alpha~ \theta_{13}^q, &~~\theta_{23} &= \alpha~ \theta_{23}^q \label{case7}.
\end{align}
\noindent
The proportionality constant $\alpha$ in the above 
Eqs.~(\ref{case1}-\ref{case7}) is taken as real parameter.
We have carried out a detailed study for these cases in this work.

We  note that there exist GUT models in the literature where proportionality 
between the quark and the leptonic mixing angles are explicitly 
shown.  For example, the proportionality relation observed between 
the leptonic mixing angle $\theta_{13}$ and 
the Cabibbo angle $\theta_C$ 
in Eq.~(\ref{eq2}) can  arise naturally  in $SU(5)$ GUTs and Pati-
Salam models.  For more details, see \cite{Antusch:2012fb}.   
Further more, it is shown in Ref.  \cite{Raidal:2004iw} that 
the relations between the quark and the 
leptonic mixing angles are possible and 
they support the idea of grand unification.  However,  non-abelian 
and abelian flavor symmetries are essential to make this happen 
\cite{Raidal:2004iw} .

There is two-fold importance of the HSMR hypothesis.   The first remarkable feature is that these relations provide a very simple way to achieve a large neutrino mixing. We shall see that predictions of these relations are easily testable in present and forthcoming experiments.  The second importance is that if predictions of HSMR hypothesis are confirmed by experimets, like neutrinoless double beta decay, this would be a strong hint of quark-lepton unification at high scale. 

The plan of the paper is as follows. In section \ref{sec2}, we present the required 
RG equations for the running of the neutrino mixing parameters.  The SUSY threshold 
corrections and the neutrino mass scale are discussed in section \ref{sec3}.  The  
results are presented in section \ref{sec4} using dimensional-5 operator as well as in the framework of type-\rom{1} seasaw. In section \ref{sec5}, for the sake of illustration, we discuss two models where HSMR hypothesis can be realised.   We summarize our results and conclude in section \ref{sec6}.

\section{RG evolution of the leptonic mixing parameters}
\label{sec2}
In this section, we briefly discuss the RG evolution of the leptonic mixing parameters.The most often studied scenario is the one where the Majorana mass term for the left handed neutrinos 
is given by the lowest dimensional operator~\cite{Antusch:2003kp}
\begin{equation}\label{eq:Kappa:Babu:1993:1}
 \mathscr{L}_{\kappa} 
 =\frac{1}{4} 
 \kappa_{gf} \, \overline{\ell_\mathrm{L}^\mathrm{C}}^g_c\varepsilon^{cd} \phi_d\, 
 \, \ell_{\mathrm{L}b}^{f}\varepsilon^{ba}\phi_a  
  +\text{h.c.} 
  \;,
\end{equation}
in the SM.  In the MSSM, it is given by
\begin{equation}\label{eq:Kappa-MSSM-s}
\mathscr{L}_{\kappa}^{\mathrm{MSSM}} 
\,=\, \mathscr{W}_\mathrm{\kappa} \big|_{\theta\theta}   +\text{h.c.}
= -\tfrac{1}{4} 
  {\kappa}^{}_{gf} \, \mathds{L}^{g}_c\varepsilon^{cd}
 \mathds{h}^{(2)}_d\, 
 \, \mathds{L}_{b}^{f}\varepsilon^{ba} \mathds{h}^{(2)}_a 
 \big|_{\theta\theta}   +\text{h.c.} \;,
\end{equation}
where $\kappa_{gf}$ has mass dimension $-1$, $\ell_\mathrm{L}^C$ is the charge 
conjugate of a lepton doublet and 
\(a,b,c,d \in \{1,2\}\) are $\mathrm{SU}(2)_\mathrm{L}$ indices.  
The double-stroke letters \(\mathds{L}\) and \(\mathds{h}\)
denote the lepton doublets and the up-type Higgs superfield in the MSSM.  
Using this mass operator, we introduce 
neutrino masses in a rather model independent way since it does not depend on the underlying mass mechanism.

The evolution of the above dimensional-5 operator below the scale where it is generated
is provided by its RG equation.  The one loop equation is as follows \cite{Chankowski:1993tx,Babu:1993qv,Antusch:2001ck,Antusch:2001vn}
\begin{equation}\label{eq:BetaKappa}
 16\pi^2 \, \Dot\kappa
 \,=\,
 C\,(Y_e^\dagger Y_e)^T\,\kappa+C\,\kappa\,(Y_e^\dagger Y_e) + \hat{\mathcal{\alpha}}\,\kappa\;,
\end{equation}
where $\Dot\kappa = \frac{d\kappa}{dt}$, $t=\ln(\mu/\mu_0)$ and $\mu$ is the renormalization scale and
\begin{eqnarray}
\label{cfc}
        C &=& 1 \hphantom{-\frac{3}{2}} \;\; \text{in the MSSM}\;,
\nonumber\\
        C &=& -\frac{3}{2} \hphantom{1} \;\; \text{in the SM}\;.
\end{eqnarray}
The parameter $\hat{\mathcal{\alpha}}$ in the SM and MSSM is given by
%\begin{subequations}
\begin{eqnarray}
\label{halpha}
        \hat{\mathcal{\alpha}}_\mathrm{SM}
& = &
        -3 g_2^2 + 2 (y_\tau^2+y_\mu^2+y_e^2) +
        6 \left( y_t^2 + y_b^2 + y_c^2 + y_s^2 + y_d^2 + y_u^2 \right) 
        + \lambda \;,
\nonumber \\
        \hat{\mathcal{\alpha}}_\mathrm{MSSM}
& = &
 -\frac{6}{5} g_1^2 - 6 g_2^2 + 6 \left( y_t^2 + y_c^2 + y_u^2 \right)
 \;.
\end{eqnarray}
%\end{subequations}
The quantities \(y_f\) (\(f\in\{e,d,u\}\)) represent the Yukawa coupling matrices
of the charged leptons, down- and up-type quarks respectively, \(g_i\) ($i = 1,2$) denote
the gauge couplings and $\lambda$ is the Higgs self coupling.  For more details 
see Ref.~\cite{Antusch:2003kp}.

We are interested in the RG evolution of parameters that are the masses, the mixing
angles and the physical phases. The mixing angles and the physical phases are described
by the PMNS matrix. This matrix is parameterized as follows
\begin{eqnarray}\label{eq:StandardParametrizationU}
 U_{PMNS} & = & V \cdot U,
\end{eqnarray}
where 
\begin{equation}
 V=\left(
 \begin{array}{ccc}
 c_{12}c_{13} & s_{12}c_{13} & s_{13}e^{-\I\delta}\\
 -c_{23}s_{12}-s_{23}s_{13}c_{12}e^{\I\delta} &
 c_{23}c_{12}-s_{23}s_{13}s_{12}e^{\I\delta} & s_{23}c_{13}\\
 s_{23}s_{12}-c_{23}s_{13}c_{12}e^{\I\delta} &
 -s_{23}c_{12}-c_{23}s_{13}s_{12}e^{\I\delta} & c_{23}c_{13}
 \end{array}
 \right),
\end{equation}
and
\begin{center}
$U= \begin{pmatrix}
 e^{-\I\varphi_1/2} & 0&0 \\
   0 & e^{-\I\varphi_2/2}&0\\
   0&0&1\\ 
 \end{pmatrix}$,
\end{center}

with \(c_{ij}\) and \(s_{ij}\) defined as \(\cos\theta_{ij}\) and \(\sin\theta_{ij}\) $(i,j = 1,2,3)$, 
respectively. The quantity $\delta$ is the Dirac phase and $\varphi_1$, $\varphi_2$ are the Majorana
phases.  The global experimental status of the leptonic mixing parameter is summarized in Table~\ref{tab1}.
\begin{table}[h]
\begin{center}
\begin{tabular}{|c|c|c|}
  \hline
  Quantity                                       & Best Fit                                   &  3$\sigma$ Range\\
  \hline
  $\Delta m^2_{21}~(10^{-5}~{\rm eV}^2)$         & $7.54$                                    &  6.99 -- 8.18       \\
  $\Delta m^2_{32}~(10^{-3}~{\rm eV}^2)$         & $2.39$                                 &  2.20 -- 2.57        \\
  $\theta_{12}^{\circ}$                        & $33.71$                                   &  30.59 -- 36.81           \\
  $\theta_{23}^{\circ}$                        & $41.38$            &  37.7 -- 52.3   \\
  $\theta_{13}^{\circ}$                        & $8.8 $                                  &  7.63 -- 9.89    \\
  \hline
     \end{tabular}
\end{center}
\caption{The global fits for the neutrino mixing parameters \cite{Capozzi:2013csa}.}
 \label{tab1}
\end{table}

Here we would like to remark that the RG equations (\ref{eq:BetaKappa}) for Yukawa couplings matrices are parametrization independent. 
 The main aim is to probe if there is any connection between the quark and the leptonic mixing.  For this purpose,  we have chosen the standard parametrization which is the most studied and also commonly used in the literature.  In principle, one could use an alternative parameterization to work and test the reality of HSMR. The results can be always interpreted as a possible indication of a connection between quark and leptonic mixing.

We now summarize the RG equations used for running the leptonic mixing parameters from high to the low scale.
For a detailed discussion of these equations, see Ref.~\cite{Antusch:2003kp}.  
These equations are derived using the lowest dimensional neutrino mass operator as discussed above
and are given by the following analytical expressions \cite{Antusch:2003kp}
\begin{eqnarray} 
\label{eq:AnalyticApproxT12}
\hspace{0.0cm}
\Dot{\theta}_{12}
& = &
        -\frac{C y_\tau^2}{32\pi^2} \,
        \sin 2\theta_{12} \, s_{23}^2\, 
        \frac{
      | {m_1}\, e^{\I \varphi_1} + {m_2}\, e^{\I  \varphi_2}|^2
     }{\Delta m^2_{21} }     
        + \mathscr{O}(\theta_{13}) \;,
         \label{eq:Theta12Dot}
\end{eqnarray} 
\begin{eqnarray} 
\label{eq:AnalyticApproxT13} 
\hspace{0.0cm}
\Dot{\theta}_{13}
& = & 
        \frac{C y_\tau^2}{32\pi^2} \, 
        \sin 2\theta_{12} \, \sin 2\theta_{23} \,
        \frac{m_3}{\Delta m^2_{32} \left( 1+\zeta \right)}
        \times
\nonumber\\
&& \quad \times
        \left[
         m_1 \cos(\varphi_1-\delta) -
         \left( 1+\zeta \right) m_2 \, \cos(\varphi_2-\delta) -
         \zeta m_3 \, \cos\delta
        \right]
        +       \mathscr{O}(\theta_{13}) \;,
 \label{eq:Theta13Dot}
\end{eqnarray}
\begin{eqnarray} \label{eq:AnalyticApproxT23} 
        \Dot{\theta}_{23}
& = &
        -\frac{C y_\tau^2}{32\pi^2} \, \sin 2\theta_{23} \,
        \frac{1}{\Delta m^2_{32}} 
        \left[
         c_{12}^2 \, |m_2\, e^{\I \varphi_2} + m_3|^2 +
         s_{12}^2 \, \frac{|m_1\, e^{\I \varphi_1} + m_3|^2}{1+\zeta}
        \right] 
        \nonumber\\
        & & {}
        + \mathscr{O}(\theta_{13}) \;,
 \label{eq:Theta23Dot}
\end{eqnarray}
where $\Dot{\theta}_{ij} = \frac{d \theta_{ij}}{dt}$ (with $i,j$ = 1, 2, 3), $t = \rm{ln}(\mu/\mu_0)$, $\mu$ being the
renormalization scale and 
\begin{equation}
        \zeta
        \,:=\,
        \frac{\Delta m^2_{21}}{\Delta m^2_{32}} \;,~~~\Delta m^2_{21} := m_2^2-m_1^2,~~~\Delta m^2_{32}  := m_3^2-m_2^2 .
\end{equation}
For the masses, the results for $y_e=y_\mu=0$ but arbitrary $\theta_{13}$ are 
\begin{subequations}\label{eq:EvolutionOfMassEigenvalues}
\begin{eqnarray}
 16\pi^2\,\Dot{m}_{1}
 & = &
        \left[
         \hat{\mathcal{\alpha}} + C y_\tau^2 \left( 2 s_{12}^2 \, s_{23}^2 + F_1 \right)
        \right] m_1 \;,
\\
 16\pi^2\,\Dot{m}_2
 & = &
        \left[
         \hat{\mathcal{\alpha}} + C y_\tau^2 \left( 2 c_{12}^2 \, s_{23}^2 + F_2 \right)
        \right] m_2
 \;,
 \\
 16\pi^2\,\Dot{m}_3
 & = &
 \left[  \hat{\mathcal{\alpha}}
 +2 C y_\tau^2 \, c_{13}^2 \, c_{23}^2
 \right] m_3
 \;,
\end{eqnarray}
\end{subequations}
where $\Dot{m}_{i} = \frac{d m_i}{dt}$ ($i$ =1, 2, 3) and  \(F_1\), \(F_2\) contain 
terms proportional to $\sin\theta_{13}$,
\begin{subequations}\label{eq:Fi}
\begin{eqnarray}
        F_1 &=&
        -s_{13} \, \sin 2\theta_{12} \, \sin 2\theta_{23} \, \cos\delta +
        2 s_{13}^2 \, c_{12}^2 \, c_{23}^2 \;,
\\
        F_2 &=&
    s_{13} \, \sin 2\theta_{12} \, \sin 2\theta_{23} \, \cos\delta +
        2 s_{13}^2 \, s_{12}^2 \, c_{23}^2 \;.
\end{eqnarray}
\end{subequations}
\noindent
In this work, we are working in the CP conserving limit which means Majorana and Dirac 
phases are assumed to be zero.  Therefore, we have not provided the RG equations for them.  The non-zero phases are expected to have non-trivial impact on the parameter space.  However, this study is beyond the scope of the present work and will be presented in a future investigation. Furthermore, we also study the effect of the new physics which could generate the above dimensional-5 operator.  For this purpose, we present our analysis within the framework of type-\rom{1} seesaw. 

Now, we briefly discuss the evolution of the leptonic mixing angles.  In the SM as can
be seen from Eq.~(\ref{halpha}), only tau Yukawa coupling will dominate the evolution
which is already very small. Hence the running of the neutrino masses is governed by a
common scaling factor and the evolution of leptonic mixing angles can only be enhanced
for QD mass pattern. In the MSSM the value of tau Yukawa coupling can be larger with
respect to the value in the SM for a large value of $\tan \beta$.  Hence the evolution
of the leptonic mixing parameters can be enhanced in addition to the enhancement coming
from the QD neutrino mass pattern as discussed below.

It is interesting to note from Eqs.~(\ref{eq:AnalyticApproxT12}, \ref{eq:AnalyticApproxT13} and \ref{eq:AnalyticApproxT23}) that  the major contribution 
to RG evolution of the mixing angles arises due to following enhancement factors
\begin{eqnarray}
\Dot{\theta}_{12}  \propto \xi_1, \quad \quad 
\Dot{\theta}_{13}, \, \Dot{\theta}_{23} \propto  \xi_2,
 \label{ang1}
\end{eqnarray}
 where 
\begin{eqnarray}
\xi_1  &= & \frac{m^2}{\Delta m^2_{21}}, \quad \quad 
\xi_2 = \frac{m^2}{\Delta m^2_{32}}, 
 \label{ang2}
\end{eqnarray}
and $m$ is the average neutrino mass with $m = (m_1 + m_2 + m_3)/3$. It is clear that we need masses of 
the neutrinos to be QD to explain 
the largeness of mixing angles at the low scale.

\section{The low energy SUSY threshold corrections and the absolute neutrino mass scale}
\label{sec3}
We discuss the required low energy SUSY threshold corrections for  the mass square 
differences and the significance of the absolute neutrino mass scale in this section.
\subsection{The low energy SUSY threshold corrections}
It is well established in the previous works on HSMU hypothesis that among the five 
mixing parameters, one of the mass square differences ($\Delta m^2_{21}$) lies
outside the 3$\sigma$ global range \cite{Mohapatra:2003tw,Mohapatra:2005gs,Mohapatra:2005pw,Agarwalla:2006dj,Abbas:2014ala}.
As shown in the previous works, this mass square difference can be brought well within the 
3$\sigma$ global limit, if the low energy SUSY threshold corrections are incorporated to the mass 
square differences \cite{Mohapatra:2003tw,Mohapatra:2005gs,Mohapatra:2005pw,Agarwalla:2006dj,Abbas:2014ala}. 
The importance of SUSY threshold corrections for  QD neutrinos is discussed in 
Refs.~\cite{Chun:1999vb,Chankowski:2000fp,Chun:2001kh,Chankowski:2001hx}. These 
corrections are given by the following equations \cite{Mohapatra:2005gs}
\begin{eqnarray} 
(\Delta m_{21}^2)_{th}
&=& 2\rm m^2\cos 2\theta_{12}[-2T_e + T_\mu + T_\tau], \nonumber \\
(\Delta m_{32}^2)_{th}
&=& 2\rm m^2\sin^2\theta_{12}[-2T_e + T_\mu + T_\tau], \nonumber \\
(\Delta m_{31}^2)_{th}
&=& 2\rm m^2\cos^2\theta_{12}[-2T_e + T_\mu + T_\tau].
\end{eqnarray}
where  $m$ is the mean mass of the QD neutrinos
and the one loop factor $T_{\hat{\alpha}} (\hat{\alpha} = e, \mu, \tau)$ is given by \cite{Chankowski:2001hx,Chun:1999vb}
\begin{eqnarray}
T_{\hat{\alpha}}&=& \frac{g^2_2}{32\pi^2} \left[\frac{x_{\mu}^2-x_{\hat{\alpha}}^2}{y_{\mu}y_{\hat{\alpha}}} 
+ \frac{(y_{\hat{\alpha}}^2-1)}{ y_{\hat{\alpha}}^2}ln(x_{\hat{\alpha}}^2) 
%\right. \nonumber \\
%&-& \left. 
- \frac{(y_{\mu}^2-1)}{y_{\mu}^2} ln(x_{\mu}^2) \right],
\label{eq23}
\end{eqnarray}
where $g_2$ is the $SU(2)$ coupling constant and $y_{\hat{\alpha}} = 1-x_{\hat{\alpha}}^2$ with 
$x_{\hat{\alpha}} = M_{\hat{\alpha}}/M_{\tilde w}$; $M_{\tilde w}$ stands for wino mass, $M_{\hat{\alpha}}$ 
represents the mass of charged sleptons.  We work with an inverted hierarchy in the charged-slepton sector
where the mass of selectron is defined through the ratio $R = \frac{M_{\tilde e}}{M_{\tilde \mu, \tilde \tau}}$.  
The mass of the wino is chosen to be $400$ GeV following the direct searches at the LHC \cite{Craig:2013cxa}.

\subsection{The absolute neutrino mass scale}
The scale of the neutrino mass is one of the open questions, ever since it has been confirmed that 
the neutrinos are massive. In case of QD and the normal hierarchical spectra, we have
\begin{equation}
m_{1} \lesssim m_{2} \lesssim m_{3} \simeq m_{0}
\end{equation}
with
\begin{equation}
m_{0}
\gg
\sqrt{\Delta{m}^{2}_{32}} \approx 5 \times 10^{-2} \, \text{eV}
.
\label{mQD}
\end{equation}
There are three complementary ways to measure the neutrino mass scale.  The first one, a model independent method, is to use the kinematics of $\beta$-decay to 
determine the effective electron (anti) neutrino mass ($m_\beta$).  It is given by 
\begin{equation} 
\label{mbeta}
m_\beta \equiv \sqrt{\sum |U_{ei}|^2 \, m_i^2 } \, . 
\end{equation}
The $m_\beta$ has an upper bound of $2$ eV from tritium beta decay \cite{Kraus:2004zw,Aseev:2011dq}.  
In future, the KATRIN experiment has sensitivity to probe $m_\beta$ as low as $0.2$ eV at $90\%$ CL \cite{Drexlin:2013lha}.  
We note that $m_0$ in the QD regime for CP conservation is approximately equal
to the effective beta decay mass $m_\beta$.  Hence QD mass pattern is well within the sensitivity of the KATRIN.

The second method to extract the neutrino mass is neutrinoless double beta decay which assumes
that neutrinos are Majorana particles \cite{Rodejohann:2012xd,Bilenky:2014uka}. The observable
parameter $M_{ee}$,  the double beta decay effective mass is given as following

\bea
\label{mee_1}
M_{ee} & =& \left| \sum U_{ei}^2 \, m_i \right|,  \nonumber \\
& = & \left| m_1 c_{12}^{2} \, 
c_{13}^{2} e^{- i \varphi_1}  + m_2 s_{12}^{2} \, c_{13}^{2}\, 
e^{-i\varphi_2} + m_3 s_{13}^{2} \, e^{-i 2 \delta}
\right| .
\eea
For quasi-degenerate  neutrinos
\bea
\label{mee_2}
M_{ee} & \approx & m_0 \left|  c_{12}^{2} \, 
c_{13}^{2} e^{- i \varphi_1}  + s_{12}^{2} \, c_{13}^{2}\, 
e^{-i\varphi_2} +s_{13}^{2} \, e^{-i 2 \delta}
\right| .
\eea
Since the contribution of $m_{3}$ is suppressed by the small 
$\sin^{2}\theta_{13}$ coefficient, we obtain
\begin{equation}
\label{quasi1}
M_{ee} \simeq m_{0} \, \sqrt{1-\sin^{2}2\theta_{12} \frac{(1-\cos (\varphi_1-\varphi_2))}{2}}.
\end{equation}
For CP conserving case where the Majorana and Dirac phases are zero, $M_{ee} \simeq m_0$.
For $M_{ee} \simeq 0.1$ eV, the above expression corresponds approximately to half-life
in the range of $10^{25}$ to $10^{26}$ yrs \cite{Rodejohann:2012xd} which makes the QD mass
scheme testable in present and future experiments .  In the QD regime, the neutrino mass can be written as~\cite{Rodejohann:2012xd}   

\be \label{eq:m0_lim}
m_0 \le  (M_{ee})_{\max}^{\rm exp} \,  \frac{1 + \tan^2 \theta_{12}}
{1 - \tan^2 \theta_{12} - 2 \, |U_{e3}|^2 } 
\equiv  (M_{ee})_{\max}^{\rm exp} \,  f(\theta_{12}, \theta_{13}) \, .
\ee
Using inputs from Table~\ref{tab1}, the function $f(\theta_{12}, \theta_{13}) $ has a range from $2.2$ to $4.1$ at $3 \sigma$.
The most stringent upper limit on the effective mass $M_{ee}$ provided by the GERDA experiment  is $0.4~\rm{eV}$~\cite{Agostini:2013mzu}.  
Hence $m_0 \leq 1.64$ eV and sum of the neutrino masses $\Sigma m_i = 3 m_0 \leq 4.91$ eV.

The third determination of neutrino masses is provided by the cosmological and astrophysical observations.
The sum of the neutrino masses, $\Sigma m_i$, has a range for upper bound to be $0.17 - 0.72$ eV at $95\%$ CL \cite{Ade:2015xua}. 
This limit is not model independent and depends on the cosmological model applied to the data.  

%%%%%%%%%%%%%%%%%%%%%%%%%%%%%%%%%%%%%%%%%%%%%%%%%%%%%%%%%%%%%%%%%%%%%%%%%%%%%%%%%%%%%%%%%%%%%%%%%%%%%%%%%%%%%%%%%%%%%%%%%%%%%%%%%%%%%%%%%%%%%%%%%%%%%%%%%%%%%%%%%%%%%%%%%%%%%%%%%%%%%%%%%%%%%%%%%%%%%%%%%%%%%%%%%%%%%%%%%%%%%%%%%%%%%%%%%%%%%%%%%%%%%%%%%%%%%%%

\section{Results}
\label{sec4}
We present our results in this section for the different cases listed in Eqs.~(\ref{case1} - \ref{case7}) and for limiting cases of the most general HSMR as shown in  Eqs.~(\ref{caseA} - \ref{caseG}).
As discussed earlier, we need MSSM as an extension of 
the SM for the implementation of HSMR and HSMU hypothesis.  In the first step, we run 
quark mixing angles, gauge couplings, Yukawa couplings of quarks and  charged leptons 
from the low scale to the SUSY breaking scale.  The evolution from the SUSY breaking scale to the unification
scale is done through the MSSM RG equations.  After evolving up to the unification scale, we obtain quark mixing angles  $\theta_{12}^q = 13.02^\circ$, $\theta_{13}^q=0.17^\circ$ and $\theta_{23}^q=2.03^\circ$.  In the next step, quark mixing angles 
are used to calculate the leptonic mixing angles using HSMR  at the unification scale.  After this, 
we run down the MSSM RG equations up to the SUSY breaking scale.  The SM RG equations take over the 
evolution of mixing parameters beyond the SUSY breaking scale.   The SUSY breaking scale is chosen 
to be 2 TeV following the direct LHC searches \cite{Craig:2013cxa}.  We also need a large $\tan \beta$ 
which is chosen to be $55$.  The unification scale where HSMR  can exist is chosen to be $10^{14}$ GeV 
which is consistent with present experimental observations \cite{Capozzi:2013csa}. We have used 
the MATHEMATICA based package REAP \cite{Antusch:2005gp} for the numerical computation of
our results.  We have done a rigorous,  thorough and comprehensive in this work.  For this pupose, we have written an interface code which together with public code can be used to scan whole parameter space. 

\subsection{RG evolution of HSMR}

\begin{figure}[ht]
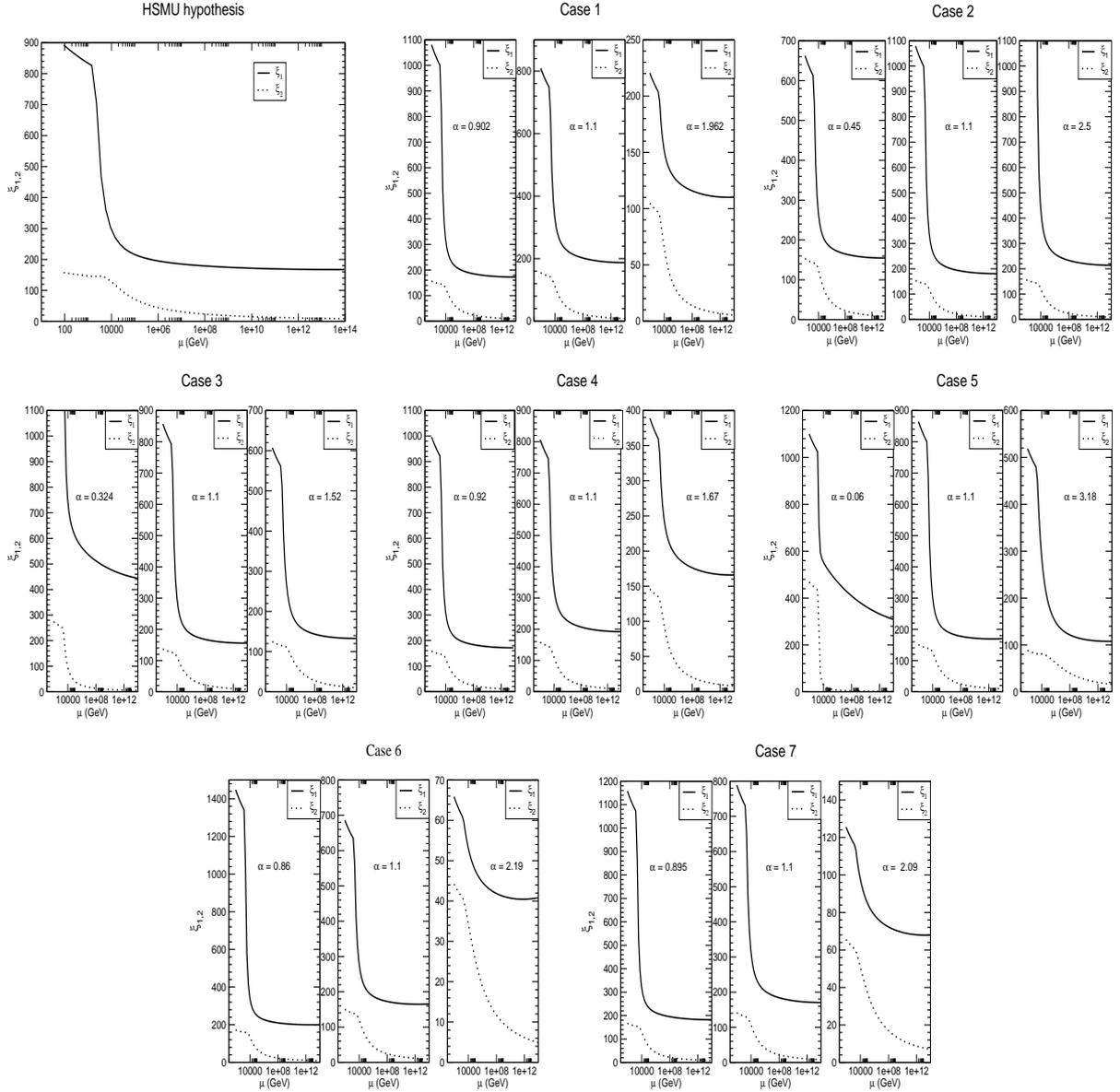

\begin{minipage}[b]{0.3\linewidth}
\vspace*{0.65cm}
\centering
\includegraphics[width=5cm, height=5cm]{ef_hsmu.eps}
\end{minipage}
\hspace{0.2cm}
\begin{minipage}[b]{0.3\linewidth}
\centering
\includegraphics[width=5cm, height=5cm]{ef_case1.eps}
\end{minipage}
\hspace{0.2cm}
\begin{minipage}[b]{0.3\linewidth}
\centering
\includegraphics[width=5cm, height=5cm]{ef_case2.eps}
\end{minipage}
\begin{minipage}[b]{0.3\linewidth}
\vspace*{0.3cm}
\centering
\includegraphics[width=5cm, height=5cm]{ef_case3.eps}
\end{minipage}
\hspace{0.2cm}
\begin{minipage}[b]{0.3\linewidth}
\centering
\includegraphics[width=5cm, height=5cm]{ef_case4.eps}
\end{minipage}
\hspace{0.2cm}
\begin{minipage}[b]{0.3\linewidth}
\centering
\includegraphics[width=5cm, height=5cm]{ef_case5.eps}
\end{minipage}
\begin{minipage}[b]{0.3\linewidth}
\vspace*{0.3cm}
\centering
\includegraphics[width=5cm, height=5cm]{ef_case6.eps}
\end{minipage}
\hspace{0.4cm}
\begin{minipage}[b]{0.3\linewidth}
\centering
\includegraphics[width=5cm, height=5cm]{ef_case7.eps}
\end{minipage}
\caption{The change in the RG evolution of the enhancement factors ($\xi_{1,2}$), \ref{ang2}, for the different cases of HSMR 
as a function of the RG scale $\mu$ when $\alpha$ deviates from unity.}
 	\label{figef}
\end{figure}
We study the RG evolution of HSMR as given in Eqs.~(\ref{case1} - \ref{case7}) and compare our results with respect to 
the HSMU hypothesis.  In Fig.~\ref{figef}, we show how enhancement factors $\xi_1$ and $\xi_2$ 
evolve from the unification scale to the low scale, as $\alpha$ deviates from unity.
The results are displayed for all the cases. It can be seen from Fig.~\ref{figef} that the evolution
in Case 1 at $\alpha = 1.1$ is similar to HSMU hypothesis.  However, as $\alpha$ approaches to lowest 
value on the left panel of Case 1, $\xi_1$ changes sufficiently. Similarly for the upper limit of 
 $\alpha = 1.962$, the evolution again becomes very different from the HSMU hypothesis. This 
explains why the RG evolution of the PMNS mixing angles change when $\alpha$ deviates from unity. 
The same argument follows for all the other cases  of HSMR and can be checked from Fig.~\ref{figef}.
\begin{figure}[ht]
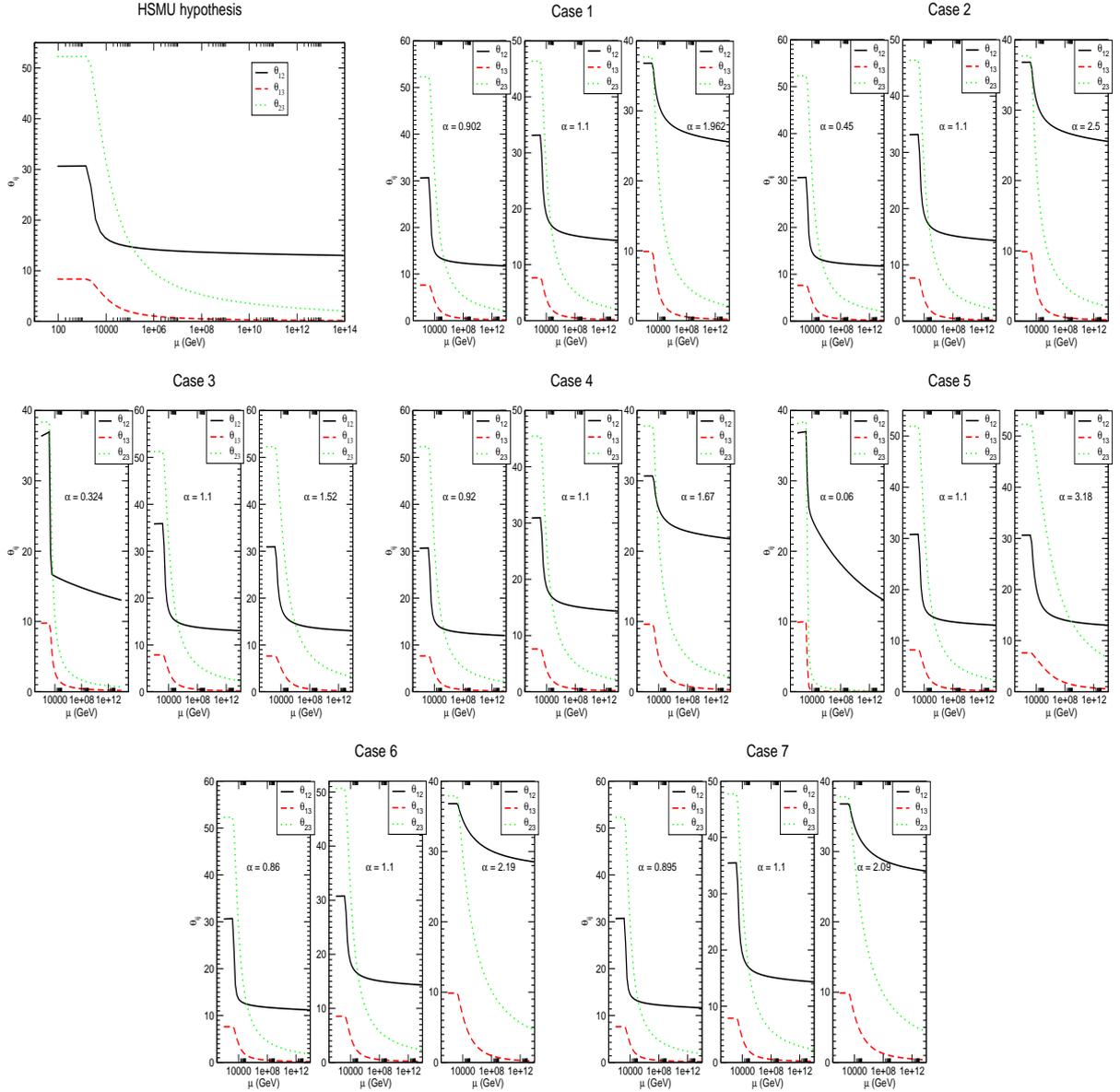

\begin{minipage}[b]{0.3\linewidth}
\vspace*{0.65cm}
\centering
\includegraphics[width=5cm, height=5cm]{angles_hsmu.eps}
\end{minipage}
\hspace{0.2cm}
\begin{minipage}[b]{0.3\linewidth}
\centering
\includegraphics[width=5cm, height=5cm]{angles_case1.eps}
\end{minipage}
\hspace{0.2cm}
\begin{minipage}[b]{0.3\linewidth}
\centering
\includegraphics[width=5cm, height=5cm]{angles_case2.eps}
\end{minipage}
\begin{minipage}[b]{0.3\linewidth}
\vspace*{0.3cm}
\centering
\includegraphics[width=5cm, height=5cm]{angles_case3.eps}
\end{minipage}
\hspace{0.2cm}
\begin{minipage}[b]{0.3\linewidth}
\centering
\includegraphics[width=5cm, height=5cm]{angles_case4.eps}
\end{minipage}
\hspace{0.2cm}
\begin{minipage}[b]{0.3\linewidth}
\centering
\includegraphics[width=5cm, height=5cm]{angles_case5.eps}
\end{minipage}
\begin{minipage}[b]{0.3\linewidth}
\vspace*{0.3cm}
\centering
\includegraphics[width=5cm, height=5cm]{angles_case6.eps}
\end{minipage}
\hspace{0.4cm}
\begin{minipage}[b]{0.3\linewidth}
\centering
\includegraphics[width=5cm, height=5cm]{angles_case7.eps}
\end{minipage}
\caption{The RG evolution of the PMNS mixing angles with respect to the RG scale $\mu$, for the different cases, 
with $\alpha$ varied in the respective allowed range. }
 	\label{figang}
\end{figure}

We next show the evolution of the mixing angles for the different cases in Fig.~\ref{figang}
along with the HSMU hypothesis.  We observe from Figs.~\ref{figef} and \ref{figang},
that the evolution of HSMR is similar to the HSMU hypothesis when $\alpha$ deviates slightly from unity.
However, when $\alpha$ is very far from unity, RG evolution undergoes dramatic changes. There is another interesting
phenomenon that can be observed from Fig.~\ref{figang}. It can be easily seen that the RG evolution of the 
mixing angles, for Cases 3 and 5 are similar, with $\theta_{12}$ and $\theta_{23}$ almost similar at the low scale
at the lower end of $\alpha$.  The difference between them at the low scale increases with the increase 
in value of $\alpha$. The pattern is exactly opposite in the other cases of HSMR, with the difference between
$\theta_{12}$ and $\theta_{23}$ at the low scale decreasing as one goes from the lower to the upper end of $\alpha$.
This in a way tells us beforehand that the phenomenology of Case 3 and 5 will be similar, which will be discussed
in detail afterwards.

\subsection{Phenomenology of HSMR}
In this subsection, we discuss in details the phenomenological implications of HSMR.  Our aim is to investigate 
the behavior of $\alpha$ as it deviates from unity and its phenomenological consequences taking into account all the experimental 
constraints of Table~\ref{tab1} and the GERDA limit \cite{Agostini:2013mzu}.  The common observation among all HSMR is 
the emergence of the strong correlations among $\Delta m^2_{32}$, $M_{ee}$, $\theta_{23}$, $\theta_{13}$ and $\Sigma m_i$.

\subsubsection{HSMU hypothesis}
\begin{figure}[htb]
\begin{minipage}[b]{0.45\linewidth}
\vspace*{0.65cm}
\centering
\includegraphics[width=6.5cm, height=5cm]{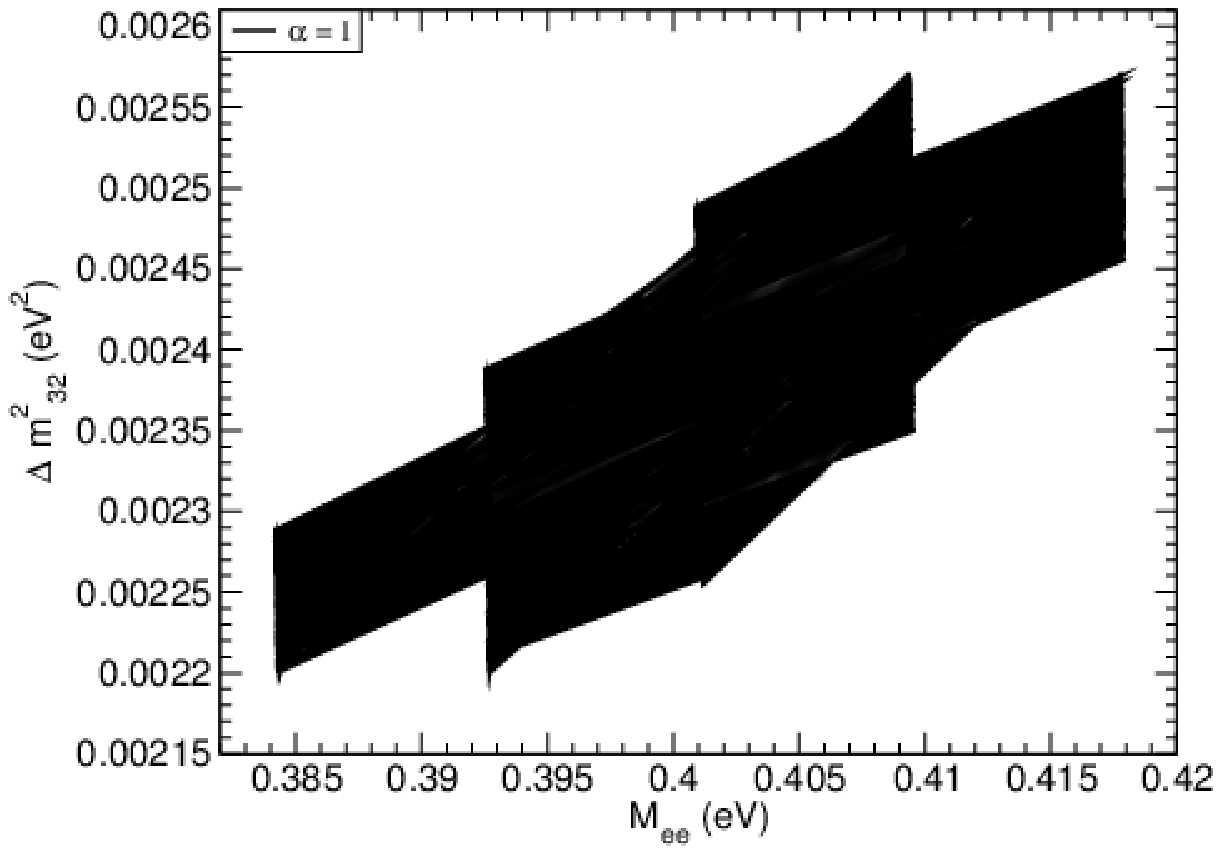}
\caption{The variation of $\Delta m^2_{32}$ with respect to $M_{ee}$, in the context of the HSMU hypothesis,
with $\alpha$ =1.}
\label{fig1}
\end{minipage}
\hspace{0.7cm}
\begin{minipage}[b]{0.45\linewidth}
\centering
\includegraphics[width=6.5cm, height=5cm]{th1323_hsmu.eps}
\caption{The variation of $\theta_{23}$ with respect to $\theta_{13}$, in the context of the HSMU hypothesis,
with $\alpha$ =1. }
\label{fig2}
\end{minipage}
\end{figure}

As observed earlier, the value $\alpha=1$ will reduce all cases of HSMR  to HSMU hypothesis.  We present a full 
parameter scan of the HSMU hypothesis using dimensional-5 operator. It  should be noted
that this analysis was absent in the previous works on  HSMU hypothesis 
\cite{Abbas:2014ala, Abbas:2015aaa, Srivastava:2015tza}
and is reported in this work for the first time. We present a correlation in Fig.~\ref{fig1}, 
which is not studied in the previous investigations.  We show here the variation of $\Delta m^2_{32}$
with respect to $M_{ee}$.  The  $M_{ee}$ has an upper bound of $0.4$ eV from the GERDA 
experiment \cite{Agostini:2013mzu}.  Using this limit, we are able to put an upper 
bound on the allowed range of $\Delta m^2_{32}$.  The allowed range for 
$\Delta m^2_{32}$ is $(2.21-2.45) \times 10^{-3}~{\rm eV}^2$  as 
observed from Fig. \ref{fig1}. The lower bound on $M_{ee}$ is $0.384$ eV
for the HSMU hypothesis.  Hence, our work on the HSMU hypothesis will be ruled out 
if GERDA  crosses this number in  the future.
The effective $\beta$ decay mass $m_\beta$  is another interesting observable
since it does not depend on whether the neutrinos are Majorana or Dirac.   The 
prediction for $m_\beta$  coincides with the effective double beta decay mass 
$M_{ee}$ in the QD regime and for CP conservation.   Hence, the allowed range for 
$m_\beta$ is identical to that of $M_{ee}$ in our work.

In Fig.~\ref{fig2}, we show the variation of  $\theta_{23}$  with respect to $\theta_{13}$. 
We observe a strong correlation between $\theta_{13}$ and  $\theta_{23}$. 
The difference between this investigation and that of presented in 
Ref.~\cite{Abbas:2014ala} is the variation of $\theta_{12}$.  In the previous work,
this correlation was reported for a chosen value of the angle $\theta_{12}$ at 
the low scale in the context of type \rom{1} seesaw.  In this work, we do not
choose any particular value of $\theta_{12}$ at the low scale.  We obtain a band 
for this correlation and previous results are a specific case of our present results.
We observe that $\theta_{23}$ is non maximal and always lies in the second octant. 
This confirms the predictions of our earlier work \cite{Abbas:2014ala}.  
The allowed  range of $\theta_{13}$ is $7.63^\circ-8.34^\circ $ and that of $\theta_{23}$ is  $49^\circ-52.3^\circ$.

\begin{figure}[htb]
\begin{minipage}[b]{0.45\linewidth}
\vspace*{0.65cm}
\centering
\includegraphics[width=6.5cm, height=5cm]{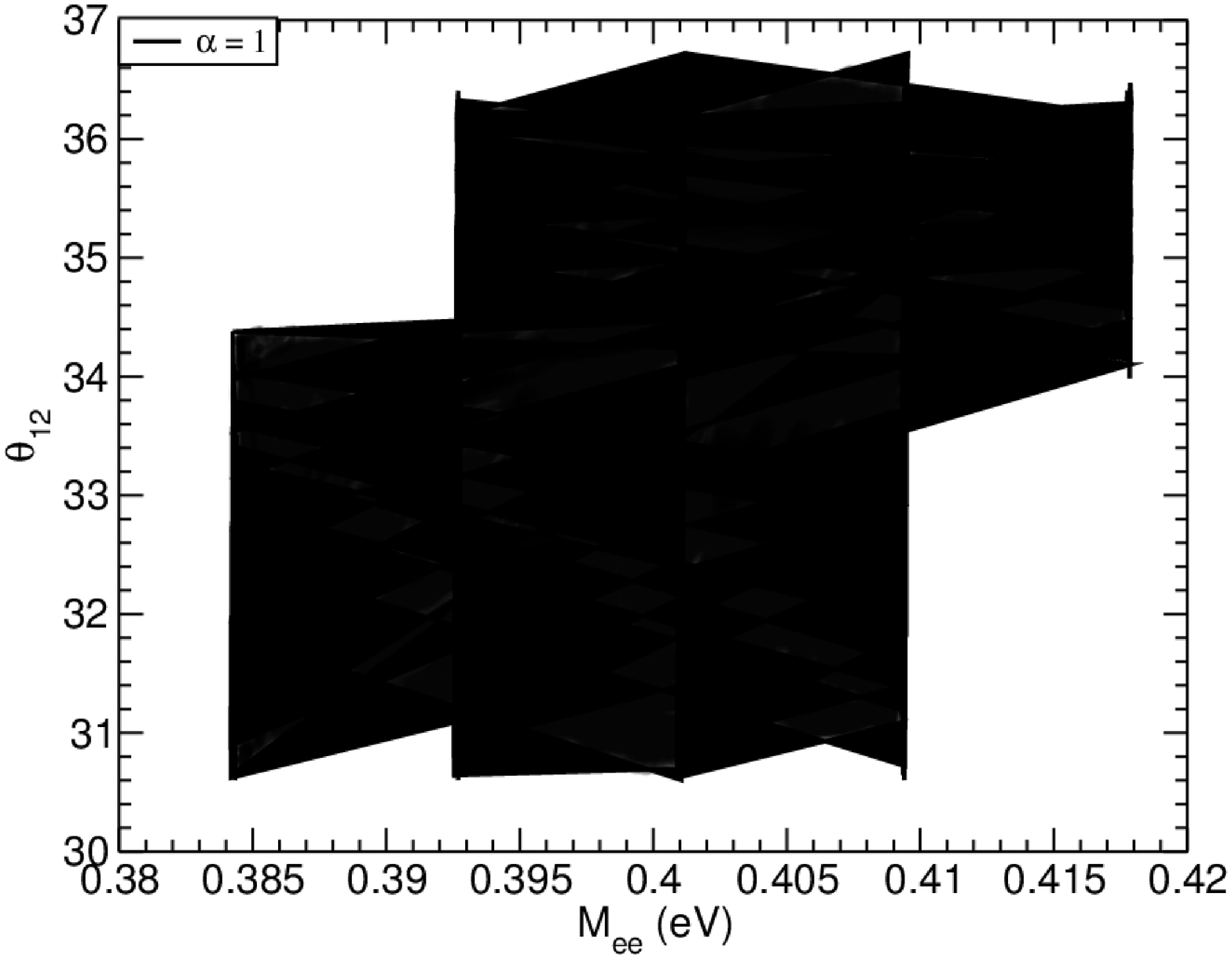}
\caption{The variation of $\theta_{12}$ with respect to $M_{ee}$, in the context of the HSMU hypothesis,
with $\alpha$ =1.}
\label{fig3}
\end{minipage}
\hspace{0.7cm}
\begin{minipage}[b]{0.45\linewidth}
\centering
\includegraphics[width=6.5cm, height=5cm]{mee_msum_hsmu.eps}
\caption{The variation of $\Sigma m_i$ with respect to $M_{ee}$, in the context of the HSMU hypothesis,
with $\alpha$ =1. }
\label{fig4}
\end{minipage}
\end{figure}
 
We next present variation of $\theta_{12}$ against $M_{ee}$ in Fig.~\ref{fig3}.  This correlation is also a new prediction of our work and do not exist in previous studies. The whole $3 \sigma$ global range for the angle $\theta_{12}$ is allowed for the $M_{ee} \leq 0.4$~eV, 
However, as can be observed from  Fig.~\ref{fig3},  the range $34.4^\circ \leq \theta_{12} \leq 36.81^\circ$
is ruled out for $0.384~{\rm eV} \leq M_{ee} \leq 0.393$~eV.  The precise predictions for all observables are provided 
in Table \ref{tab2}. In the end, we also have a new correlation between the sum of neutrino masses and $M_{ee}$ which is not studied previously. This correlation  is shown in Fig. \ref{fig4}.  Our prediction for sum of neutrino masses is $1.16-1.2$ eV using the upper bound on $M_{ee}$ given by the GERDA.  

\subsubsection{The most general HSMR within the same generations}
 
The most general HSMR  within the same generations for $(k_1,k_2,k_3)=(1,1,1)$ as defined
before is given by the following equation
 
 \begin{equation}
\theta_{12} = \alpha_{1} ~\theta_{12}^q, ~~ \theta_{13} = \alpha_{2}~ \theta_{13}^q, ~~\theta_{23} =\alpha_{3}~~\theta_{23}^q. 
\end{equation}

We present the results for the maximum and the minimum values of
$\alpha_i$ for Eqs.~(\ref{caseA}-\ref{caseH}), taking into account 
all the experimental constriants. In the Table \ref{tab_mgr}, we
present the allowed values of $\alpha_i$ along with the
the respective physical masses and the mixing angles.

\begin{table}[h]
\large
\begin{center}
\resizebox{\textwidth}{!}{\begin{tabular}{| c|c|c|c|c|c|c|c|c|c|c|c|c|c|c|c|}
  \hline
            & $\alpha_1$    &$\alpha_2$ & $\alpha_3$   &  \multicolumn{3}{ c| }{Masses at unification scale (eV)}  &  $\Sigma m_i$ (eV) & $\theta_{12}^\circ$ & $\theta_{13}^\circ$ 
                  & $\theta_{23}^\circ$& $\Delta m^2_{32}$ $(10^{-3} {\rm eV}^2)$& $\Delta m^2_{21}$ $(10^{-5} {\rm eV}^2)$&$M_{ee}$ (eV)
                  & Lightest neutrino mass: $m_1$ (eV)& R\\          
                  \cline{5-7}
                &   &   &      & $m_1$ & $m_2$ &    $m_3$             &           &                 &          
                      &      &       &       &                      &           &  \\ \hline
            &      &    &      &       &       &      &     &     &     &      &       &       &   &   &  \\ 
Case A & 1.46 & 2.54  &   1.19    & 0.4583  & 0.461      &   0.5186   & 1.16
                           & 36.52    & 9.88    & 41.14 & 2.5      & 8.06       & 0.385       & 0.3850      & 2.29\\  
          &   &       &      &       &       &      &     &     &     &      &       &       &   &   &  \\         
       \hline  
        &      &    &      &       &       &      &     &     &     &      &       &       &   &   &  \\               
Case B  & 1.45    &   1.68  & 0.91   &  0.4757     &  0.478    &   0.5380   &  1.20   &  30.61   & 8.79    &   37.97   &  2.25     &   8.12    &0.40   &  0.3997 & 1.8 \\
         
  &   &       &      &       &       &      &     &     &     &      &       &       &   &   &  \\         
       \hline    
        &      &    &      &       &       &      &     &     &     &      &       &       &   &   &  \\   
Case C         & 1.38    &   0.71  & 1.28  &  0.489     &  0.493   &   0.5527   &  1.24   &  36.8  &  9.87   & 50.83     &  2.22     &   8.14    &0.411   &  0.4106 & 5.3 \\ 
        
  &   &       &      &       &       &      &     &     &     &      &       &       &   &   &  \\         
       \hline
       &      &    &      &       &       &      &     &     &     &      &       &       &   &   &  \\   
Case D        & 1.14     &   0.92 & 0.94   &  0.4754     &  0.478    &   0.5370   &  1.20   &  31.18   & 7.70    &   45.31   & 2.20     &  8.14    & 0.40   &  0.3994 & 1.69 \\  
  &   &       &      &       &       &      &     &     &     &      &       &       &   &   &  \\         
       \hline   
       &      &    &      &       &       &      &     &     &     &      &       &       &   &   &  \\   
Case E        & 0.8    &   2.2  & 1.15   &  0.4096     &  0.412   &   0.4625   &  1.04   &  32.77  & 7.65    &   48.13   &  2.35     &   7.01    & 0.344   &  0.3442 & - \\ 
  &   &       &      &       &       &      &     &     &     &      &       &       &   &   &  \\         
       \hline    
       &      &    &      &       &       &      &     &     &     &      &       &       &   &   &  \\   
Case F        & 0.89    &   1.61  & 0.82   &  0.4751     &  0.477   &   0.5361   &  1.20   &  30.6   & 7.65    &   43.66   &  2.22     &   7.37    &0.40   &  0.3993 & 1.06 \\ 
       
  &   &       &      &       &       &      &     &     &     &      &       &       &   &   &  \\         
       \hline   
       &      &    &      &       &       &      &     &     &     &      &       &       &   &   &  \\   
Case G        &  0.92    &   0.98  & 1.03   &  0.4421     &  0.445   &   0.4989   &  1.12   &  32.37  & 7.64    &   52.19   &  2.22     &   7.86    & 0.372   & 0.3714 & 1.48 \\ 
  &   &       &      &       &       &      &     &     &     &      &       &       &   &   &  \\
  \hline
  &      &    &      &       &       &      &     &     &     &      &       &       &   &   &  \\ 
Case H         &  0.88    & 0.95    & 0.86     &  0.4764     & 0.479     & 0.5372     &  1.20  &  30.99   &   7.63  &   51.97   &   2.22    & 7.55      &0.40   &  0.4003 & 1.29 \\
  &   &       &      &       &       &      &     &     &     &      &       &       &   &   &  \\ 
  \hline                                                                                
\end{tabular}}
\end{center}
\caption{The allowed predictions for the different cases of the most general HSMR for minimum and maximum  allowed values of $\alpha_i$, Eqs.~(\ref{caseA}-\ref{caseG}).}
 \label{tab_mgr}
\end{table}   

It is remarkable that in the Case E, all the mixing parameters are within $3 \sigma $ global range without adding threshold corrections.  If we add threshold corrections, the predictions are $\Delta m^2_{32}$ = 2.35 $\times$ $(10^{-3} {\rm eV}^2)$ and $\Delta m^2_{21}$= 7.01 $\times$ $(10^{-5} {\rm eV}^2)$ for $R=1.0$.  Thus, threshold corrections at this point are effectively negligible.  We further notice that the different combinations of the allowed end points of $\alpha_i$, leads to $M_{ee}$ around 0.35 eV - 0.4 eV. This most general case with different $\alpha_i$, alone will not suffice, when the value of $M_{ee}$ will be further constrained by the future experiments. We then have to look for more specific cases, where the $\alpha_i$'s will not be different, but have some relations among them. We consider the simplified scenario, where the $\alpha_i$'s are equal. We have carried out a detailed analysis for all the possible seven cases in this scenario in the next subsections.

\subsubsection{Case 1: $\theta_{12}= \alpha ~ \theta_{12}^q,~~\theta_{13} =  \theta_{13}^q,~~\theta_{23} =  \theta_{23}^q$}

\begin{figure}[htb]
\begin{minipage}[b]{0.45\linewidth}
\vspace*{0.65cm}
\centering
\includegraphics[width=6.5cm, height=5cm]{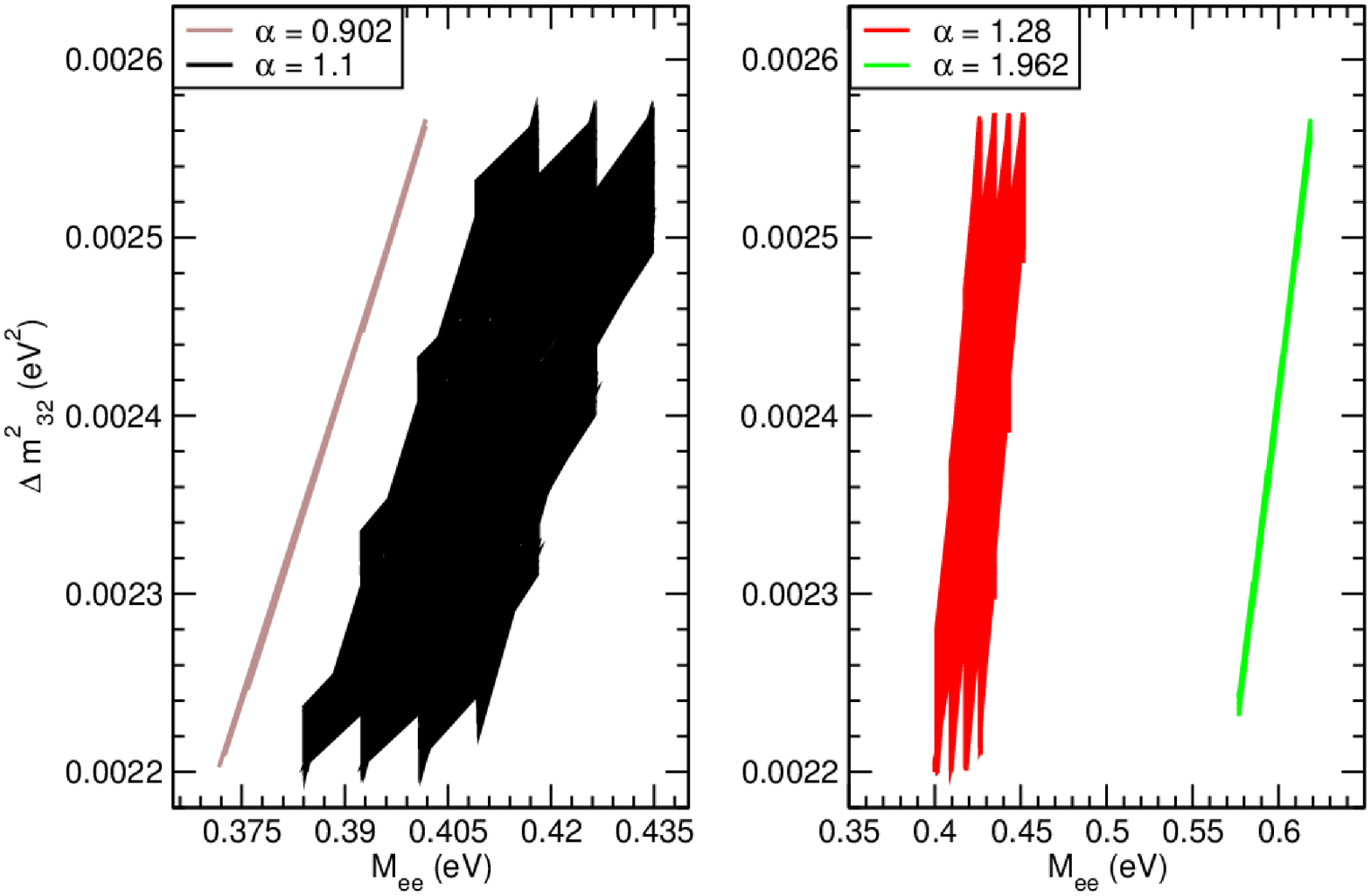}
\caption{ The variation of $\Delta m^2_{32}$ with respect to $M_{ee}$ for Case 1 of HSMR.}
\label{fig5}
\end{minipage}
\hspace{0.7cm}
\begin{minipage}[b]{0.45\linewidth}
\centering
\includegraphics[width=6.5cm, height=5cm]{th1323_case1.eps}
\caption{The variation of $\theta_{23}$ with respect to $\theta_{13}$ for Case 1 of HSMR. }
\label{fig6}
\end{minipage}
\end{figure}

The first case of HSMR is the one where leptonic mixing angle $\theta_{12}$ is proportional to $\theta_{12}^q$ and the other  two angles are identical.  
In Fig.~\ref{fig5}, we show how the correlation between $\Delta m^2_{32}$ and $M_{ee}$ changes as $\alpha$ deviates 
from unity.   We observe on the left panel of Fig.~\ref{fig5} that the lowest allowed value of 
$\alpha$ is $0.902$.  This value is derived by the $3 \sigma$ global limit of the leptonic mixing angles.  On the right panel of  Fig.~\ref{fig5}, the upper bound on $\alpha$ is shown. 
For  the upper bound on $\alpha$, in principle, one can go up to $1.962$  with all mixing parameters within 
the global range.  This value of $\alpha$ belongs to $M_{ee} > 0.4$~eV and hence is ruled out by the GERDA limit.  
The  allowed upper bound on $\alpha$ is $1.28$ which is derived using the GERDA limit.

We compare Fig.~\ref{fig5} with Fig.~\ref{fig1} of the HSMU hypothesis ($\alpha = 1$)  to study the 
phenomenological behavior of $\alpha$.  As obvious from the left panel of Fig.~\ref{fig5} , $M_{ee}$ has 
its maximum allowed range at the lowest value of $\alpha$.  This is because the absolute neutrino mass decreases 
for $\alpha < 1$ and increases for $\alpha > 1$ in the case under study.  Hence,  at $\alpha =0.902$ on the left panel of Fig.~\ref{fig5}, 
we obtain $0.365~{\rm eV} \leq M_{ee} \leq 0.40$~eV corresponding to whole $3 \sigma$  global range of $\Delta m^2_{32}$.  The same 
prediction for the HSMU hypothesis in Fig.~\ref{fig1},  ($\alpha =1$) is $0.385~{\rm eV} \leq M_{ee} \leq 0.418$ eV 
which belongs to $\Delta m^2_{32}=(2.21-2.45) \times 10^{-3}~{\rm eV}^2$.  The prediction when $\alpha $ 
slightly deviates from unity ($\alpha=1.1 $)  is $0.384~{\rm eV} \leq M_{ee} \leq 0.435$ eV corresponding 
to $\Delta m^2_{32}=(2.22-2.57) \times 10^{-3}~{\rm eV}^2$.  At the upper allowed value of $\alpha=1.28$, 
we have  $0.4~{\rm eV} \leq M_{ee} \leq 0.45$ eV  which belongs to $\Delta m^2_{32}=(2.20-2.57) \times 10^{-3}~{\rm eV}^2$. 
We observe that the uppermost value of $\alpha =1.962$ has $0.571~{\rm eV} \leq M_{ee} \leq 0.625$ eV belonging 
to $\Delta m^2_{32}=(2.23-2.57) \times 10^{-3}~{\rm eV}^2$.  This value of $\alpha$ is already ruled out by the GERDA limit.

This case can be ruled out if GERDA  reaches $M_{ee} < 0.365$~eV.   There is an apparent overlap between predictions of the case under
study and the HSMU hypothesis. This can be discriminated using the SUSY ratio $R$.  For a clear picture of the phenomenological 
consequences, we provide values of  mixing parameters and other observables belonging to  minimum
and maximum allowed values of $\alpha$  for each case and the HSMU hypothesis  in Table~\ref{tab2}.
 
\begin{figure}[htb]
\begin{minipage}[b]{0.45\linewidth}
\vspace*{0.65cm}
\centering
\includegraphics[width=6.5cm, height=5cm]{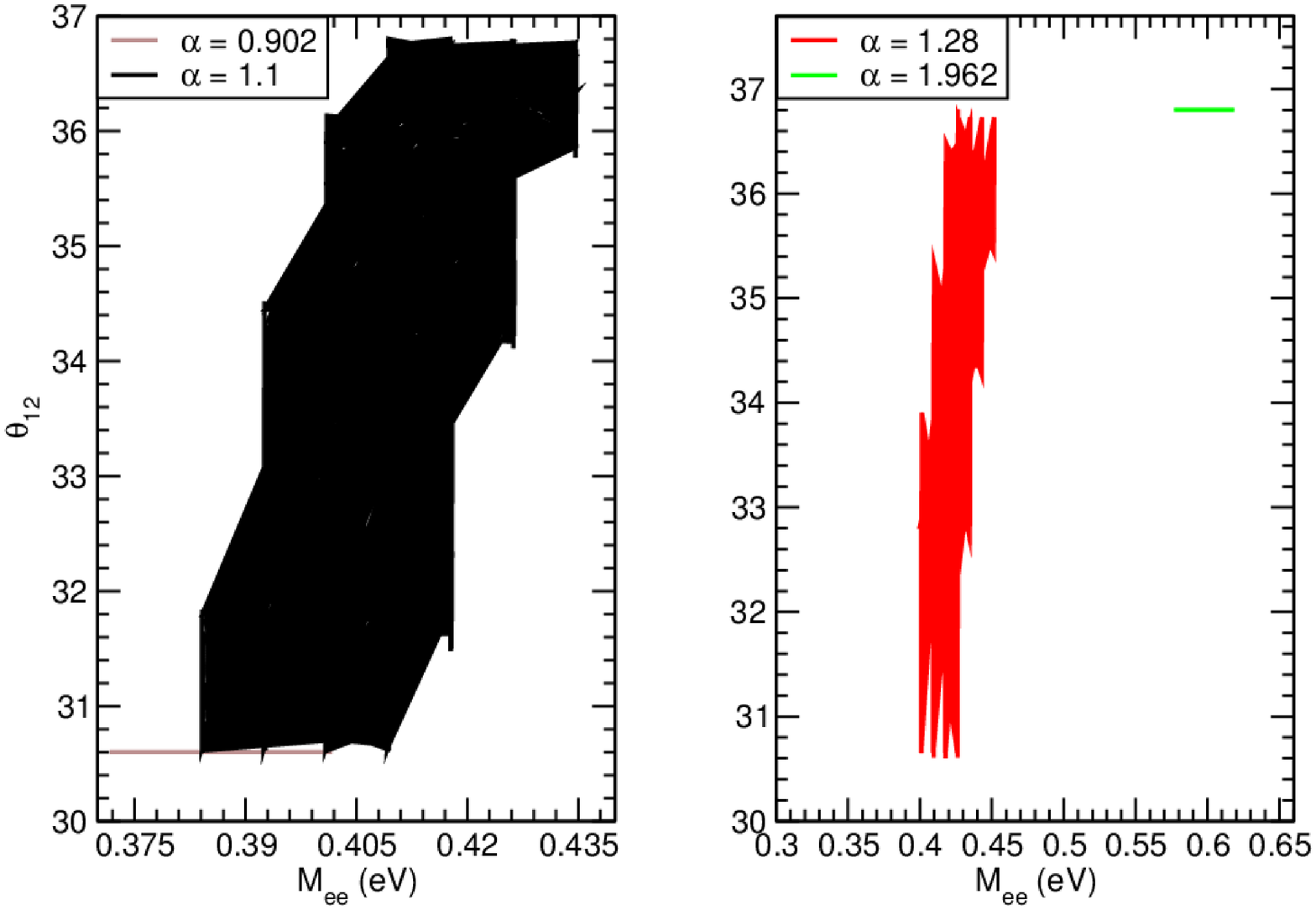}
\caption{ The variation of $\theta_{12}$ with respect to $M_{ee}$ for Case 1 of HSMR.} 
	\label{fig7}
\end{minipage}
\hspace{0.7cm}
\begin{minipage}[b]{0.45\linewidth}
\centering
\includegraphics[width=6.5cm, height=5cm]{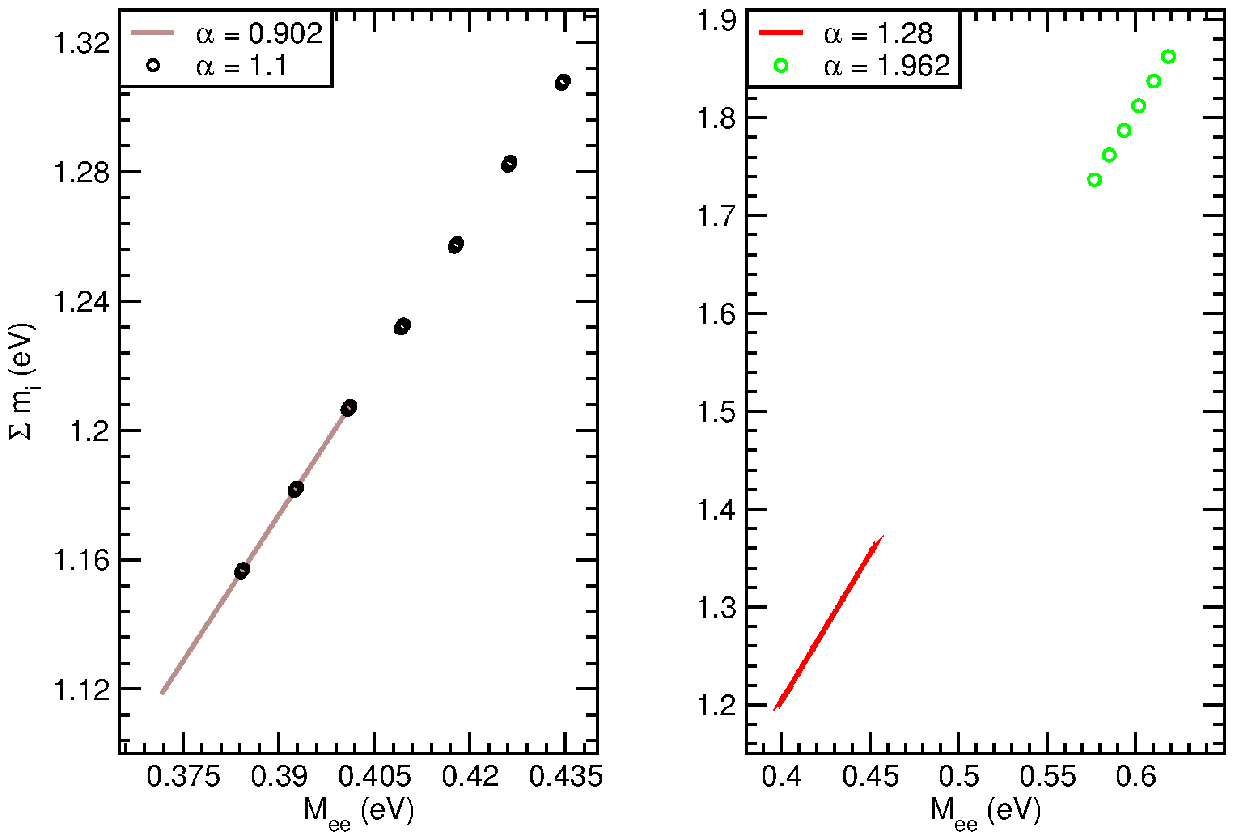}
\caption{The variation of $\Sigma m_i$ with respect to $M_{ee}$ for Case 1 of HSMR. }
	\label{fig8}
\end{minipage}
\end{figure}

The variation of  $\theta_{23}$  with respect to $\theta_{13}$ is shown in Fig.~\ref{fig6}.  The mixing angles reach their 
$3\sigma$ limits at their lower and upper ends.  For example, at $\alpha=0.902$, $\theta_{13}$ is at its minimum of the $3\sigma$ global limit while $\theta_{23}$ is at its maximum independent of the upper bound of $M_{ee}$.   On the other hand, at $\alpha=1.962$,
the predictions are reversed but this value is already rejected by the GERDA limit of $M_{ee}$.  
The allowed ranges of $\theta_{13}$ and $\theta_{23}$, at $\alpha=1.1$, are $7.62^\circ-9.1^\circ$ 
and $46.09^\circ-52.2^\circ$ respectively.  Compared to Fig.~\ref{fig2} for the HSMU hypothesis where 
the allowed range of $\theta_{23}$ is always in the second octant, $\theta_{23}$ has its minimum value $44.04^\circ$ at $\alpha=1.28$ which
belongs to $M_{ee}=0.4$~eV and lies in the first octant. The corresponding value of $\theta_{13}$ 
is $8.16^\circ$.

We next show the behavior of $\theta_{12}$ with respect to $M_{ee}$ in Fig.~\ref{fig7}. 
We observe that at $\alpha = 0.902$, $\theta_{12}$  is at its global minimum $30.6^\circ$.  On the left panel of  Fig.~\ref{fig7}, 
at $\alpha = 1.1 $, $\theta_{12}$ has an allowed range of $30.6^\circ-35.65^\circ$ for $M_{ee} \leq 0.4$~eV.  For the HSMU hypothesis
in Fig.~\ref{fig3}, $\theta_{12}$ has the whole $3 \sigma$ global range with some higher values ruled out for $M_{ee} \leq 0.393$~eV. 
For $\alpha = 1.28 $, the value of $\theta_{12}$ is $32.82^\circ$ for $M_{ee}=0.4$~eV as 
can be seen from the right panel of the figure.  For  $\alpha = 1.962$, $\theta_{12}$ reaches to the maximum of its $3\sigma$ global limit.

Finally in Fig.~\ref{fig8},  the variation of sum of the neutrino masses 
with respect to $M_{ee}$ is presented.  For lowest value of $\alpha=0.902$,
on the left panel,  the range of $\Sigma m_i $ is $1.12-1.2$~eV  for $M_{ee} \leq 0.4$~eV. 
At $\alpha=1.1$, it is $1.16-1.2$~eV  for $M_{ee} \leq 0.4$~eV. On the right panel,
the value of sum at $\alpha=1.28$ is $1.2$~eV and the region $\Sigma m_i > 1.2$~eV belongs to $M_{ee} > 0.4$~eV.

\subsubsection{Case 2: $\theta_{12}= \theta_{12}^q,~~\theta_{13} =  \alpha~\theta_{13}^q,~~\theta_{23} =  \theta_{23}^q$}

The second case which we consider has  leptonic mixing angle $\theta_{13}$  proportional to $\theta_{13}^q$. 
In this case, the lower bound on $\alpha$ is $0.45$ which is derived using global limits on the mixing angles.  
The $\alpha$ on the upper side, however, is remarkably bounded by the ratio $R$.  This theoretical bound 
arises because we work with an inverted hierarchy in the charged-slepton sector and at $\alpha=2.5$, 
we have $R=1$.  In principle, $\alpha$ has a range up to $3.5$ satisfying all experimental constraints with $R<1$.

\begin{figure}[htb]
\begin{minipage}[b]{0.45\linewidth}
\vspace*{0.65cm}
\centering
\includegraphics[width=6.5cm, height=5cm]{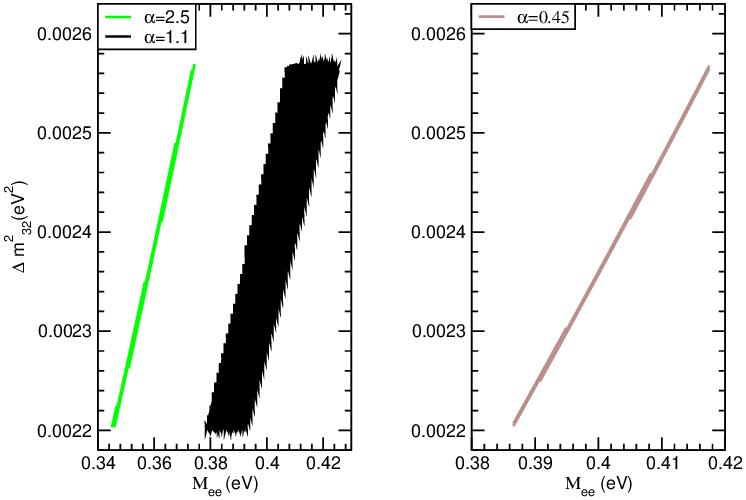}
\caption{The variation of $\Delta m^2_{32}$ with respect to $M_{ee}$ for Case 2 of HSMR.}
\label{meematm_case2}
\end{minipage}
\hspace{0.7cm}
\begin{minipage}[b]{0.45\linewidth}
\centering
\includegraphics[width=6.5cm, height=5cm]{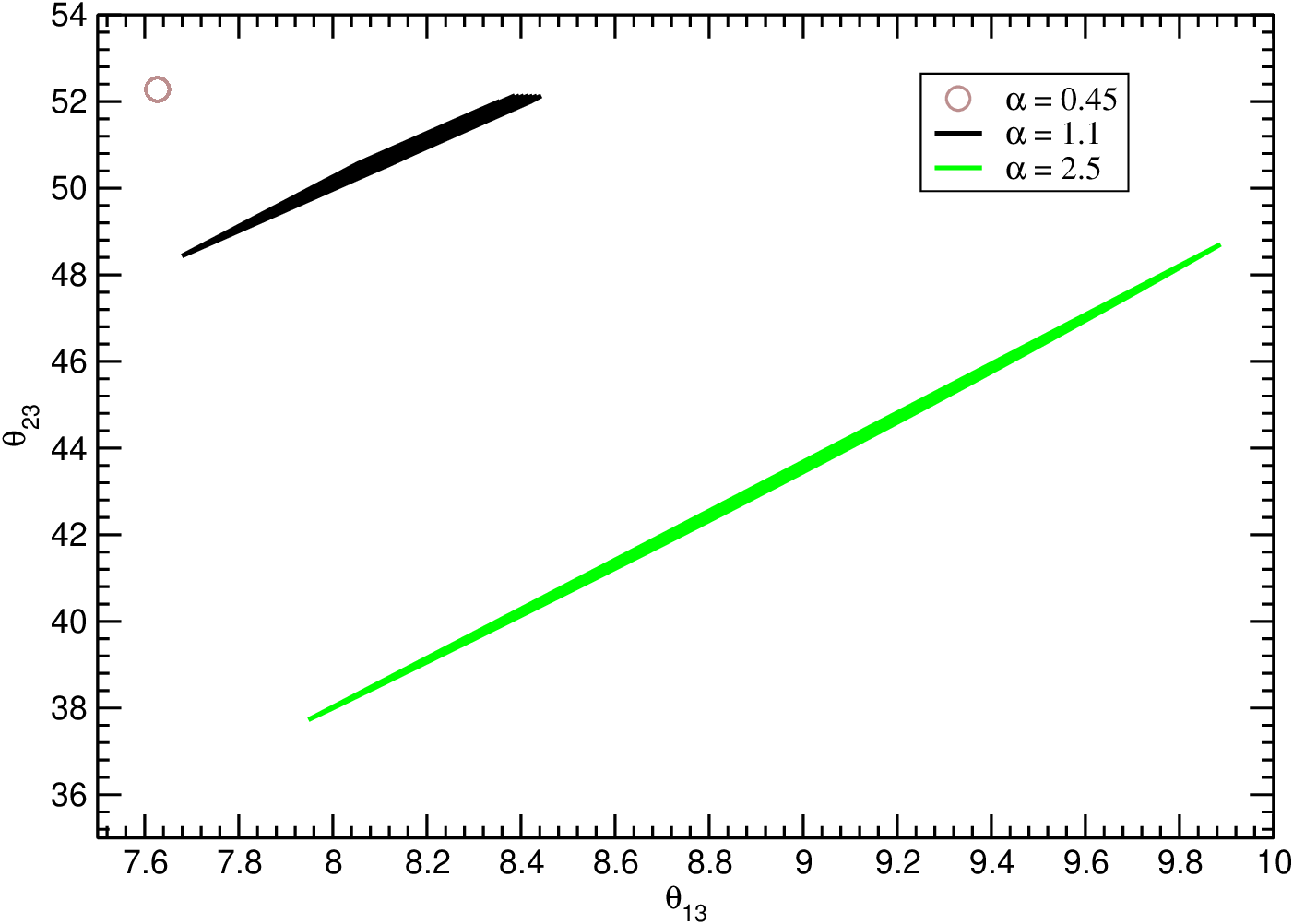}
\caption{The variation of $\theta_{23}$ with respect to $\theta_{13}$ for Case 2 of HSMR.}
\label{th13th23_case2}
\end{minipage}
\end{figure}

In Fig.~\ref{meematm_case2}, we show the behavior of $\Delta m^2_{32}$ versus $M_{ee}$ for
different values of $\alpha$.  For the $\alpha=0.45$ on the right panel,  we have 
$0.382~{\rm eV} \leq M_{ee} \leq 0.418$ eV  
which corresponds to the whole $3 \sigma$ global range of $\Delta m^2_{32}$. In case of $\alpha=1.1$ on 
the left panel,  $0.38~{\rm eV} \leq M_{ee} \leq 0.428$~eV belongs to  $\Delta m^2_{32} = (2.2-2.57)\times 10^{-3}~{\rm eV}^2$.  
The range of $ M_{ee} $ at the upper end $\alpha=2.5$ is $0.342~{\rm eV} \leq M_{ee}\leq 0.378$ eV corresponding 
to $\Delta m^2_{32} = (2.2-2.53)\times 10^{-3}~{\rm eV}^2$.  A remarkable
feature emerges in this case.  Unlike case 1, the absolute neutrino mass scale
increases for $\alpha < 1$ and decreases for $\alpha > 1$.
Now, at the upper allowed value of $\alpha=2.5$, $M_{ee}$ is sufficiently  below the GERDA limit.  
We would emphasize that one of the main observations of this case is that $\alpha$ is not 
constrained by the GERDA limit on either side.  These results can easily be tested by GERDA in the near future.

We show the  variation of $\theta_{23}$  with respect to  $\theta_{13}$ in Fig.~\ref{th13th23_case2}. As can be seen, for the lowest
possible value of $\alpha= 0.45$, the allowed range is just a point which is
located at $\theta_{13}=7.65^\circ$ and $\theta_{23}=52.5^\circ$.
As  $\alpha=1.1$, the range of $\theta_{13}$ is  $7.65^\circ- 8.4^\circ$ and that of $\theta_{23}$ is $48.5^\circ-52.5^\circ$. 
Finally for the highest value of $\alpha=2.5$,  $\theta_{13}$ has almost the whole  $3 \sigma$ range $7.92^\circ-9.88^\circ$  
and the range of $\theta_{23}$ is  $36.8^\circ-48^\circ$.  These results can be contrasted to case $1$ where the minimum of the mixing
angle $\theta_{23}=44.04^\circ$ also happens for the upper
value of $\alpha$ namely $\alpha=1.28$ and in both cases, the value
of $\theta_{23}$ can be in the first octant, contrary to the HSMU hypothesis.  Also in both cases,
the maximum of the mixing angle of $\theta_{23}$ corresponds to the lower value of alpha.
\begin{figure}[htb]
\begin{minipage}[b]{0.45\linewidth}
\vspace*{0.65cm}
\centering
\includegraphics[width=6.5cm, height=5cm]{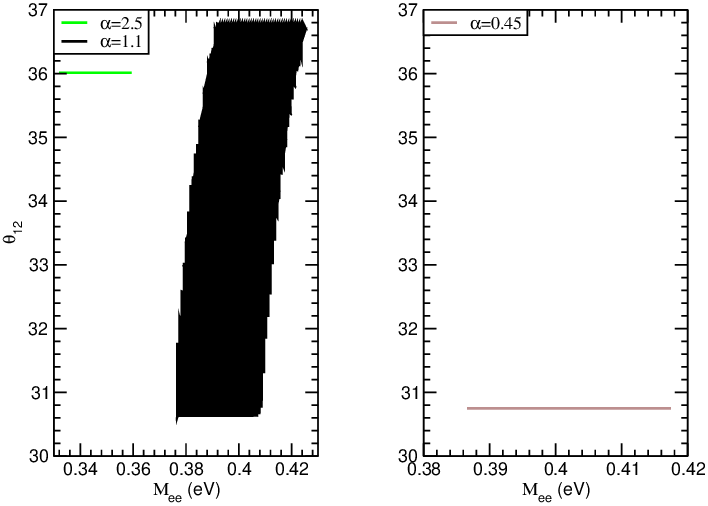}
\caption{ The variation of $\theta_{12}$ with respect to $M_{ee}$ for Case 2 of HSMR.}
    \label{meeth12_case2}
\end{minipage}
\hspace{0.7cm}
\begin{minipage}[b]{0.45\linewidth}
\centering
\includegraphics[width=6.5cm, height=5cm]{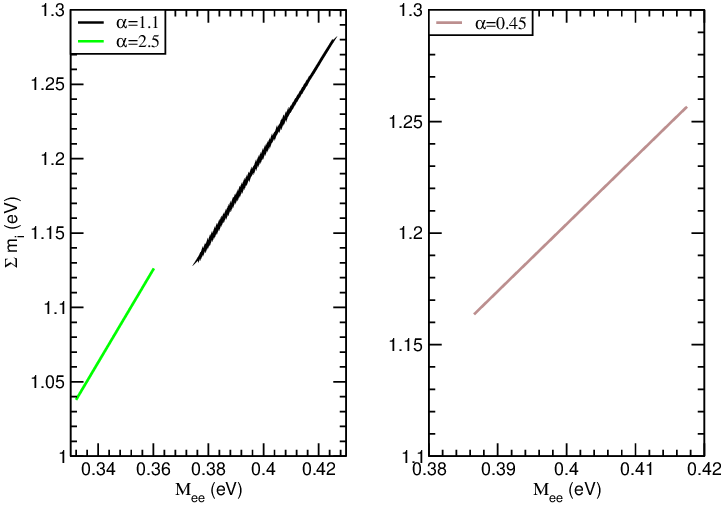}
\caption{The variation of $\Sigma m_i$ with respect to $M_{ee}$ for Case 2 of HSMR. }
    \label{meemTOT_case2}
\end{minipage}
\end{figure}

The next to be considered is the variation of $\theta_{12}$ versus $M_{ee}$ as
is shown in Fig.~\ref{meeth12_case2}. In the right panel it can be observed that
$\alpha=0.45$ corresponds to the minimum $\theta_{12}=30.8^\circ$ and $0.384~{\rm eV} \leq M_{ee} \leq 0.42$~eV.  For $\alpha=1.1$ on the left panel,
the whole $3 \sigma$ range for $\theta_{12}$ is allowed for
$0.39~{\rm eV} \leq M_{ee} \leq 0.4$~eV and for $M_{ee} \leq 0.39$ eV the allowed range of $\theta_{12}$ decreases.  
The upper value of $\alpha=2.5$, on the left panel, corresponds  $\theta_{12} = 36.02^\circ$  while $0.344~{\rm eV} \leq M_{ee} \leq 0.366$~eV.

In Fig.~\ref{meemTOT_case2}, we show the behavior of  the sum of the neutrino
masses with respect to $M_{ee}$. As can be seen in the right
panel, for the lowest value of $\alpha= 0.45$, $\Sigma m_i$ lies in the range $1.16-1.2$ eV
which corresponds to $0.383~{\rm eV} \leq M_{ee}\leq0.4$~eV.   For $M_{ee}>0.4$, the range of  
$\Sigma m_i$ is  $1.2-1.26$~eV. For the upper value of $\alpha= 2.5$,  we have $\Sigma m_i= 1.05-1.12$ eV while $0.342~{\rm eV} \leq M_{ee}\leq 0.366$~eV. 
On the left panel, for $\alpha= 1.1$, the range of $\Sigma m_i$ is $1.13-1.2$~eV  which corresponds to  
$0.36~{\rm eV} \leq M_{ee}\leq 0.4$ eV and the rest of the data point corresponds to $M_{ee}>0.4$~eV.

\subsubsection{Case 3: $\theta_{12}= \theta_{12}^q,~~\theta_{13} =  \theta_{13}^q,~~\theta_{23} 
= \alpha ~  \theta_{23}^q$}

\begin{figure}[htb]
\begin{minipage}[b]{0.45\linewidth}
\vspace*{0.65cm}
\centering
\includegraphics[width=6.5cm, height=5cm]{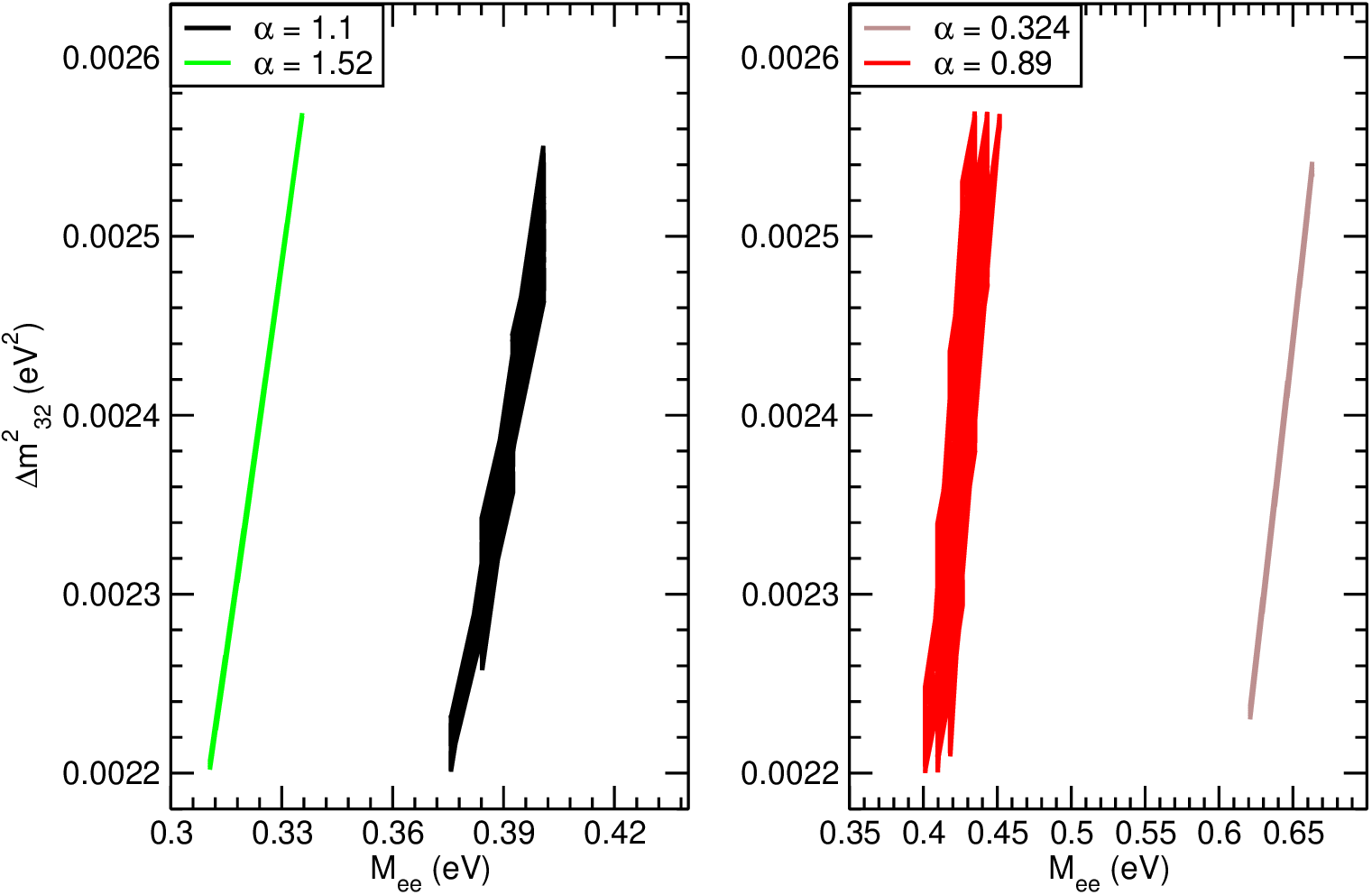}
\caption{ The variation of $\Delta m^2_{32}$ with respect to $M_{ee}$ for Case 3 of HSMR.}
\label{case3:fig1}
\end{minipage}
\hspace{0.7cm}
\begin{minipage}[b]{0.45\linewidth}
\centering
\includegraphics[width=6.5cm, height=5cm]{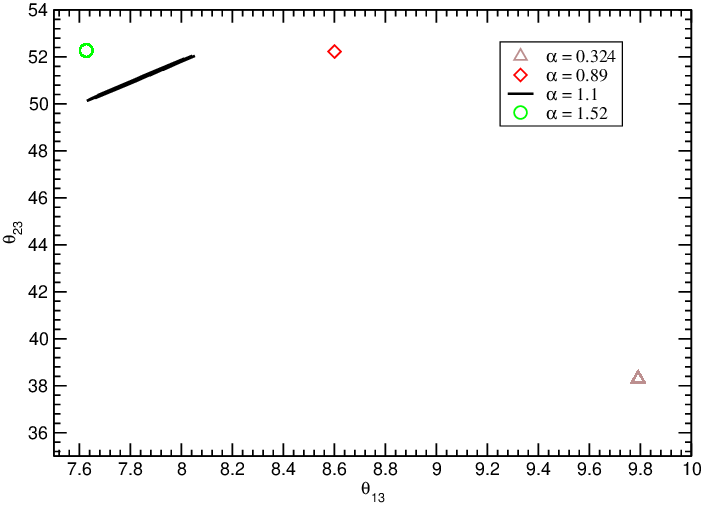}
\caption{The variation of $\theta_{23}$ with respect to $\theta_{13}$ for Case 3 of HSMR. }
\label{case3:fig2}
\end{minipage}
\end{figure}

We now consider the final case where two of the leptonic mixing angles $\theta_{12},~\theta_{13}$ are identical to the 
quark mixing angle $\theta_{12}^q,~\theta_{13}^q$ and the third leptonic mixing angle $\theta_{23}$ is proportional 
to the quark mixing angle $\theta_{23}^q$.  The correlation between $\Delta m_{32}^2$ and $M_{ee}$ is shown in Fig.~\ref{case3:fig1}.  The minimum allowed value of $\alpha$, with all the mixing parameters 
within the global range, is 0.324. However in this case as can be seen from the
right panel of Fig.~\ref{case3:fig1}, we have   $0.62~{\rm eV} \leq M_{ee} \leq 0.66$ eV.  
This value of $\alpha$ corresponds to the entire 3$\sigma$ range of $\Delta m_{32}^2$ and violates the upper limit from GERDA. 
Therefore we also consider the lower value of $\alpha$ = 0.89, with all the mixing parameters within the global range and 
$0.4~{\rm eV} \leq M_{ee} \leq 0.43$~eV.  This value corresponds to the entire 3$\sigma$ range of $\Delta m_{32}^2$.  
The prediction of $M_{ee} $ at $\alpha = 1.1 $ is $0.372~{\rm eV} \leq M_{ee} \leq 0.41$ eV corresponding to the 
entire 3$\sigma$ range of $\Delta m_{32}^2$.  The upper allowed value of $\alpha$ in this case is 1.52, (left panel of Fig.~\ref{case3:fig1})
with $0.3~{\rm eV} \leq M_{ee} \leq 0.34$~eV and having the entire 3$\sigma$ range of $\Delta m_{32}^2$.  Hence, 
the allowed range of $\alpha$ in this case covers the entire 3$\sigma$ range of $\Delta m_{32}^2$.   The absolute neutrino mass scale
increases for $\alpha < 1$ and decreases for $\alpha > 1$ similar to Case 2. The behavior of $\alpha$ in this case, 
Fig.~\ref{case3:fig1} is different from Case 1, Fig.~\ref{fig5}, with the lower end of $\alpha$ constraining $M_{ee}$. 
In this case it is possible to reach values of $M_{ee}$ as low as 0.3 eV
compared to Cases 1, 2 and the HSMU hypothesis, and will only be ruled out if 
the limit from GERDA reaches $M_{ee} <$ 0.3 eV.

We next show the correlation of $\theta_{23}$ with respect of $\theta_{13}$ in Fig.~\ref{case3:fig2}. 
The $\theta_{23}$ and $\theta_{13}$ reach their 3$\sigma$ global limits at the lowest and upper most end of $\alpha$.  
The value of  $\theta_{13}$ is $8.56^\circ$ and that of $\theta_{23}$ is $52.4^\circ $ for lower allowed value of $\alpha$ = 0.89, 
corresponding to $M_{ee}=0.4$ eV. The allowed ranges of $\theta_{13}$ and $\theta_{23}$ for $\alpha$ = 1.1 in this case are much more constrained 
compared to Cases 1, 2 and the HSMU hypothesis.  They  are  $7.62^\circ-8.05^\circ$ and $50.1^\circ-52.1^\circ $ respectively.  The upper end of $\alpha$ = 1.52, results in a minimum value of $\theta_{13}$,
whereas $\theta_{23}$ is at maximum with $\theta_{23}$ = $52.4^\circ$.  The behavior of $\alpha$ here is different from Cases 1 and 2 with the lower end of $\alpha$ resulting in the upper end point of $\theta_{13}$ and lower end point of $\theta_{23}$.  
\begin{figure}[htb]
\begin{minipage}[b]{0.45\linewidth}
\vspace*{0.65cm}
\centering
\includegraphics[width=6.5cm, height=5cm]{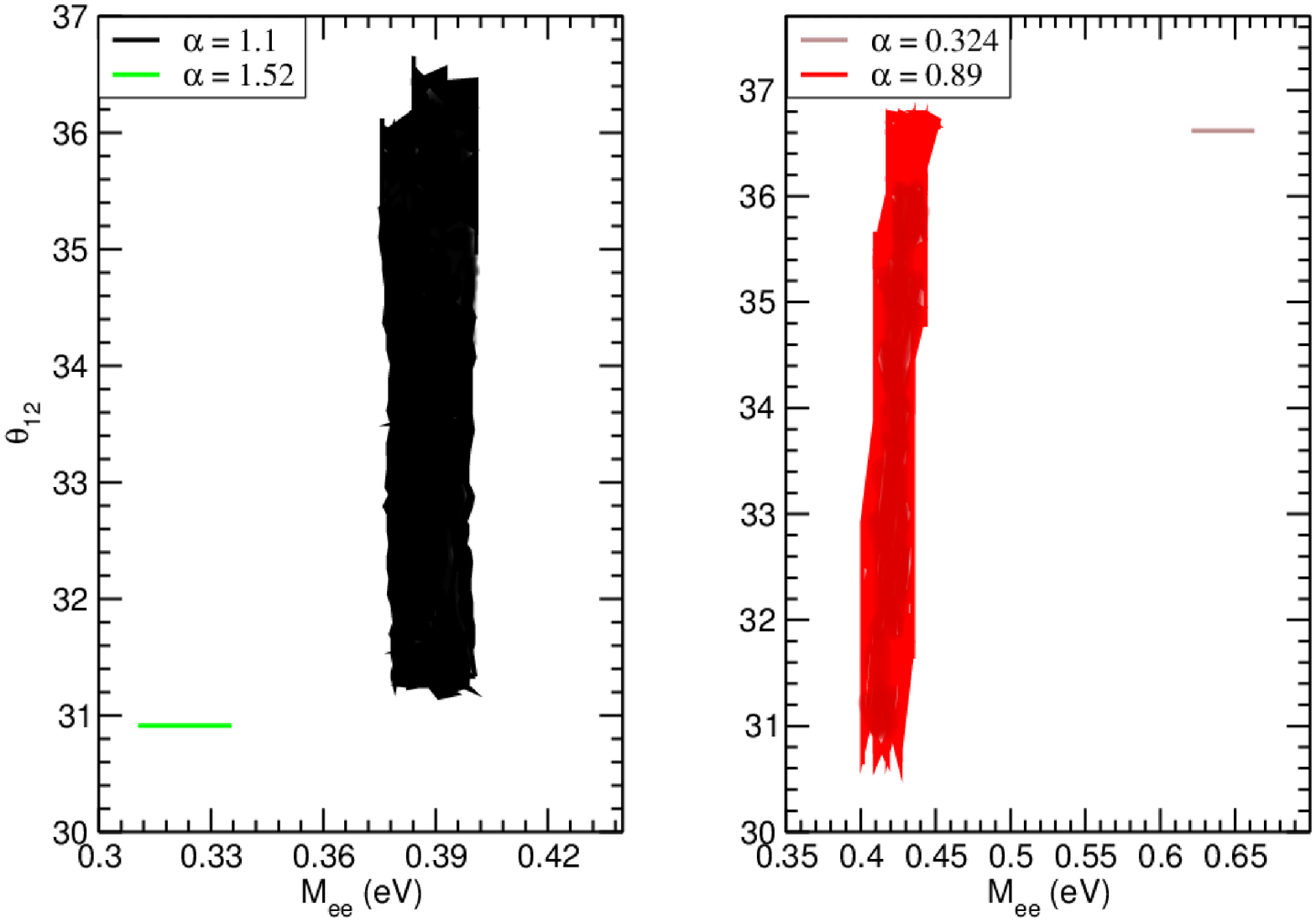}
\caption{ The variation of $\theta_{12}$ with respect to $M_{ee}$ for Case 3 of HSMR.} 
	\label{case3:fig3}
\end{minipage}
\hspace{0.7cm}
\begin{minipage}[b]{0.45\linewidth}
\centering
\includegraphics[width=6.5cm, height=5cm]{mee_msum_case3.eps}
\caption{The variation of $\Sigma m_i$ with respect to $M_{ee}$ for Case 3 of HSMR. }
	\label{case3:fig4}
\end{minipage}
\end{figure}

In Fig.~\ref{case3:fig3},  we show the variation of $\theta_{12}$ with $M_{ee}$. We observe that the 
lower ($30.78^\circ $) and upper ($36.7^\circ $) 3$\sigma$ global limits of $\theta_{12}$, are reached at
the upper most and the lowest ends of $\alpha$ respectively. In case of $\alpha$ = 0.89, the value of 
$\theta_{12}$ is $30.62^\circ$  which belongs to $M_{ee}=0.4$.   The whole 3$\sigma$ global range of $\theta_{12}$ 
is allowed for  $\alpha = 1.1$. 

Finally we show in Fig.~\ref{case3:fig4}, the variation of the 
sum of the neutrino masses with respect to $M_{ee}$. The region with $M_{ee} \geq$ 0.4 eV for $\alpha$ = 0.324, 
has $\Sigma m_i$ in the range 
of $1.84-2$ eV and for $\alpha = 0.89$ it is 
in the range  $1.2 - 1.38$ eV. In case of $\alpha$ = 1.1, with $M_{ee} <$ 0.4 eV, $\Sigma m_i$ is in the range $1.12 - 1.2$ eV.  
The upper end of $\alpha$ = 1.52 has the sum in the range of $0.93 - 1.01$ eV.  It is seen 
that $\Sigma m_i$ and $M_{ee}$ is much more relaxed compared to the HSMU and Cases 1, 2. 

%%%%%%%%%%%%%%%%%%%%%%%%%%%%%%%%%%%%%%%%%%%%%%%%%%%%%%%%%%%%%%%%%%%%%%%%%%%%%%%%%%%%%%%%%%%%%%%%%%%%%%%%%%%%%%%%%%%%%%%%%%%%%%%%%%%%%%%%%

\subsubsection{Case 4: $\theta_{12}= \alpha ~ \theta_{12}^q,~~\theta_{13} = \alpha ~ \theta_{13}^q,~~\theta_{23} 
=  \theta_{23}^q$}
\begin{figure}[htb]
\begin{minipage}[b]{0.45\linewidth}
\vspace*{0.65cm}
\centering
\includegraphics[width=6.5cm, height=5cm]{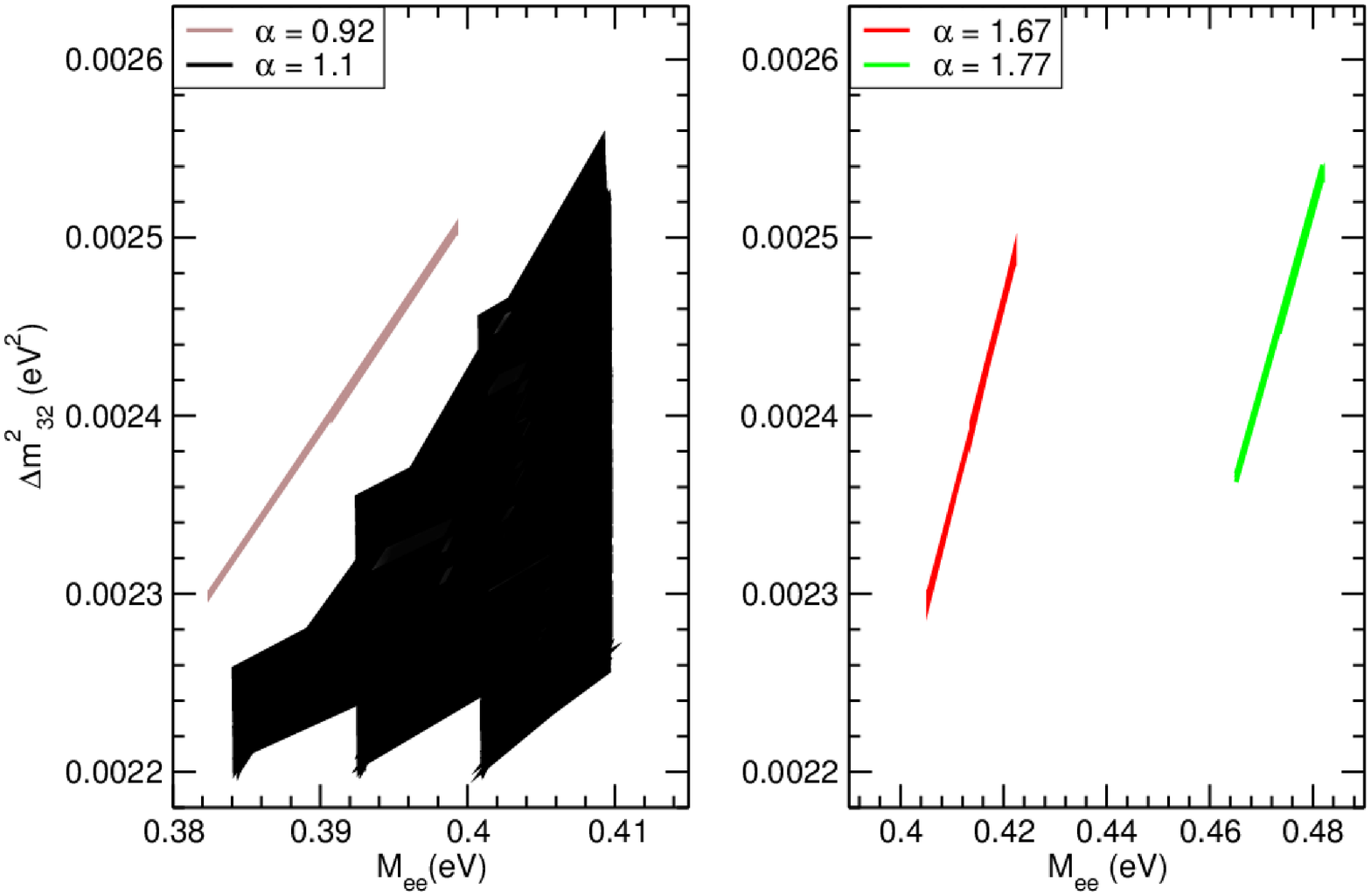}
\caption{ The variation of $\Delta m^2_{32}$ with respect to $M_{ee}$ for Case 4 of HSMR.}
\label{fig4.4}
\end{minipage}
\hspace{0.7cm}
\begin{minipage}[b]{0.45\linewidth}
\centering
\includegraphics[width=6.5cm, height=5cm]{th1323_case4.eps}
\caption{The variation of $\theta_{23}$ with respect to $\theta_{13}$ for Case 4 of HSMR. }
\label{fig4.1}
\end{minipage}
\end{figure}

We now consider the case where the leptonic mixing angles $\theta_{12}$, $\theta_{13}$ are proportional
to the corresponding quark mixing angles $\theta_{12}^q, \theta_{13}^q$ and the leptonic mixing 
angle $\theta_{23}$ is identical to the quark mixing angle $\theta_{23}^q$. The lowest allowed value of $\alpha$ 
for Case 4 is 0.92 which is derived using the $3 \sigma$ global limits on mixing angles.  The upper allowed value of $\alpha$, 
respecting the GERDA limit, is 1.67. When we relax the GERDA limit then $\alpha$ turns out to be 1.77 satisfying 
the $3 \sigma$ global limits.   We show, the correlation between $\Delta m^2_{32}$ and $M_{ee}$ in Fig.~\ref{fig4.4}. 
The lowest value of $\alpha = 0.92$, covers the range $0.38~{\rm eV} \leq M_{ee} \leq 0.4$ eV which corresponds to 
$\Delta m^2_{32} = (2.30-2.50)\times 10^{-3}~{\rm eV}^2$ (cf. left panel of Fig.~\ref{fig4.4}).  The prediction 
of $M_{ee}$ for  $\alpha =1.1$ is $0.384~{\rm eV} \leq M_{ee} \leq 0.41$ eV corresponding to the whole $3 \sigma$  
global range of $\Delta m^2_{32}$.  On the right panel, the  upper allowed end of $\alpha=1.67$ has  
$0.4~{\rm eV} \leq M_{ee} \leq 0.42$ eV with $\Delta m^2_{32} = (2.30-2.49)\times 10^{-3}~{\rm eV}^2$.  The upper 
most end  $\alpha=1.77$ where the GERDA limit is not satisfied, has $ 0.46~{\rm eV} \leq M_{ee} \leq 0.48$ eV 
corresponding to $\Delta m^2_{32} = (2.37-2.53)\times 10^{-3}~{\rm eV}^2$, as shown in the right panel 
of Fig.~\ref{fig4.4}. The behavior of $\alpha$ is similar to Case 1 with the upper values of $\alpha$ 
being constrained by the GERDA limit. The first
distinction that this case offers, with the others considered before is that the whole 3$\sigma$ range of $\Delta m^2_{32}$ 
is not covered in Case 4 for all the allowed values of $\alpha$. 

Next, we show the correlation between $\theta_{23}$ and $\theta_{13}$ as illustrated in Fig.~\ref{fig4.1}. 
The lower end of  $\alpha = 0.92$  reaches to the minimum of its  
3$\sigma$ global limit  for $\theta_{13}$  and the maximum of the 3$\sigma$ limit
for $\theta_{23}$.   The situation for the upper most end of $\alpha = 1.77$, is just opposite to the lower end, i.e. 
$\theta_{13}$ is at the maximum of the 3$\sigma$ global limit  whereas $\theta_{23}$ is at 
its global minimum.  This observation is just opposite to Case 3, where  $\theta_{13}$ ($\theta_{23}$)
reaches the global minimum (maximum), at the upper end of
$\alpha$. The allowed ranges of $\theta_{13}$ and $\theta_{23}$, at $\alpha = 1.1$ for 
this case are $7.62^\circ-9.2^\circ$ and $45.41^\circ-52.17^\circ$, respectively. 
The value of $\theta_{13}$ is $9.59^\circ$ and that of $\theta_{23}$ is $37.71^\circ-37.76^\circ$, for $\alpha = 1.67$.

\begin{figure}[htb]
\begin{minipage}[b]{0.45\linewidth}
\vspace*{0.65cm}
\centering
\includegraphics[width=6.5cm, height=5cm]{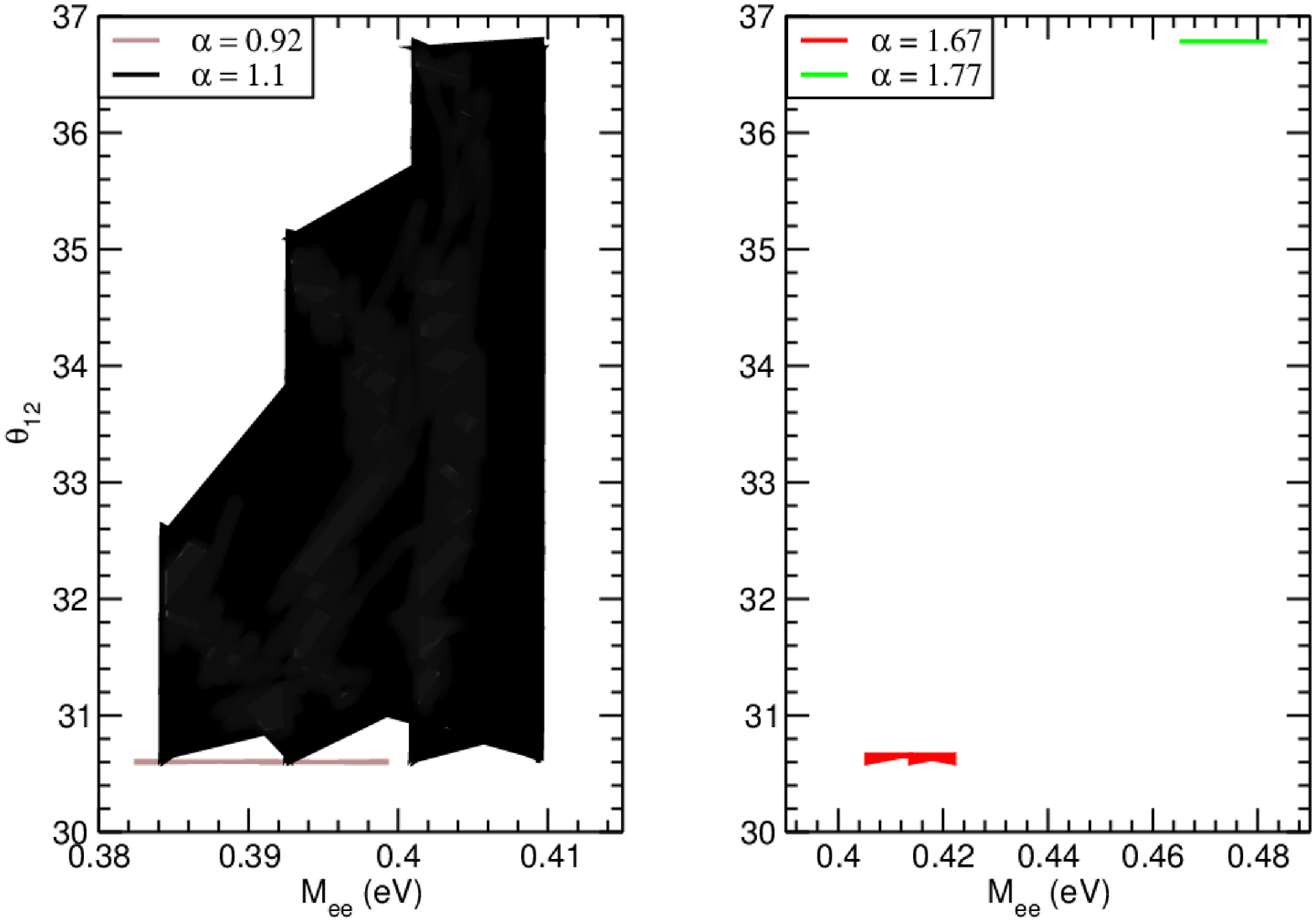}
\caption{ The variation of $\theta_{12}$ with respect to $M_{ee}$ for Case 4 of HSMR.} 
\label{fig4.6}
\end{minipage}
\hspace{0.7cm}
\begin{minipage}[b]{0.45\linewidth}
\centering
\includegraphics[width=6.5cm, height=5cm]{mee_msum_case4.eps}
\caption{The variation of $\Sigma m_i$ with respect to $M_{ee}$ for Case 4 of HSMR. }
	\label{fig4.2}
\end{minipage}
\end{figure}
The variation of $\theta_{12}$ with respect to $M_{ee}$ is shown in Fig.~\ref{fig4.6}. 
The lower and upper 3$\sigma$ global limits of $\theta_{12}$ are obtained  at the lower and upper most end of $\alpha$
respectively.  This observation is in contrast with Case 3 and Case 5 (to be discussed later). 
We get the full range of  $\theta_{12}$ (cf. Fig.~\ref{fig4.6} for details) for $\alpha = 1.1$, with some higher values 
of  $\theta_{12}$ being ruled out for $M_{ee} \leq 0.394$ eV.  The value of $\theta_{12}$ at $\alpha=1.67$ is $30.59^\circ-30.66^\circ$.

Finally, we show the variation of the sum of neutrino masses with respect to $M_{ee}$ in Fig.~\ref{fig4.2}. We find that 
for the lowest value of $\alpha = 0.92$, the sum of neutrino mass ranges between $1.15 - 1.20$ eV for  
$0.38~{\rm eV} \leq M_{ee} \leq 0.4$ eV. In case of $\alpha = 1.1$, $\Sigma m_i$ has a range of $1.157 - 1.23$ eV corresponding 
to $0.38~{\rm eV} \leq M_{ee} \leq 0.41$ eV as can be 
seen from the left panel of Fig.~\ref{fig4.2}.  A close look at the right panel of Fig.~\ref{fig4.2} reveals that for $\alpha = 1.67$, 
$\Sigma m_i$ is in the range $1.22-1.27$ eV corresponding to $0.4~{\rm eV} \leq M_{ee} \leq 0.42$ eV.   
The range for $\Sigma m_i$ turns out to be $1.4 - 1.45$ eV for $\alpha = 1.77$ which corresponds to $M_{ee}>0.4$ eV. 

%%%%%%%%%%%%%%%%%%%%%%%%%%%%%%%%%%%%%%%%%%%%%%%%%%%%%%%%%%%%%%%%%%%%%%%%%%%%%%%%%%%%%%%%%%%%%%%%%%%%%%%%%%%%%%%%%%%%%%%%%%%%%%%%%%%%%%%%%%%%%%%%%%%%%%%%%%%%%%%%%%%%%%%%%%%%%%%%%%%%%%%%%%%%%%%%%%%%%%%%%%%%%%%%%%%%%%%%%%%%%%%%%%

\subsubsection{Case 5: $\theta_{12}= \theta_{12}^q,~~\theta_{13} = \alpha~\theta_{13}^q,~~\theta_{23} 
= \alpha ~  \theta_{23}^q$}

%%%%%%%%%%%%%%%%%%%%%%%%%%%%%%%%%%%%%%%%%%%%%%%%%%%%%%%%%%%%%%%%%%%%%%%%%%%%%%%%%%%%%%%%%%%%%%%%%%%%%%%%%%%%%%%%%%%%%%%%%%%%%%%%%%%%%%%%%%%%%%%%%%%%%%%%%%%%%%%%%%%%%%%%%%%%%%%%%%%%%%%%%%%%%%%%%%%%%%%%%%%%%%%%%%%%%%%%%%%%%%%%%%

\begin{figure}[htb]
\begin{minipage}[b]{0.45\linewidth}
\vspace*{0.65cm}
\centering
\includegraphics[width=6.5cm, height=5cm]{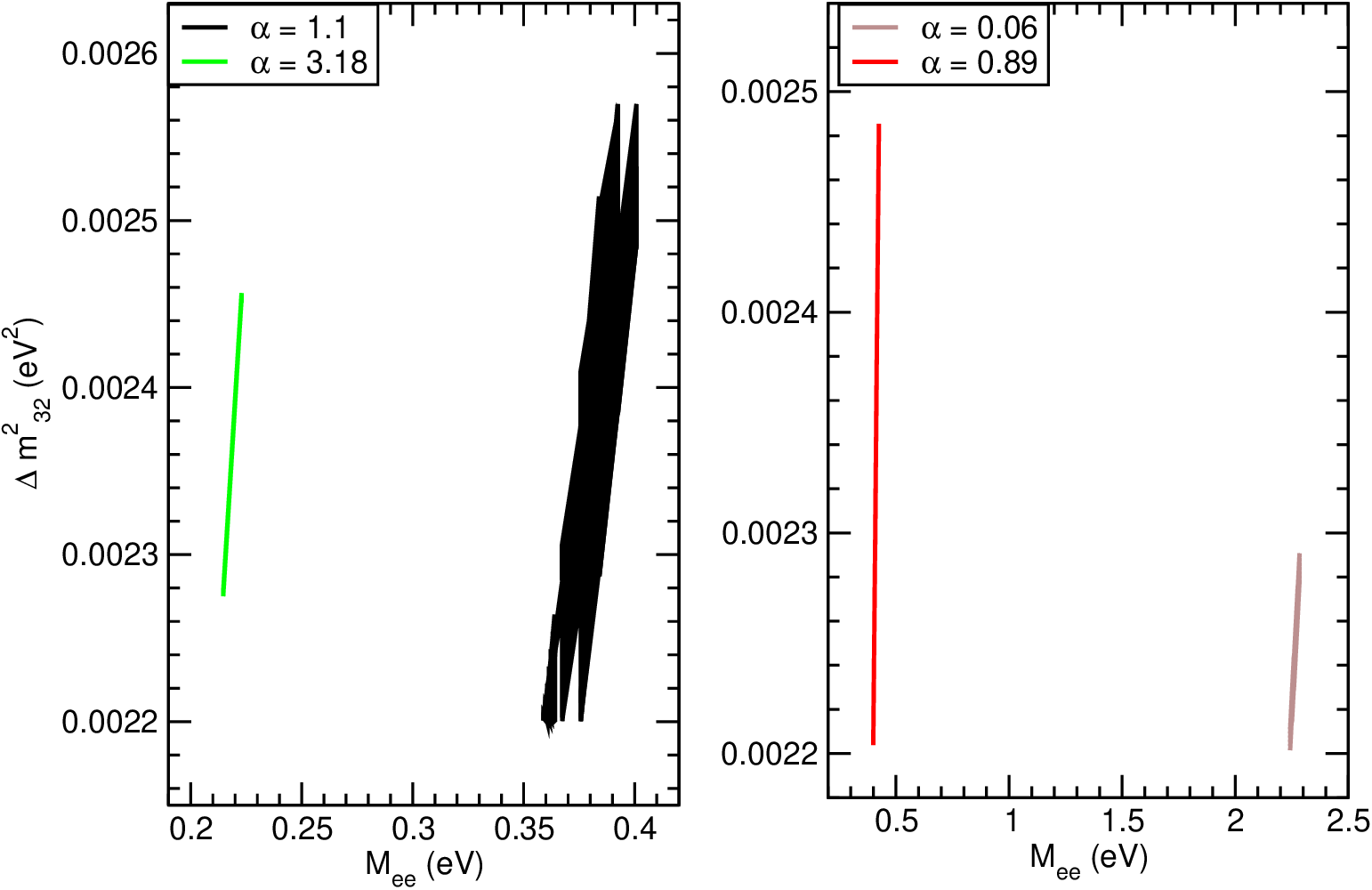}
\caption{ The variation of $\Delta m^2_{32}$ with respect to $M_{ee}$ for Case 5 of HSMR.}
\label{case5:fig1}
\end{minipage}
\hspace{0.7cm}
\begin{minipage}[b]{0.45\linewidth}
\centering
\includegraphics[width=6.5cm, height=5cm]{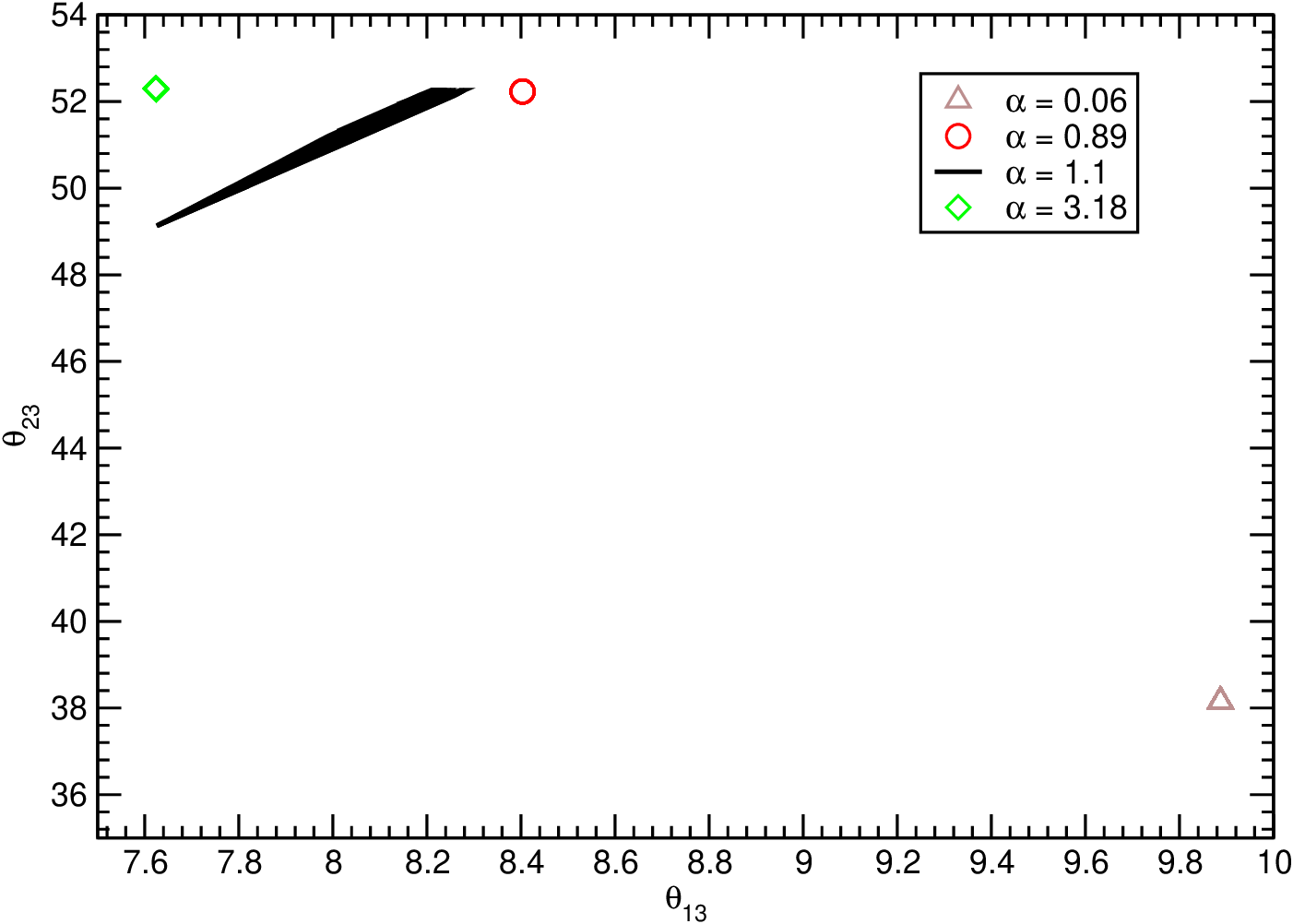}
\caption{The variation of $\theta_{23}$ with respect to $\theta_{13}$ for Case 5 of HSMR. }
\label{case5:fig2}
\end{minipage}
\end{figure}

We now look at the case of the leptonic mixing angle $\theta_{12}$ being identical with its CKM counterpart and 
the other two leptonic mixing angles being proportional to the quark mixing angles.  The correlation between 
$\Delta m_{32}^2$ and $M_{ee}$ is shown in Fig.~\ref{case5:fig1} as $\alpha$ deviates from unity.  
The minimum allowed value of $\alpha$, with all the mixing parameters 
within the global range, is 0.06. However in this case as can be seen from the
right panel of Fig.~\ref{case5:fig1}, we have  $2.24~{\rm eV} \leq M_{ee}\leq 2.28$ eV, which violates the 
upper limit from GERDA. Therefore including the constraints of GERDA, the lowest possible value of $\alpha$ becomes 0.89.
For $\alpha = 0.89$, as can be seen from the right panel of Fig.~\ref{case5:fig1}, 
we obtain $0.40~{\rm eV} \leq M_{ee} \leq 0.42$ eV corresponding to $\Delta m^2_{32}=(2.20-2.48) \times 10^{-3}~{\rm eV}^2$.  
The prediction of $M_{ee}$ for $\alpha$=1.1 from the left panel of Fig.~\ref{case5:fig1} is  $0.36~{\rm eV} \leq M_{ee} \leq 0.41$ eV 
which belongs to  the whole 3$\sigma$ range of $\Delta m^2_{32}$. The upper allowed value of
$\alpha$ in this case is 3.18, (left panel of Fig.~\ref{case5:fig1})
with $0.214~{\rm eV} \leq M_{ee} \leq 0.223$ eV corresponding to $\Delta m^2_{32}=(2.28-2.46) \times 10^{-3}~{\rm eV}^2$. 
The absolute neutrino mass scale increases for $\alpha < 1$ and decreases for $\alpha > 1$ similar to Cases 2 and 3. 
The behavior of $\alpha$ in this case is similar to Cases 2 and 3,
with the lower end of $\alpha$ being constrained by the GERDA limit. It can also
be seen from Fig.~\ref{case5:fig1}, that as we move towards the upper and lower ends
of $\alpha$, the whole 3$\sigma$ range of $\Delta m_{32}^2$ is not covered.

We next show the correlation of $\theta_{23}$ with respect of $\theta_{13}$ in Fig.~\ref{case5:fig2}. The 3$\sigma$
global end point limits of $\theta_{23}$ and $\theta_{13}$, are reached at the lowest and upper ends of $\alpha$.  The values of $\theta_{13} $ and 
$\theta_{23}$ for  lowest value of $\alpha$ = 0.89 are $8.40^\circ$ and $52.23^\circ$ respectively.  
These values belong to $M_{ee}=0.4$ eV.  The allowed ranges of $\theta_{13}$ and $\theta_{23}$, at 
$\alpha$ = 1.1 for this case are $7.62^\circ - 8.31^\circ$ and $49.0^\circ - 52.3^\circ$ respectively. 
The upper end of $\alpha$ = 3.18, results in a global minimum value of $\theta_{13}$ and a global maximal value of $\theta_{23}$,
similar to Case 3. The lower end of $\alpha$ results in a global maximum value of $\theta_{13}$ and a global minimum value of $\theta_{23}$. 

\begin{figure}[htb]
\begin{minipage}[b]{0.45\linewidth}
\vspace*{0.65cm}
\centering
\includegraphics[width=6.5cm, height=5cm]{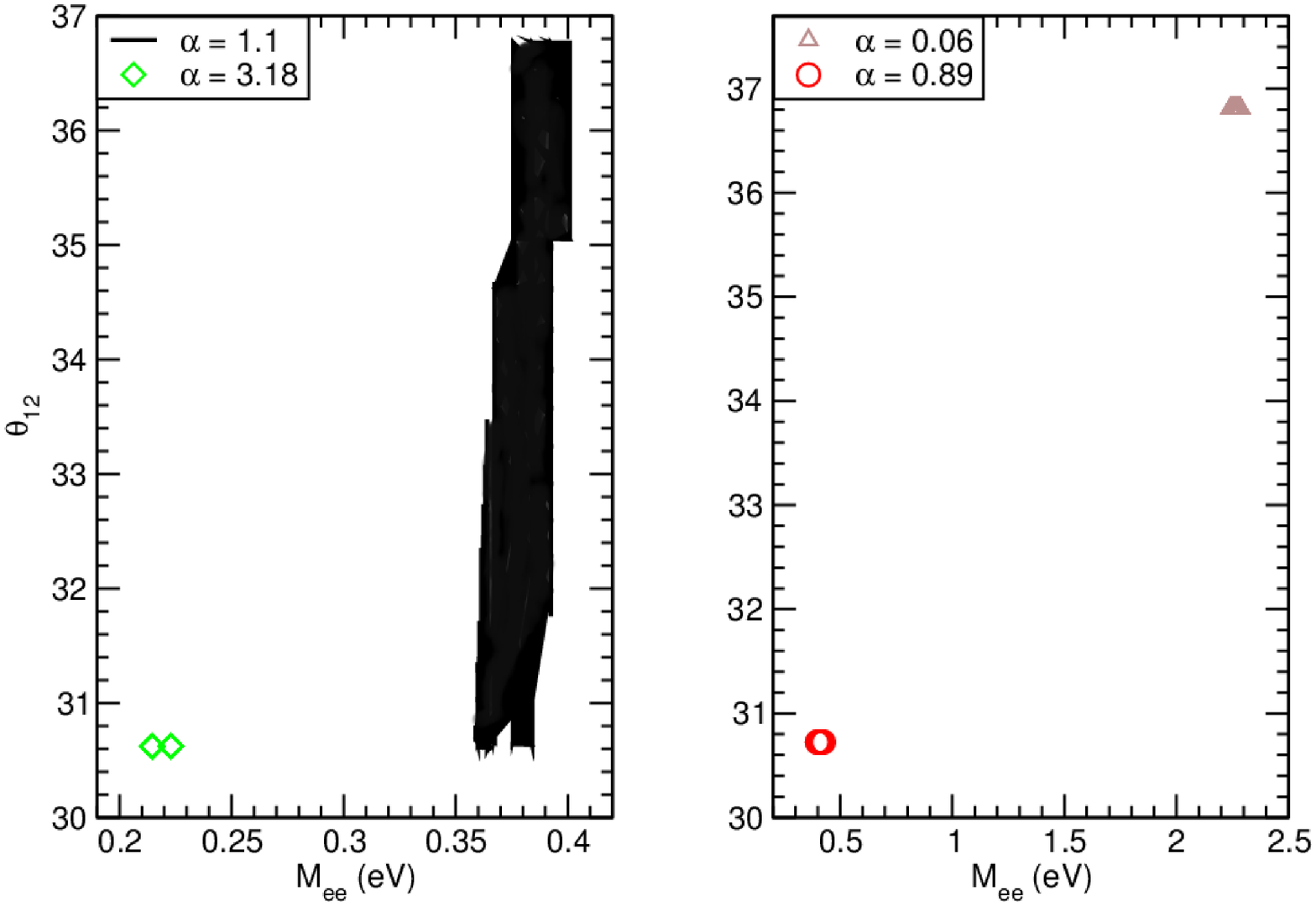}
\caption{ The variation of $\theta_{12}$ with respect to $M_{ee}$ for Case 5 of HSMR.} 
	\label{case5:fig3}
\end{minipage}
\hspace{0.7cm}
\begin{minipage}[b]{0.45\linewidth}
\centering
\includegraphics[width=6.5cm, height=5cm]{mee_msum_case5_v1.eps}
\caption{The variation of $\Sigma m_i$ with respect to $M_{ee}$ for Case 5 of HSMR. }
	\label{case5:fig4}
\end{minipage}
\end{figure}

We next show the variation of $\theta_{12}$ with $M_{ee}$ in Fig.~\ref{case5:fig3}. 
The lower ($30.62^\circ$) and the upper ($36.81^\circ$) 3$\sigma$ global limits of $\theta_{12}$, correspond
to the upper and the lowest ends of $\alpha$. However, for case of $\alpha$ = 1.1, the whole 3$\sigma$ range of 
$\theta_{12}$ ($30.62^\circ-36.81^\circ$) is covered. 

Finally we show in Fig.~\ref{case5:fig4}, the variation of the 
sum of the neutrino masses with respect to $M_{ee}$. The region with $M_{ee} \geq$ 0.4 eV,
for $\alpha$ = 0.06, has $\Sigma m_i$ in the range of $6.73 - 6.85$ eV and for $\alpha = 0.89$ it is 
in the range  $1.20 - 1.28$ eV. For $\alpha$ = 1.1, with $M_{ee} <$ 0.4 eV, $\Sigma m_i$ is in the range $1.07 - 1.22$ eV.  
The upper end of $\alpha$ = 3.18 has the sum in the range of $0.65 - 0.67$ eV which is remarkably 
lower than the previous cases.  We note that the $\Sigma m_i$ is below the cosmological upper
bound~\cite{Ade:2015xua}. The further discussion on the cosmological constraints on 
our work will be provided in the last section of this paper. It is observed that 
Case 5 behaves almost similarly to Case 3, and is the most relaxed one in terms of 
$M_{ee}$. We can go to values of $M_{ee}$ as low as 0.21 eV, consistent with the upper 
end of $\alpha$.  Hence, this case is partially beyond the reach of GERDA sensitivity which is maximum $0.3$ eV.  However, this is well within the reach of KATERIN experiment \cite{Drexlin:2013lha}.   
%%%%%%%%%%%%%%%%%%%%%%%%%%%%%%%%%%%%%%%%%%%%%%%%%%%%%%%%%%%%%%%%%%%%%%%%%%%%%%%%%%%%%%%%%%%%%%%%%%%%%%%%%%%%%%%%%%%%%%%%%%%%%%%%%%%%%%%%%%%%%%%%%%%%%%%%%%%%%%%%%%%%%%%%%%%%%%%%%%%%%%%%%%%%%%%%%%%%%%%%%%%%%%%%%%%%%%%%%%%%%%%%%%

\subsubsection{Case 6: $\theta_{12}= \alpha~\theta_{12}^q,~~\theta_{13} =  \theta_{13}^q,~~\theta_{23} 
= \alpha ~  \theta_{23}^q$}

%%%%%%%%%%%%%%%%%%%%%%%%%%%%%%%%%%%%%%%%%%%%%%%%%%%%%%%%%%%%%%%%%%%%%%%%%%%%%%%%%%%%%%%%%%%%%%%%%%%%%%%%%%%%%%%%%%%%%%%%%%%%%%%%%%%%%%%%%%%%%%%%%%%%%%%%%%%%%%%%%%%%%%%%%%%%%%%%%%%%%%%%%%%%%%%%%%%%%%%%%%%%%%%%%%%%%%%%%%%%%%%%%%

\begin{figure}[htb]
\begin{minipage}[b]{0.45\linewidth}
\vspace*{0.65cm}
\centering
\includegraphics[width=6.5cm, height=5cm]{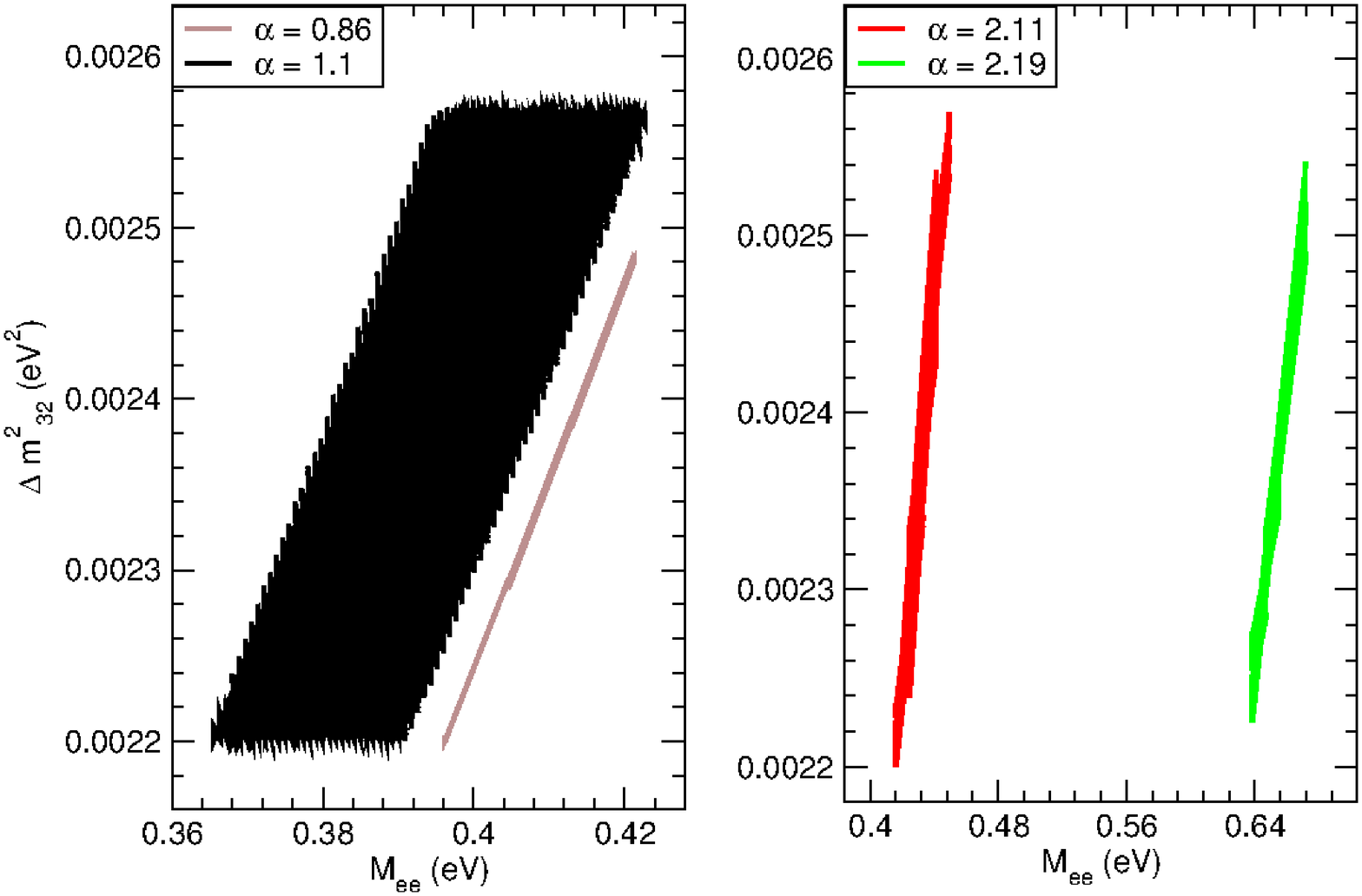}
\caption{ The variation of $\Delta m^2_{32}$ with respect to $M_{ee}$ for Case 6 of HSMR.}
\label{case6:fig1}
\end{minipage}
\hspace{0.7cm}
\begin{minipage}[b]{0.45\linewidth}
\centering
\includegraphics[width=6.5cm, height=5cm]{th1323_case6_v2.eps}
\caption{The variation of $\theta_{23}$ with respect to $\theta_{13}$ for Case 6 of HSMR. }
\label{case6:fig2}
\end{minipage}
\end{figure}

We next consider the case where the leptonic mixing angle $\theta_{13}$ is identical with its CKM counterpart and 
the other two leptonic mixing angles are proportional to the quark mixing angles.  
The correlation between $\Delta m_{32}^2$ and $M_{ee}$ is shown in Fig.~\ref{case6:fig1}.  
The minimum allowed value of $\alpha$, with all the mixing parameters 
within the global range, is 0.86.  It can be seen from the left panel of Fig.~\ref{case6:fig1}, for this value of $\alpha$,
$M_{ee}$ has a range $0.397~\rm{eV} \leq M_{ee}\leq 0.42$ eV corresponding to $\Delta m^2_{32}=(2.20-2.48) \times 10^{-3}~{\rm eV}^2$. 
In case of $\alpha=1.1$, we have $0.36~{\rm eV} \leq M_{ee}\leq 0.424$ eV which corresponds to the whole 3$\sigma$ range of  $\Delta m^2_{32}$.  
At the upper allowed value of $\alpha=2.11$ as seen from the right panel of Fig.~\ref{case6:fig1}, we have 
$0.41~{\rm eV} \leq M_{ee}\leq 0.45$ eV with the whole 3$\sigma$ 
range of  $\Delta m^2_{32}$ covered.  We have $0.64~{\rm eV} \leq M_{ee}\leq 0.67$ eV corresponding to 
$\Delta m^2_{32}=(2.22-2.54) \times 10^{-3}~{\rm eV}^2$ for the uppermost value of $\alpha=2.19$.  This end is already rejected by the GERDA limit.  
In this case, it is worth mentioning that the absolute neutrino mass scale increases for both $\alpha < 1$ and $\alpha > 1$.

The behavior of $\alpha$ in this case Fig.~\ref{case6:fig1} is different from Case 5, 
Fig.~\ref{case5:fig1}. Unlike Case 5 considered before, in this case for both 
lower and upper end values of $\alpha$ we get $M_{ee}$ close to its upper limit. 
This is because unlike the previous cases, in this case the limit on lower value of $\alpha$ comes not from $M_{ee}$ but 
from the neutrino oscillation parameters. 
Also, although the upper limit of $\alpha = 2.11$ is constrained by $M_{ee}$ but this value is quite close to the 
upper limit of 2.19 obtained without the $M_{ee}$ constraint.

In Fig.~\ref{case6:fig2}, we show the 
correlation of $\theta_{23}$ with respect of $\theta_{13}$. The 3$\sigma$
global end point limits of $\theta_{23}$ and $\theta_{13}$, are reached at the lowest and uppermost end of $\alpha$.  
The allowed ranges of $\theta_{13}$ and $\theta_{23}$, at $\alpha$ = 1.1 for this case are 
$7.62^\circ-8.90^\circ$ and $47.0^\circ-52.3^\circ$ respectively.   At  $\alpha=2.11$, 
the value of $\theta_{13}$ is $9.4^\circ-9.76^\circ$ and that of $\theta_{23}$ is $38.7^\circ-38.8^\circ$.

\begin{figure}[htb]
\begin{minipage}[b]{0.45\linewidth}
\vspace*{0.65cm}
\centering
\includegraphics[width=6.5cm, height=5cm]{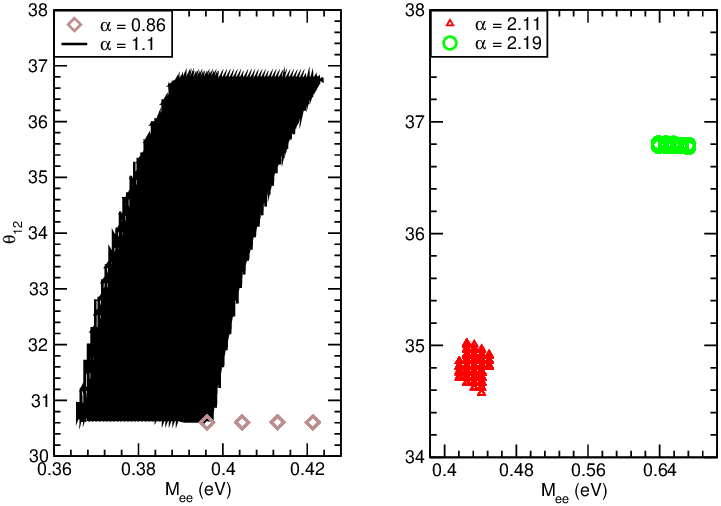}
\caption{ The variation of $\theta_{12}$ with respect to $M_{ee}$ for Case 6 of HSMR.} 
	\label{case6:fig3}
\end{minipage}
\hspace{0.7cm}
\begin{minipage}[b]{0.45\linewidth}
\centering
\includegraphics[width=6.5cm, height=5cm]{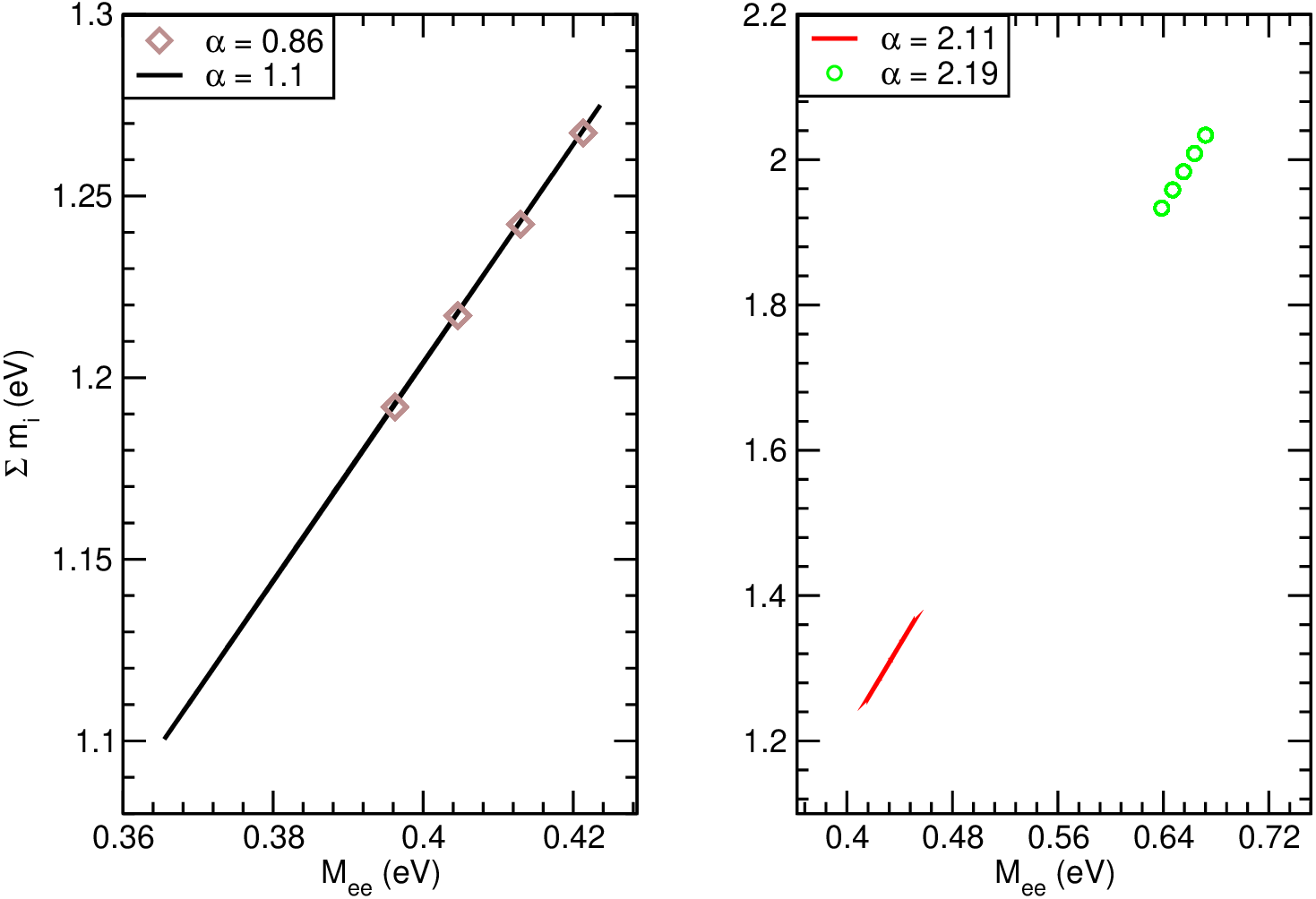}
\caption{The variation of $\Sigma m_i$ with respect to $M_{ee}$ for Case 6 of HSMR. }
	\label{case6:fig4}
\end{minipage}
\end{figure}

We next show the variation of $\theta_{12}$ with $M_{ee}$ in Fig.~\ref{case6:fig3}. 
The lower ($30.60^\circ$) and upper ($36.81^\circ$) 3$\sigma$ global limits of $\theta_{12}$, 
is reached at the lowest and the uppermost end of $\alpha$. This behavior is quite the opposite of the behavior 
shown in Fig.~\ref{case5:fig3} for Case 5. In case of $\alpha$ = 1.1, 
the whole 3$\sigma$ range of 
$\theta_{12}$ $(30.60^\circ - 36.81^\circ)$ is covered. The value of $\theta_{12}$ at $\alpha$ = 2.11 is $34.57^\circ-35.02^\circ$.

Finally we show in Fig.~\ref{case6:fig4}, the variation of the 
sum of the neutrino masses with respect to $M_{ee}$. In case of $\alpha$ = 0.86 the sum of neutrino masses $\Sigma m_i$ is
in the range of $1.19-1.27$ eV.  Next for $\alpha = 1.1$,  $\Sigma m_i$ has a range of $1.1-1.27$ eV for $M_{ee} \leq 0.4$ eV
and when $\alpha = 2.11$ it is in the range  $1.26 - 1.36$ eV for $0.4~{\rm eV} \leq M_{ee} \leq 0.48$ eV. 

\subsubsection{Case 7: $\theta_{12}= \alpha ~ \theta_{12}^q,~~\theta_{13} =  \alpha ~ \theta_{13}^q,~~\theta_{23} 
= \alpha ~  \theta_{23}^q$}

We finally consider the case where all the leptonic mixing  angles are proportional to the quark mixing 
angle by the same proportionality constant ($\alpha$). We find that
the upper bound on $\alpha$, is constrained by the mass limit ($M_{ee}$) from GERDA, whereas the lower 
limit on $\alpha$ is constrained by the $3 \sigma$ global limit of the leptonic mixing angles.  The 
lowest value of $\alpha$ is 0.89 and the highest value of $\alpha$ relaxing the GERDA limit is 2.09, 
whereas by taking into account the $M_{ee}$ limit, the highest value is 2.  

\begin{figure}[htb]
\begin{minipage}[b]{0.45\linewidth}
\vspace*{0.65cm}
\centering
\includegraphics[width=6.5cm, height=5cm]{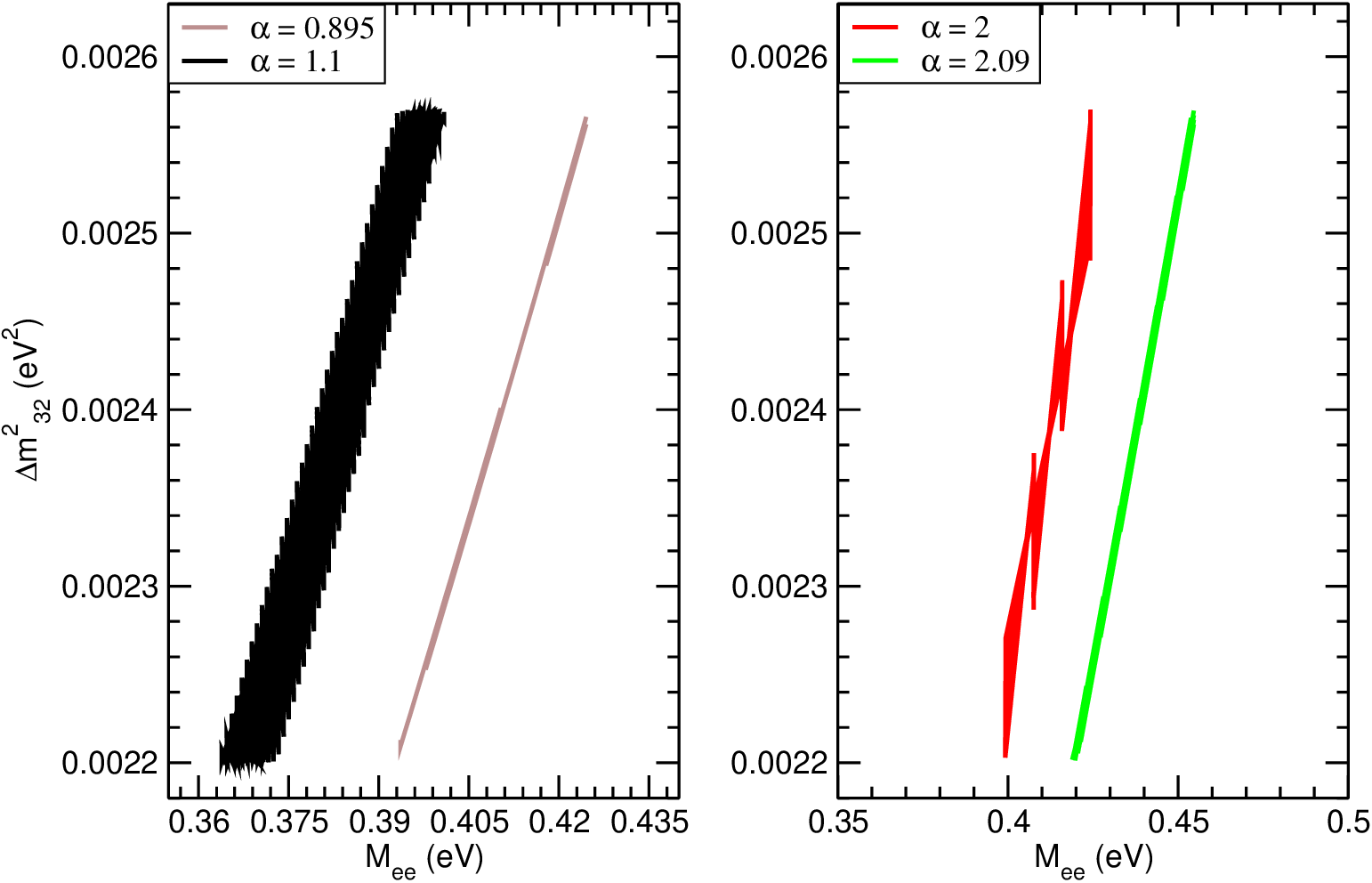}
\caption{ The variation of $\Delta m^2_{32}$ with respect to $M_{ee}$ for Case 7 of HSMR.}
\label{case7:fig1}
\end{minipage}
\hspace{0.7cm}
\begin{minipage}[b]{0.45\linewidth}
\centering
\includegraphics[width=6.5cm, height=5cm]{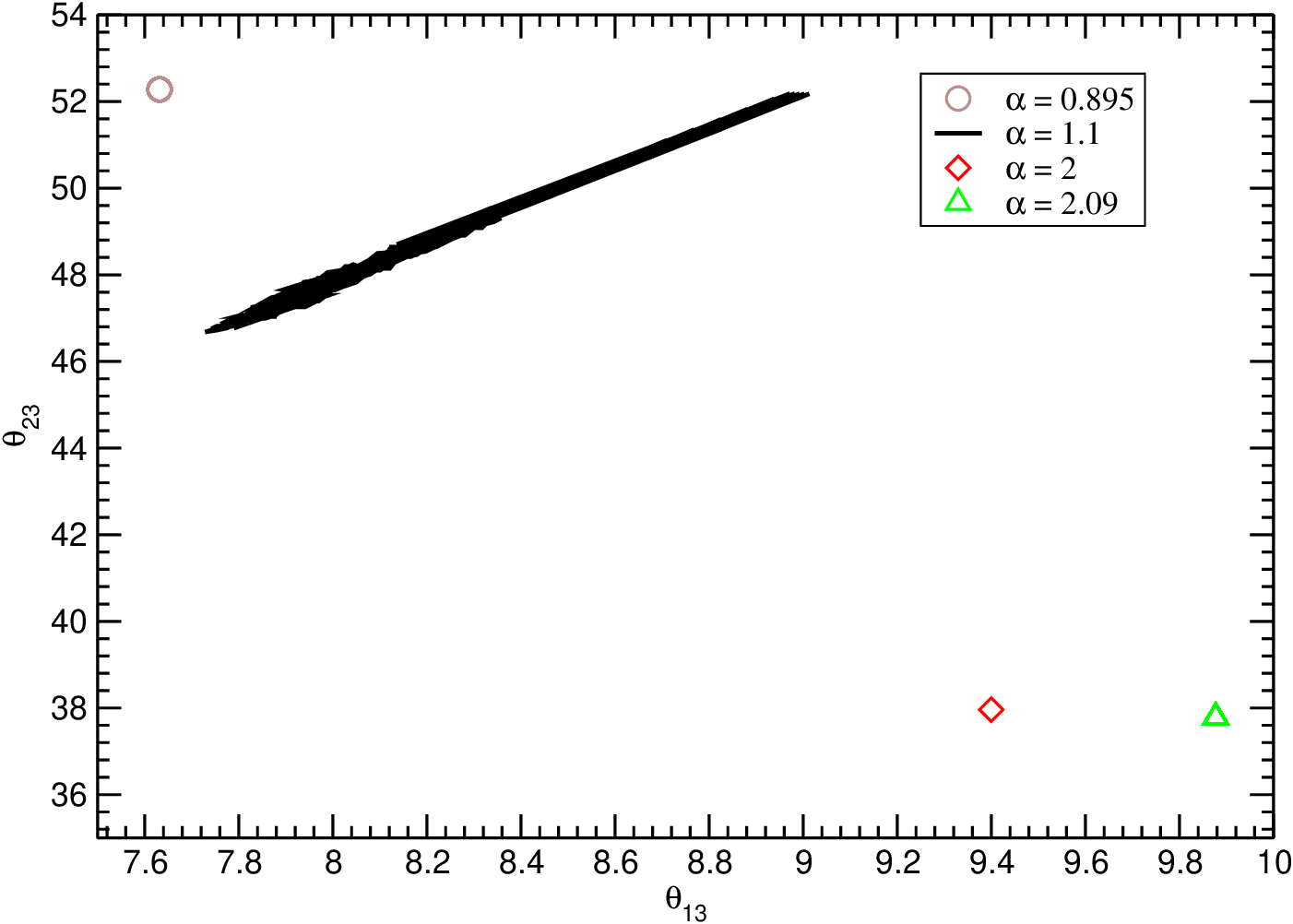}
\caption{The variation of $\theta_{23}$ with respect to $\theta_{13}$ for Case 7 of HSMR. }
\label{case7:fig2}
\end{minipage}
\end{figure}

We next discuss the 
behavior of the neutrino mass and mixing parameters in Case 7, with the variation of $\alpha$ in the 
allowed range. Firstly like all the previous cases, the variation of $\Delta m_{32}^2$ 
with $M_{ee}$ is shown in Fig.~\ref{case7:fig1}.  As seen from the left panel of Fig.~\ref{case7:fig1},
for the lowest value of $\alpha = 0.895$, 
we have  $0.391~{\rm eV} \leq M_{ee}\leq 0.425$ eV which corresponds to the whole 3$\sigma$ global range of  $\Delta m^2_{32}$. 
At $\alpha =1.1$, the range of $M_{ee}$ is  $0.362~{\rm eV} \leq M_{ee}\leq 0.405$ eV which again corresponds to 
the whole 3$\sigma$ global range of  $\Delta m^2_{32}$.  The range of $M_{ee}$ at $\alpha =2$ is 
$0.4~{\rm eV} \leq M_{ee}\leq 0.42$ eV which corresponds to the whole 3$\sigma$ global range of  $\Delta m^2_{32}$.  
The uppermost end of $\alpha =2.09$ has $0.42~{\rm eV} \leq M_{ee}\leq 0.452$ eV which corresponds to the 
whole 3$\sigma$ global range of  $\Delta m^2_{32}$ and is rejected by the GERDA limit.  Hence, 
the entire allowed range of $\alpha$ covers the  whole 3$\sigma$ range of $m_{32}^2$.  The 
absolute neutrino mass scale increases for both $\alpha < 1$ and $\alpha > 1$ similar to Case 6.  
The behavior of $\alpha$ resembles to Case 6 with the upper
and lower ends of $\alpha$ having values close to $M_{ee}$. 

We show the range of $\theta_{23}$ and $\theta_{13}$ covered by the different allowed values of
$\alpha$ in Fig.~\ref{case7:fig2}.   The 3$\sigma$ global limits on the mixing angles are reached at the lower and uppermost 
ends of $\alpha$.  The allowed ranges of $\theta_{13}$ and $\theta_{23}$ for $\alpha$ = 1.1 are 
$7.76^\circ - 9.02^\circ$  and $46.3^\circ - 52.26^\circ$ respectively.  For $\alpha=2.0$, the value 
of $\theta_{13}$ is $9.44^\circ$ and that of $\theta_{23}$ is $37.8^\circ$ which belongs to $M_{ee}=0.4$ eV.  
The uppermost end of $\alpha$ = 2.09, gives the value of $\theta_{13}$ at its global upper limit, 
whereas $\theta_{23}$ is kept at its global lower limit. The converse is true for the lower end 
of $\alpha$ with $\theta_{13},~\theta_{12}$ at its lower value and 
$\theta_{23}$ at its maximum. 

\begin{figure}[htb]
\begin{minipage}[b]{0.45\linewidth}
\vspace*{0.65cm}
\centering
\includegraphics[width=6.5cm, height=5cm]{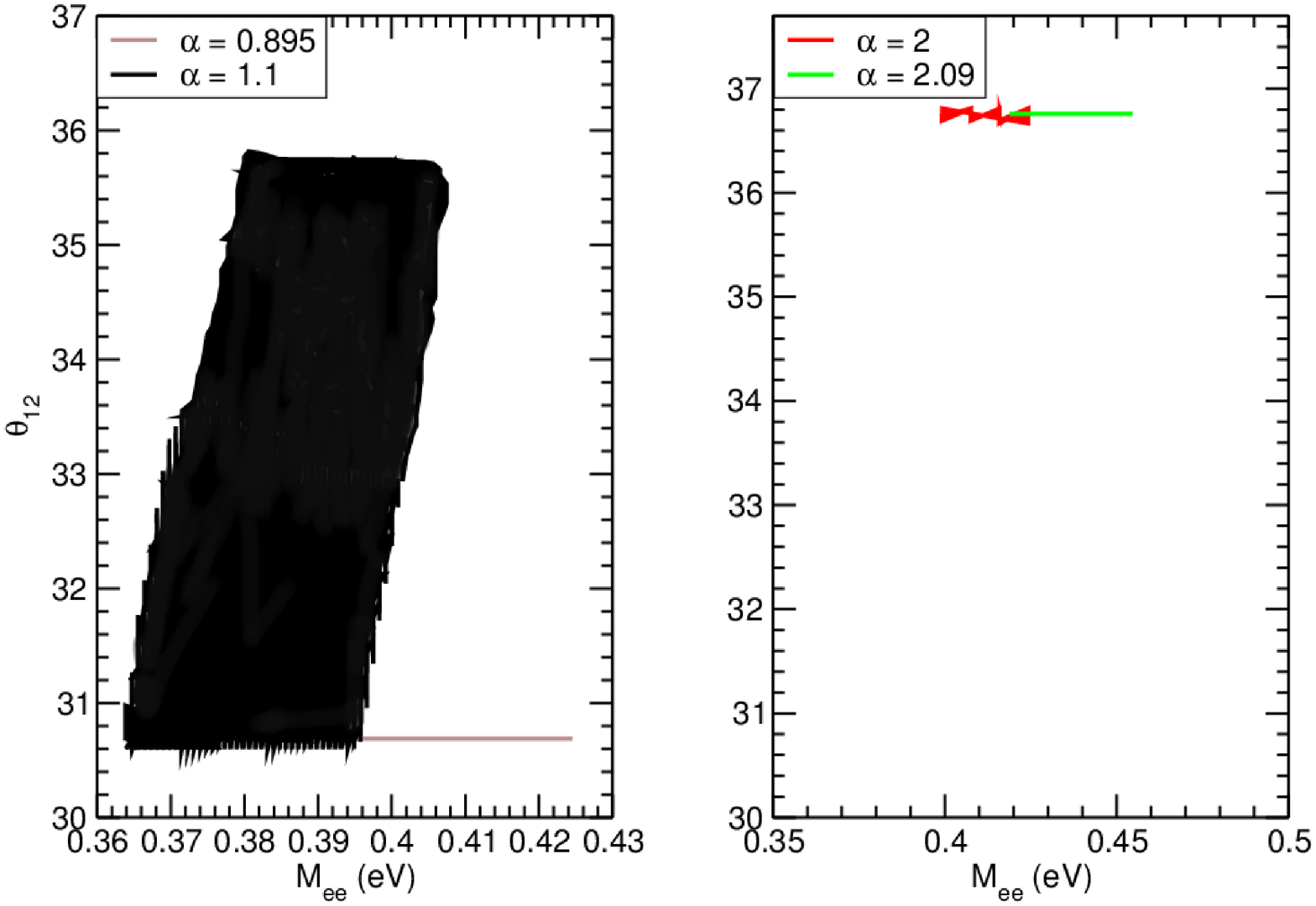}
\caption{ The variation of $\theta_{12}$ with respect to $M_{ee}$ for Case 7 of HSMR.} 
	\label{case7:fig3}
\end{minipage}
\hspace{0.7cm}
\begin{minipage}[b]{0.45\linewidth}
\centering
\includegraphics[width=6.5cm, height=5cm]{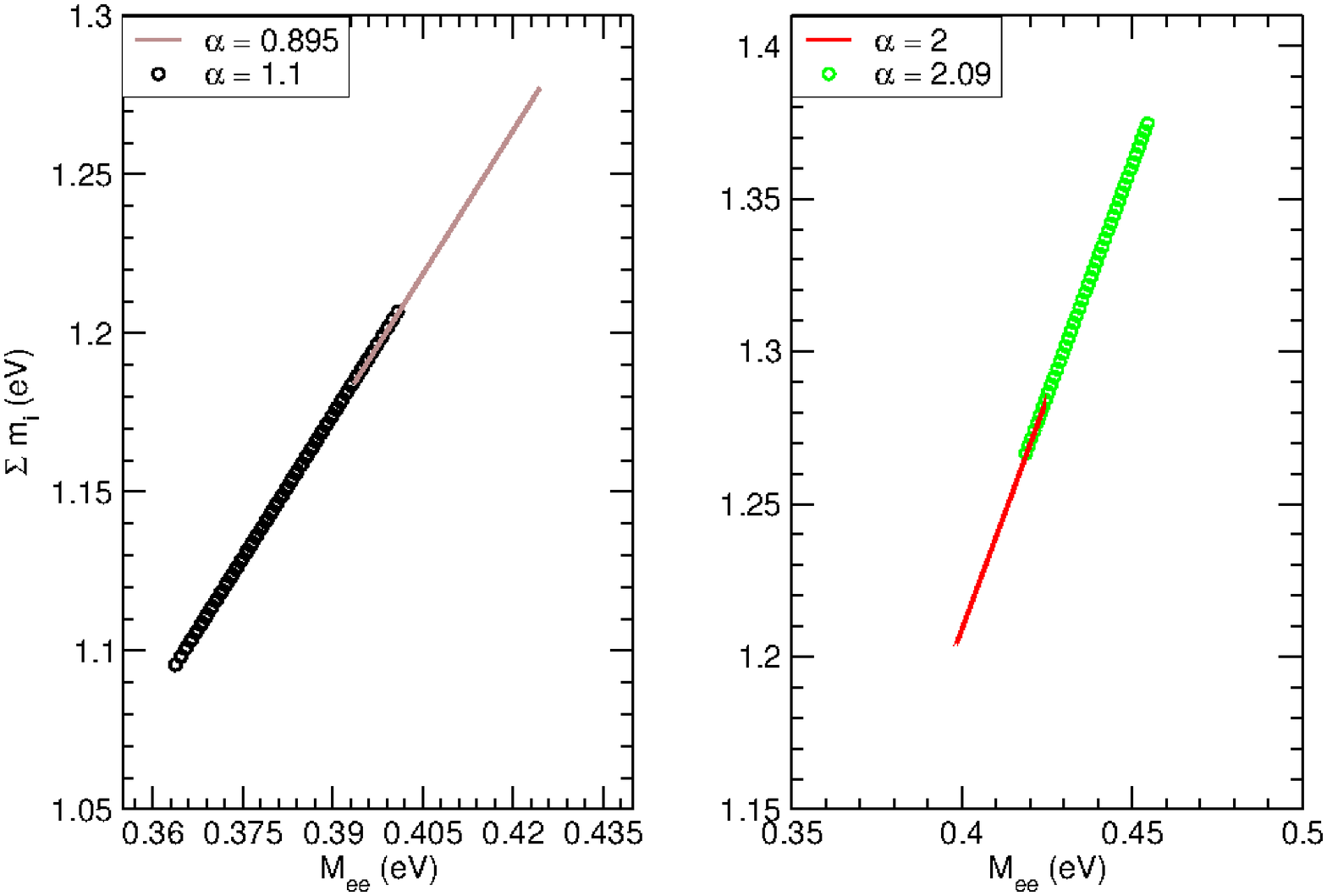}
\caption{The variation of $\Sigma m_i$ with respect to $M_{ee}$ for Case 7 of HSMR. }
	\label{case7:fig4}
\end{minipage}
\end{figure}

 The variation of the third mixing angle $\theta_{12}$ with respect to $M_{ee}$ is next
plotted in Fig.~\ref{case7:fig3}. The pattern obtained is similar to Case 6, with the lower and upper 
end of $\alpha$ giving the 3$\sigma$ global end points of $\theta_{12}$ respectively. 
The whole 3$\sigma$ global range of  $\theta_{12}$ is allowed, for $\alpha$ = 1.1. 
The value of $\theta_{12}$ at $\alpha$ = 2 is $36.6^\circ$.
 
Finally we plot the sum of the neutrino masses as a function of $M_{ee}$ in Fig.~\ref{case7:fig4}.   
For $\alpha$ = 0.895, range of $\Sigma m_i$ is $1.18-1.28$ eV corresponding to $0.391 \leq M_{ee} \leq 0.425$ eV.  
The range of $\Sigma m_i$ at $\alpha=1.1$ is  $1.09-1.18$ eV for $M_{ee} \leq 0.4$ eV.  At the upper allowed 
value of $\alpha=2.0$, it is $1.2-1.28$ eV for $0.4 \leq M_{ee} \leq 0.425$ eV.  
The sum of the neutrino masses is $1.27-1.38$ eV for $0.42 \leq M_{ee} \leq 0.452$ eV, in case of the uppermost value of $\alpha=2.09$.  
This range is not allowed by the GERDA limit.

Lastly as a completion, in order to give a clear picture of all the cases
discussed here along with their phenomenological consequences, we summarize our
results in Table~\ref{tab2}. The upper and lower ends of $\alpha$ allowed by 
the experiments for all the cases are presented along with the
corresponding values of masses and mixing angles of the neutrino sector.

\begin{table}[h]
\large
\begin{center}
\resizebox{\textwidth}{!}{\begin{tabular}{ |c|c|c|c|c|c|c|c|c|c|c|c|c|c|}
  \hline
                  & $\alpha$   &  \multicolumn{3}{ c| }{Masses at unification scale (eV)}  &  $\Sigma m_i$ (eV) & $\theta_{12}^\circ$ & $\theta_{13}^\circ$ 
                  & $\theta_{23}^\circ$& $\Delta m^2_{32}$ $(10^{-3} {\rm eV}^2)$& $\Delta m^2_{21}$ $(10^{-5} {\rm eV}^2)$&$M_{ee}$ (eV)
                  & Lightest neutrino mass: $m_1$ (eV)& R\\          
                  \cline{3-5}
                      &      & $m_1$ & $m_2$ &    $m_3$             &           &                 &          
                      &      &       &       &                      &           &  \\ \hline
                      &      &       &       &      &     &     &     &      &       &       &   &   &  \\ 
HSMU hypothesis   &   1    & 0.45700-0.47686   & 0.46-0.48      &0.51573-0.53817      & 1.16-1.2
                           & 30.59-36.81    & 7.63-8.34      &49-52.3 &2.21-2.45      &6.99-8.18         &0.384-0.4       &0.38397      &1.5-1.8 \\  
                    &      &       &       &      &     &     &     &      &       &       &   &   &  \\         
       \hline                
                    &      &       &       &      &     &     &     &      &       &       &   &   &  \\ 
 \multirow{2}{*}{Case 1}   &  0.902   &    0.44214-0.47791    &  0.445-0.481     & 0.49891- 0.53927        &1.12-1.2  
                                       & 30.6        & 7.62& 52.29         & 2.20-2.55        & 6.99-8.18       & 0.365-0.4      & 0.40155      & 1.39-1.42 \\
  \cline{2-14}
                          &  1.28   &  0.47573 &  0.47849 &  0.53772  &     1.2   & 32.82     &8.16      & 44.05       &2.20      & 7.196     & 0.4  & 0.39968  &2.1\\ 
                   &      &       &       &      &     &     &     &      &       &       &   &   &  \\        
  \hline
                   &      &       &       &      &     &     &     &      &       &       &   &   &  \\ 
 \multirow{2}{*}{Case 2} &  $0.45$ & 0.41 - 0.445& 0.411 - 0.448& 0.464 - 0.54 &1.16-1.2 & 30.8& 7.65& 52.3&2.20-2.57& 7.00-8.18&0.38-0.42 & 0.345-0.375 &1.70-1.95\\
  \cline{2-14}
  &  $2.5$ &0.415 - 0.444&  0.416 - 0.445& 0.468 - 0.5 & 1.05-1.12 & 31.89-36.6 &7.92-9.88 & 37.7-48&2.20-2.53& 7.00-8.18 &0.342-0.378 & 0.348-0.373& 1\\
              &      &       &       &      &     &     &     &      &       &       &   &   &  \\ 
  \hline  
              &      &       &       &      &     &     &     &      &       &       &   &   &  \\ 
  \multirow{2}{*}{Case 3}           &  0.89    & 0.4771& 0.4800& 0.5384 & 1.2  & 30.6  & 8.56 & 52.3 & 2.20  & 6.99 &0.4 &0.4009 &1.44 \\
 \cline{2-14}
                                   & 1.52     & 0.3693-0.3990&  0.3730-0.4030&  0.4179-0.4515  & 0.93-1.01   
                                   &30.78  &7.62 &52.3   &2.20-2.56 &6.99-8.18 &  0.31-0.336 &0.3102-0.3352 & 2.16-2.21 \\                        
  &      &       &       &      &     &     &     &      &       &       &   &   &  \\ 
  \hline
  &      &       &       &      &     &     &     &      &       &       &   &   &  \\ 
  \multirow{2}{*}{Case 4}  & 0.92  & 0.4550-0.4749&  0.458-0.478&  0.5135-0.5359 & 1.15-1.20 
                           & 30.6  & 7.65 & 52.28 & 2.30-2.50& 6.99-8.01 & 0.382-0.40& 0.3823-0.3990&1.463-1.479\\
  \cline{2-14}             & 1.67  & 0.4819-0.5018&  0.485-0.505 &  0.5460-0.5685  & 1.22 -1.27
                           & 30.59-30.66 & 9.59 & 37.71-37.76 & 2.29-2.49 & 7.08-8.18 & 0.405-0.422 &  0.4049-0.4216 & 2.82-2.86\\
  &      &       &       &      &     &     &     &      &       &       &   &   &  \\ 
  \hline
  &      &       &       &      &     &     &     &      &       &       &   &   &  \\ 
\multirow{2}{*}{Case 5}    &  0.89    &0.475-0.505& 0.478-0.508& 0.536-0.570  & 1.20-1.28 
                                      & 30.72  & 8.40 & 52.23 & 2.20-2.48  & 7.00-7.97 &0.4-0.42 &0.399-0.424 &1.465-1.485 \\
 \cline{2-14}
                          & 3.18      &0.255-0.265& 0.26-0.27&  0.29-0.30   & 0.65-0.67   
                          &30.62  &7.62 &52.3 &2.27-2.45 &6.99-8.12 &0.214-0.223 &0.214-0.222 & 3.542-3.672 \\                        
                          &      &       &       &      &     &     &     &      &       &       &   &   &  \\ 
  \hline
  &      &       &       &      &     &     &     &      &       &       &   &   &  \\ 
\multirow{2}{*}{Case 6}  &  0.86    &0.47-0.50& 0.474-0.504& 0.531-0.565  & 1.19-1.27
                                   & 30.61  & 7.62 & 52.3 & 2.20-2.48  & 6.99-8.17 &0.396-0.421 &0.3961-0.4211&1.241-1.268 \\
 \cline{2-14}
                          & 2.11     &0.494-0.533& 0.502-0.542& 0.567-0.612 & 1.26-1.36   
                          &34.57-35.02  &9.64-9.76 &37.70-37.8 &2.20-2.57 &6.99-8.18 &  0.416-0.450 &0.414-0.448 & 246.942-251.842 \\ 
  &      &       &       &      &     &     &     &      &       &       &   &   &  \\ 
  \hline
  &      &       &       &      &     &     &     &      &       &       &   &   &  \\ 
 \multirow{2}{*}{Case 7}  &  0.895     & 0.4683-0.5051& 0.4710-0.5080& 0.5282-0.5696  & 1.185-1.22  
                                       & 30.68   &7.63   & 52.28  & 2.20-2.31 & 6.99-8.18 & 0.393-0.4 &0.3935-0.4044 & 1.34-1.37   \\
 \cline{2-14}
                          & 2         & 0.4738& 0.4800& 0.5413  &1.18 & 36.6 &9.44 &37.8 &2.20  & 7.248 & 0.39  &0.3977 & 81.7-82.11 \\     
                          &      &       &       &      &     &     &     &      &       &       &   &   &  \\ 
                          \hline                        
\end{tabular}}
\end{center}
\caption{The allowed predictions for HSMU and the different cases of the HSMR for lower and upper allowed values of $\alpha$, Eqs.~(\ref{case1}-\ref{case7}).}
 \label{tab2}
\end{table}    

\subsubsection{The effects of new physics within type-\rom{1} seesaw framework}
In this sub-section, we discuss the possible effects of the new physics which could generate dimensional-5 operator.  For sake of illustration we take 
type-\rom{1} seesaw as the mechanism responsible for generating the effective dimensional-5 operator. The RG equations for type-\rom{1} seesaw can be found in the Ref.~\cite{Antusch:2005gp}.  In table \ref{tab3}, we show results for different cases of HSMR with a seesaw scale equals $4 \times 10^{13}$ GeV which is slightly lower than the scale of the dimensional-5 operator. We have chosen this scale to demonstrate and differentiate the effects of type-\rom{1} mechanism from the results obtained using dimesional-5 operator. Since the major part of RG magnification happens at scales much lower than the typical seesaw scales, the results obtained from dimension-5 operator and those obtained form type-I seesaw mechanism are not very different over a large range of parameters\cite{Mohapatra:2003tw,Mohapatra:2005gs,Mohapatra:2005pw,Agarwalla:2006dj,Abbas:2013uqh, Abbas:2014ala}. Our results for type-I seesaw are as shown in Table \ref{tab3}. To take the effect of type-I seesaw thoroughly,  we have done the RG running from the GUT scale ($2 \times 10^{16}$) GeV to the seesaw scale using the full RG equation for type-I seesaw mechanism. Below the seesaw scale the right handed neutrinos are integrated out and as before, the subsequent RG running is done with effect dimension-5 operator. Here we will like to remark that since in this case the RG running is done from a higher scale i.e. GUT scale so we expect small deviations from the previous results primarily due to the larger range of RG running. The dependence of RG evolution on the chosen high scale is studied in \cite{Abbas:2013uqh, Abbas:2014ala}.  

We observe from comparing Tables \ref{tab3} and \ref{tab2} that for case 1 of the HSMR,  the lower allowed end of $\alpha$ effectively does not change. As expected, there are slight changes in the value of the observables. For example, the $M_{ee}$ decreases and reaches to the value $0.349$ eV compare to the prediction given in table \ref{tab2}.  Similar observation for the mass of the lightest neutrino. The upper end of $\alpha$ changes after introducing seesaw scale, primarily due to increased RG running range.  In Table \ref{tab2}, the upper allowed end for case 1 is $1.28$.  As we observe in Table \ref{tab3}, it is now $1.71$ and parameter space is bit expanded. However, there is no significant qualitative change in our results which are same as before.  

Similarly, for case $2$, one can observe from the Tables  \ref{tab3} and \ref{tab2} that the lower end of $\alpha$ does not change much. The results are stable and similar to Table \ref{tab2}.  The upper end of $\alpha$ changes slightly and is 2.59 now.  Again as before, there is no significant qualitative change in our results which are same as before.  

In case $3$ of Table \ref{tab3}, the lower value of $\alpha$ has shifted a bit from that of Table\ref{tab2}, that is from  $0.89$ to $0.75$, but the higher value remains intact being $1.52$. The parameters also cover more or less the same span as before and show a stable situation .

The observed pattern in case $4$ is same as the case 1, as we see that the lower end does not change while the upper end changes 
from $1.67$ to $1.76$ after the inclusion of type-I seesaw (see Table \ref{tab3} and Table \ref{tab2} for comparison). The value of $M_{ee}$ and 
the mass of lightest neutrino decreases compare to the values given in Table \ref{tab2} and attain the values $0.3508$ eV and $0.3506$ eV, respectively.

The observed pattern in case 5 is also same as that obtained in Table~\ref{tab2}. Comparing the results with Table~\ref{tab2}, we find that the value of  $\alpha$ at the lower end changes slightly. The lower end saturates the bound for $M_{ee}$, whereas at the upper end the value of $M_{ee}$ turns out to be 0.205 eV, which is slightly smaller than the value quoted in Table~\ref{tab2}.

In case 6, the upper value of $\alpha$ changes very slightly compared to Table~\ref{tab2}, whereas the lower value remains the same. In this case both the lower and upper end saturates the bound for $M_{ee}$. 
 
As expected, the results for case $7$ are stable as can be observed from  Tables~\ref{tab2} and~\ref{tab3}, where the lower end of $\alpha$ is now $0.896$ instead of $0.895$ and the higher end remains the same, namely $\alpha=2$.
 
\begin{table}[h]
\large
\begin{center}
\resizebox{\textwidth}{!}{\begin{tabular}{ |c|c|c|c|c|c|c|c|c|c|c|c|c|c|}
  \hline
                  & $\alpha$   &  \multicolumn{3}{ c| }{Masses at unification scale (eV)}  &  $\Sigma m_i$ (eV) & $\theta_{12}^\circ$ & $\theta_{13}^\circ$ 
                  & $\theta_{23}^\circ$& $\Delta m^2_{32}$ $(10^{-3} {\rm eV}^2)$& $\Delta m^2_{21}$ $(10^{-5} {\rm eV}^2)$&$M_{ee}$ (eV)
                  & Lightest neutrino mass: $m_1$ (eV)& R\\          
                  \cline{3-5}
                      &      & $m_1$ & $m_2$ &    $m_3$             &           &                 &          
                      &      &       &       &                      &           &  \\ \hline
                      &      &       &       &      &     &     &     &      &       &       &   &   &  \\ 
 \multirow{2}{*}{Case 1}   &  0.903   &   0.46547    &  0.469     & 0.53833        &1.04 
                                       & 31.06        & 7.63& 52.30         & 2.209        & 8.00       & 0.349     & 0.34419     & 1.48 \\
  \cline{2-14}
                          &  1.71   &  0.54038-0.54199 &  0.545-0.546 &  0.62793-0.62968  &    1.203- 1.205  & 30.82 - 33.38     &8.68 - 9.87      & 37.75 - 40.25       &2.201-2.26     & 7.05-7.96     & 0.399-0.4 & 0.39815  &4.89\\ 
                   &      &       &       &      &     &     &     &      &       &       &   &   &  \\        
  \hline
   &      &       &       &      &     &     &     &      &       &       &   &   &  \\ 
 \multirow{2}{*}{Case 2} &  $0.47$ & 0.486067-0.486077 & 0.49 & 0.56241 - 0.562434 &1.08-1.0821& 30.66-30.92& 7.62& 52.15-52.19&2.2-2.23& 7.65-8&0.359 & 0.359022-0.359308 &1.96-1.99\\
  \cline{2-14}
  &  $2.59$ &0.478-0.480 &  0.481- 0.483 & 0.55 - 0.554 & 1.06 & 30.68-31.2&9.1-9.16 & 42.96-43.51&2.5-2.56& 7-7.24 &0.354 & 0.353-0.354& 1\\
              &      &       &       &      &     &     &     &      &       &       &   &   &  \\ 
  \hline  
&      &       &       &      &     &     &     &      &       &       &   &   &  \\ 
 \multirow{2}{*}{Case 3} &  $0.75$ & 0.541797 - 0.542934 & 0.545 - 0.546 &  0.625782 - 0.627092 &1.20171 - 1.20407& 30.6074 - 32.7328&  8.74619 - 8.95191& 51.3341 - 52.2853&2.2-2.24& 6.99-8.16&0.399 - 0.4 & 0.399 - 0.4 &1.351 - 1.374\\
  \cline{2-14}
  &  $1.52$ &0.392477 & 0.3971 & 0.455448 & 0.879869 & 30.7133& 7.63895 & 52.2976&2.28& 7.05 & 0.291584 & 0.291263 &  2.52\\
              &      &       &       &      &     &     &     &      &       &       &   &   &  \\ 
  \hline  
&      &       &       &      &     &     &     &      &       &       &   &   &  \\ 
   \multirow{2}{*}{Case 4}   &  0.92   & 0.4744  & 0.478   & 0.5487  & 1.06  & 30.98 & 7.64   & 52.27   &  2.266  & 8.081    & 0.3508   & 0.3506   & 1.55  \\
  \cline{2-14}
                            &  1.76   & 0.5484-0.5544  & 0.552-0.558   & 0.6366-0.6435  & 1.22-1.23  & 36.09-36.55 & 9.80-9.87   & 37.71-37.83   &  2.204-2.207  & 7.101-7.837    & 0.4045-0.4088   & 0.4039-0.4083   & 4.7-4.85 \\ 
                   &      &       &       &      &     &     &     &      &       &       &   &   &  \\        
  \hline
&      &       &       &      &     &     &     &      &       &       &   &   &  \\         
       
\multirow{2}{*}{Case 5}    &  0.80    &0.537& 0.540& 0.621  & 1.191 
                                      & 32.68  & 7.79 & 49.17 & 2.20  &8.09 &0.396 &0.396 &1.532 \\
 \cline{2-14}
                          & 3.18      &0.274& 0.280&  0.321   & 0.623   
                          &30.65  &7.63 &52.3 & 2.45 & 7.22 &0.205 &0.205 & 4.582 \\                        
                          &      &       &       &      &     &     &     &      &       &       &   &   &  \\ 
  \hline
  &      &       &       &      &     &     &     &      &       &       &   &   &  \\ 
\multirow{2}{*}{Case 6}  &  0.86    &0.497& 0.500& 0.574  & 1.104
                                   & 31.09  & 7.62 & 52.3 & 2.20  & 7.86 &0.367 &0.367&1.301 \\
 \cline{2-14}
                          & 2.14     &0.544& 0.555& 0.642 & 1.22   
                          &35.06  &9.88 &37.77 &2.21 &8.18 &  0.402 &0.400 & 957 \\ 
  &      &       &       &      &     &     &     &      &       &       &   &   &  \\ 
  \hline
   &      &       &       &      &     &     &     &      &       &       &   &   &  \\ 
\multirow{2}{*}{Case 7}  &  0.896    & 0.49& 0.50& 0.57  & 1.104
                                   & 30.65 - 31.41   & 7.62 & 52.26 & 2.24  & 7.1-8.1 &0.36&0.36&1.42 \\
 \cline{2-14}
                          & 2     &0.5& 0.49-0.51& 0.58 & 1.12   
                          &34.30 - 36.69 &9.52 - 9.84 &37.80 - 38.07&2.28 - 2.47 &7.08 - 7.21 &  0.37 &0.37 & 155.5 \\ 
  &      &       &       &      &     &     &     &      &       &       &   &   &  \\ 
  \hline
                                                                
\end{tabular}}
\end{center}
\caption{The allowed predictions for the different cases of the HSMR for lower and upper allowed values of $\alpha$, Eqs.~(\ref{case1}-\ref{case7}) within the framework of type-\rom{1} seesaw for sea-saw scale $4 \times 10^{13}$ GeV.  It should be noted that  RG evolution begins from GUT scale which is $2 \times 10^{16}$ GeV.}
 \label{tab3}
\end{table}    

Thus, as expected the results obtained with the framework of type-I seesaw mechanism are qualitatively same as those obtained using only dimension-5 effective operator. The general observation is that the absolute mass scale is decreasing due to the RG evolution starting from GUT scale ($2 \times 10^{16}$ GeV) which is higher than the scale of dimensional-5 operator.  This leads to a slight change in the allowed end of  $\alpha$ that is constrained by the observable $M_{ee}$.  That is why, we observe a slight change in the values of mixing angles.  We remark that if the high scale from where RG evolution begins, is chosen to be $10^{14}$ GeV with a seesaw scale equals $4 \times 10^{13}$ GeV, we recover the results  obtained with dimensional-5 operator which is naturally expected.

\section{Theoretical models for high scale mixing relations}
\label{sec5}
In this section, we address the theoretical implementation of HSMR hypothesis from the model building point of view.  The only aim of this section is to illustrate that the HMSR hypothesis can be simply realized  in models based on flavor symmetries.  We follow the same line of argument as presented in Ref. \cite{Mohapatra:2003tw}. 

Now we discuss a simple realization of HSMR relations using abelian  $Z_7$ flavor symmetry.
To realize the HSMR relations we add three $SU(2)$ triplet 
scalars $\xi_i$; $i = 1, 2, 3$ to the particle content of MSSM. The 
smallness of neutrino masses can then be explained by the type-\rom{2}
seesaw mechanism. Let the quarks and leptons and scalars transform 
under $Z_7$ as follows

\begin{eqnarray}
 Q^1_L & \sim & 1 , \quad Q^2_L \, \sim \, \omega , \quad Q^3_L  \, \sim \, \omega^3 , \quad u_R, d_R \, \sim \, 1 , \quad c_R, s_R \, \sim \, \omega , \quad t_R, b_R \, \sim \, \omega^3  \nonumber \\
 L^1_L & \sim & 1, \quad  L^2_L \, \sim \, \omega , \quad L^3_L \, \sim \, \omega^3 , \quad e_R \, \sim \, 1 , \quad \mu_R \, \sim \, \omega , \quad \tau_R \, \sim \, \omega^3 \nonumber \\
 H_u, H_d & \sim & 1 , \quad \xi_1 \, \sim \, 1 , \quad \xi_2 \, \sim \, \omega^2 , \quad \xi_3 \, \sim \, \omega^6
 \label{z7rep}
\end{eqnarray}

where $\omega = e^{\frac{2 \pi i}{7}}$ is the seventh root of 
unity. In the above equation $Q^i_L, L^i_L$; $i=1,2,3$ are the 
quark and the lepton doublets respectively whereas $u_R, d_R,$ 
$c_R, s_R,$ $t_R, b_R, e_R, \mu_R, \tau_R$ are the  quark and the 
charged lepton singlets. Moreover, $H_u, H_d$ are the two scalar 
doublets required to give mass to the up and down type quarks 
respectively. 

It is easy to see from Eq.~(\ref{z7rep}) that the $Z_7$ symmetry 
leads to diagonal mass matrices for both the quarks and the leptons 
leading to $U_{CKM} = U_{PMNS} = I$. To obtain the  realistic CKM 
and PMNS matrices as well as the HSMR relations,  we allow for 
small $Z_7$ symmetry breaking terms as done in 
Ref. \cite{Ma:2002yp} albeit for $A_4$ symmetry. Such corrections 
can arise from soft supersymmetry breaking sector as shown 
in Ref. \cite{Babu:1998tm, Babu:2002dz, Gabbiani:1996hi}. Allowing 
for symmetry breaking terms of the form $|h'''_i|<< |h''_i| << |h'_i| << |h_i|$ where $h_i$ are the terms invariant under $Z_7$ 
symmetry and $h'_i$, $h''_i$ and $h'''_i$ are the symmetry breaking 
terms transforming as $\omega$, $\omega^2$ and $\omega^3$ 
respectively under $Z_7$ symmetry.  Following the approach 
of Ref. \cite{Ma:2002yp} one can then easily realize the HSMR 
relations. Here we want to emphasize that owing to quite different 
masses of quarks and charged leptons, this analysis will in general 
lead to HSMR relations and not to HSMU relations. To obtain HSMU 
relations from such an approach one has to invoke a symmetry or 
mechanism to ensure that the symmetry breaking terms are exactly 
same in both quark and lepton sectors.

Before, ending this section we would like to further remark that 
although we have only discussed realization of HSMR relations 
through the $Z_7$ symmetry, they can also be quite easily and 
naturally realized using other flavor symmetries and also using 
other type of seesaw mechanisms. For example, one can also realize 
HSMR relations within the framework of type-\rom{1} seesaw mechanism  
using $Z_7$ symmetry. For this, instead of adding triplet scalars 
we add three right handed neutrinos which transform as $N_{1R} \sim 1, N_{2R} \sim \omega, N_{3R} \sim \omega^3$ under the $Z_7$ 
symmetry. We also add three heavy singlet scalars 
$\phi_i$; $i = 1,2,3$ transforming as 
$\phi_i \sim 1, \phi_2 \sim \omega^5, \phi_3 \sim \omega$ under $Z_7$ symmetry. Following computations analogous 
to those done above, one can again easily obtain the HSMR 
relations.  Thus, it is clear that HSMR relations are very natural 
and can be easily realized using discreet flavor symmetries. In 
this work we do a model independent analysis of the consequences of 
the HSMR relations assuming they are realized at the high scale by 
appropriate flavor symmetries.

\section{Summary}\label{sec6}
The very small mass of the neutrinos along with a large mixing among them is arguably 
a remarkable observation. This phenomenon is starkly different from the mixing in the
quark sector which is small in the SM. The quest to understand the origin of a large mixing 
among the neutrinos and a small mixing among the quarks has led to many interesting 
theoretical ideas.  Many beyond the standard model scenarios have been constructed,
trying to understand the major theoretical challenge posed by the neutrino mixing. 
GUT theories with the quark-lepton unification have been extensively used in the 
literature to understand the neutrino sector at low energies. The postulated HSMR in 
this work is another effort to understand this extraordinary observation of neutrino mixing. We have shown from a model building point of view, how the HSMR can be naturally realized
using different flavor symmetries and seesaw mechanisms. We have first considered the most general relation among the leptonic and the quark mixing angles, with different proportionality constants ($\alpha_i$).  We then list the different possible cases which arise, for the maximum and minimum allowed values of $\alpha_i$. It is found that for the allowed range of $\alpha_i$, $M_{ee}$ is between 0.35 eV - 0.4 eV. The future experiments from GERDA will severely constrain these scenarios. We then look into more simplified cases to have a clear physical picture and therefore consider the $\alpha_i$ to be equal for the three generations and vary $k_i$ to 0 or 1.  We then list the seven possible
ways the quark and the leptonic mixing angles can be proportional to each other 
(cf. Eqs.~(\ref{case1}-~\ref{case7})).  It is remarkable that these relations naturally
explain the difference between  $V_{CKM}$ and $U_{PMNS}$ at the low scale.  Furthermore,
the QLC relation and the observation in Eq.~(\ref{eq2}) can be easily recovered by 
these relations.

We have thoroughly investigated the implications and the phenomenological consequences of all the 
possible cases, taking into account the latest experimental constraints. The whole analysis has been
done with the assumption of normal hierarchy and QD mass pattern. In general, we have discovered three new correlations among $\Delta m_{32}^2$, $M_{ee}$, $\theta_{12}$ and sum of neutrino masses.  These correlations are not investigated in previous studies.

We first discuss about the HSMU scenario, which is a  special case of all the HSMR scenarios in the $\alpha$ =1 limit. 
The behavior of the neutrino masses and the mixing parameters at the low energy scale is discussed in
detail for all the cases in HSMR with the value of $\alpha$ deviating from unity in the allowed range.
The allowed range of $\alpha$ is bounded  by the recent experimental results listed in 
Table~\ref{tab1} and the upper limit on  $M_{ee}$ provided by GERDA~\cite{Agostini:2013mzu}.
It is seen that for all the cases except Case 2, the $M_{ee}$ constraint from the GERDA results
in either upper (Cases 1, 4, 6 and 7) or lower (Cases 3 and 5) limit of $\alpha$. 
Otherwise the allowed value of $\alpha$ is mostly constrained by the $3 \sigma$ global 
limits on neutrino mixing parameters. 

An interesting feature is observed in Case 2, where the lower end is constrained by 
the $3 \sigma$ global limits of neutrino mixing parameters but the upper end is 
constrained by the value of the ratio $R$, which contributes through threshold corrections.
We have worked here in the inverted hierarchy scheme in the charged slepton sector, forcing
the ratio to be either greater than or equal to one.  A common behavior has been 
observed for all the cases, where we always find a strong correlation between 
$\theta_{23}$ and $\theta_{13}$, for all the allowed values of $\alpha$ except at the
end points which corresponds to a point in the $\theta_{23}-\theta_{13}$ plane. 
It is also seen that among all the experimental constraints $M_{ee}$ is the most 
interesting one as it mostly constrains the different cases as well as differentiates 
among them. If in the future the upper limit from GERDA goes down to 0.35 eV, then HSMU,
Case 1, Case 4, Case 6 and Case 7 will be ruled out. The ones who will survive will be 
Cases 2, 3 and 5 which allows $M_{ee}$ as low as 0.2 eV but with the value of $\alpha > 1$.  
The constraint on $M_{ee}$ can automatically be reverted to the sum of the neutrino masses. 
It will show a similar behavior while discriminating the various cases. We also notice
that if we take into account the GERDA limit of 0.4 eV, then the allowed range of 
$\alpha$ in Cases 1, 3 and 5 is limited to a small region in the 
$\theta_{23}-\theta_{13}$ plane (Figs.~\ref{fig6},~\ref{case3:fig2},~\ref{case5:fig2}). 
Therefore these cases along with HSMU will be ruled out, if the best fit value of 
$\theta_{23}$ becomes less than $44^\circ$ or that of $\theta_{13}$ becomes greater 
than $8.55^\circ$  in the future. We further see that Cases 2 and 5 can survive longer, 
and the region of $M_{ee} = 0.2$ eV is beyond the sensitivity of GERDA which is maximun $0.3$ eV.  The region of  $M_{ee} = 0.2$ eV will easily be probed by KATRIN\cite{Drexlin:2013lha} since $m_\beta$ is approximately identical to $M_{ee} $ in this work.  Here, we pause to comment on the cosmological limit
on the sum of the neutrino masses \cite{Ade:2015xua}.  Our predictions in all cases except 
Case 5 are slightly above the upper cosmological bound of $0.72$ eV.  As commented earlier, 
this bound is model dependent.  Hence, it is preferred to test predictions of this 
work in a laboratory based experiment, like GERDA~\cite{Agostini:2013mzu}.

We also observe that Cases 3 and 5 show similar behavior, this is mainly because both consider 
the framework, where the neutrino mixing angle $\theta_{23}$ is equal to the quark mixing angle 
by a proportionality constant $\alpha \theta_{23}^q$. Although Case 5 also has the condition 
of $\theta_{13}~=~\alpha~\theta_{13}^q$, but at the GUT scale $\theta_{13}^q \ll \theta_{23}^q$, 
therefore the effect of $\theta_{23}^q$ dominates. The same pattern can be observed for Case 1 
and Case 4, explained through the same argument, $\theta_{13}^q \ll \theta_{12}^q$ at the GUT scale.
Continuing the same argument as expected we find that Case 7 displays similar behavior as Case 6.
The effect of the neutrino mixing angle $\theta_{13}$ being proportional to the quark mixing
has many interesting results, as it leads to the most optimistic case. However once the other 
angles become proportional, this effect is subdued. Finally we note that all these 
interpretations have been done with the assumption that the Dirac and the Majorana phases of the PMNS 
matrix are zero and phenomenological consequences can change  with nonzero phases.  The overall scenario 
depicting a quark-lepton symmetry at a high scale  through HSMR  can be narrowed down to a particular case or completely ruled out, 
only from the future improved experimental constraints. These 
constraints can be from the neutrinoless double beta decay \cite{Agostini:2013mzu}, or the LHC constraints on the SUSY spectrum.

The different scenarios of the HSMR can be discriminated through measurement of various observables like $M_{ee}$ and by precise determination of the values of the mixing angles, particularly $\theta_{13}$ and $\theta_{23}$ mixing angles. As we have shown in the figures for various cases as well in the tables, the allowed ranges for $M_{ee}$ and the angles are different for different cases and a precise determination of these observables can be used as a way to distinguish various cases of HSMR. In addition to neutrino observables one can also use other process like lepton-flavor violation to distinguish the different allowed cases.  The mass-splitting in the charged-slepton sector is given by the ratio $R = \frac{M_{\tilde e}}{M_{\tilde \mu, \tilde \tau}}$. We observe from tables  \ref{tab_mgr} and \ref{tab2}  that the ratio $R$ almost discriminate every scenario.  Hence the processes like $\mu \rightarrow e \gamma$, $\mu \rightarrow e e e $ and anomalous magnetic moment of the electron.  For example, the SUSY contribution to the anomalous magnetic moment of the electron directly depends on the ratio $R$ \cite{Stockinger:2006zn}.  The detail study of this aspect of the work is not possible in this paper. 

Furthermore, for sake of completion, we also present our results in the framework of the type-\rom{1} seesaw.  The aim is to show how the predictions do not change in any significant way and that the analysis done with effective dimension-5 operator is quite robust. As argued before, this is not surprising as the major part of RG magnification happens only at much lower scales closer to SUSY breaking scale. At such low scales, the effective dimension-5 operator provides a very good approximation to the high scale seesaw mechanisms.  The mass scale of the right-handed neutrinos is chosen $4 \times 10^{13}$ GeV which is close to the scale of the dimensional-5 operator.  We notice that parameter spcae increases very slowly as we decrease the scale of new physics primarily due to increased span of RG running.  However predictions do not change in any significant manner and are quite robust.

We also comment on a general theoretical view which is more general
than the HSMU hypothesis and the HSMR.    Assuming that at some high scale, both the mixing matrices
(CKM and PMNS) are approximately unit matrices, but some perturbation can mix the generations leading to the Wolfenstein form of the  mixing in both the quark and lepton sectors. This results in the mixing between  the first and the second generations  to be $\lambda$ (a small number of order 0.2), the second and the third generations mixing to be second order in $\lambda$ i.e. $\sin \theta_{23} \sim \lambda^2$ while the first and the third generations mixing to be third order order in $\lambda$ i.e. $\sin \theta_{13} \sim \lambda^3$. Now after RG evolution the CKM mixing angles do not change much but the PMNS mixing angles are dramatically magnified for the reasons already mentioned in this as well as our earlier papers \cite{Abbas:2014ala,Srivastava:2015tza,Abbas:2015aaa}. 

Finally, in short, crux of our paper is following.

\begin{itemize}
\item
We have proposed and studied the HSMR hypothesis which is a more general framework than the HSMU hypothesis.

\item
The HSMR hypothesis provides a very simple explanation of the observed large neutrino mixing.  The present and future neutrino experiments can easily test  predictions of our work.  If our predictions are confirmed by experiments, like GERDA, it would be a good hint of quark-lepton unification at high scale.  

\item
We observe that the HSMU hypothesis represents $\alpha=1$ limit of the HSMR hypothesis and is constrained by the lowest allowed value of $M_{ee} $ which is $0.384$ MeV.  Therefore, if the HSMU hypothesis is ruled out by experiments, like GERDA, the other HSMR cases with $\alpha \neq 1$ may survive and their confirmation would be itself a strong hint of the proportinality between quark and leptonic mixing angles which is the basis of the HSMR hypothesis.

\item
We have done a rigorous, thorough and comprehensive study with the HSMR hypothesis which does not exist in the literature. All results reported in the literature using the HSMU hypothesis, are very small subset of our results with the HSMR hypothesis presented in our paper.  Moreover, we have also thoroughly compared HSMR hypothesis with respect to the HSMU.  

\item
In our work, we have discovered new strong correlations among different experimental observables for every limit of the HSMR hypthesis. These correlations do not exist in the literature and are easily testable in present ongoing experiments.  For example, there is a strong correlation between $\Delta m^2_{32}$ and $M_{ee}$.  This correlation can be easily tested by GERDA  experiment.  There are two more such correlations namely among $\theta_{12}$, $M_{ee}$, $\Sigma m_i$ and $M_{ee}$  discussed in our work which are completely new and unexplored in the literature. 

\item
Furthermore, we have comprehensively studied a  strong correlation between $\theta_{23}$  and $\theta_{13}$ and predictions can be easily tested in present ongoing experiments.  This correlation was studied in a previous study in a specific limit.  Since we have done a comprehensive full parameter scan, this correlation has become a robust band now.  

\end{itemize}

\begin{acknowledgments}
It is a pleasure to thank Antonio Pich for his valuable suggestions and comments on the manuscript.  GA dedicates this paper to I. Sentitemsu Imsong.  The work of GA has been supported in part by the Spanish Government and ERDF funds from the EU Commission
[Grants No. FPA2011-23778, FPA2014-53631-C2-1-P No. and CSD2007-00042 (Consolider Project CPAN)].  The work of
SG is supported by Conselho Nacional de Desenvolvimento Cient\'{i}fico e Tecnol\'{o}gico (CNPq) Brazil grant 151112/2014-2.
\end{acknowledgments}

%%%%%%%%%%%%%%%%%%%%%%%%%%%%%%%%%%%%%%%%%%%%%%%%%%%%%%%%%%%%%%%%%%%%%%%%%%%%%%%%%%%%%%%%%%%%%%%%%%%%%%%%%%%%%%%%%%%%%%%%%%%%%%%%%%%%%%%%%%%%%%%%%%%%%%%%%%%%%%%%%%%%%%%%%%%%%%%%%%%%%%%%%%%%%%%%%%%%%%%%%%%%%%%%%%%%%%%%%%%%%%%%%%%%%%%%%%%%%%%%%%%%%%%%%%%%%%%

\end{document}